\documentclass{aa}

\usepackage[english]{babel}
\usepackage{graphicx}
\usepackage{txfonts}
\usepackage{twoopt}
\usepackage{longtable}
\usepackage{rotating}
\usepackage{verbatim}
\usepackage{enumitem}
\usepackage{color}
\usepackage{threeparttablex}
\usepackage{hyperref}
\usepackage{multirow}
\usepackage{booktabs}
\usepackage[table]{xcolor}

\graphicspath{{./}{figures/}}

\defcitealias{Desomma2021}{DS21}
\defcitealias{Desomma2020apj}{DS20b}
\defcitealias{Desomma2020mnras}{DS20a}
\defcitealias{Desomma2022}{DS22}
\defcitealias{Desomma2024}{DS24}
\defcitealias{Riess20248sigma}{R24}

\makeatletter
\newcommand{\DT@size}{}
\newcommand{\DT@cols}{}

\newenvironment{deluxetable*}[1]{%
  \begin{table*}\centering\def\DT@cols{#1}\renewcommand{\DT@size}{}%
}{%
  \end{tabular}\end{table*}%
}

\newcommand{\tabletypesize}[1]{\renewcommand{\DT@size}{#1}}
\newcommand{\tablehead}[1]{\DT@size\begin{tabular}{\DT@cols}#1}

\newcommand{\tablewidth}[1]{}

\newcommand{\tablenotetext}[2]{}
\newcommand{\tablecomments}[1]{}
\makeatother

\begin{document}

\title{Classical Cepheids in the JWST and Roman Era: A New Theoretical Framework and its Metallicity Dependence}
\titlerunning{Theoretical CC relations for JWST and Roman}

\author{
Giulia De Somma\inst{1,2}
\and
Marcella Marconi\inst{1}
\and
Vincenzo Ripepi\inst{1}
\and
Roberto Molinaro\inst{1}
\and
Ilaria Musella\inst{1}
\and
Silvio Leccia\inst{1}
\and
Teresa Sicignano\inst{3,4,1,2}
\and
Erasmo Trentin\inst{1}
\and
Mami Deka\inst{1}
\and
Laura Salmeri\inst{1,5}
}

\institute{
INAF -- Osservatorio Astronomico di Capodimonte,
Via Moiariello 16,
80131 Napoli, Italy
\and
Istituto Nazionale di Fisica Nucleare (INFN) -- Sezione di Napoli,
Complesso Universitario di Monte Sant'Angelo,
Edificio G, Via Cinthia,
80126 Napoli, Italy
\and
European Southern Observatory,
Karl-Schwarzschild-Strasse 2,
85748 Garching bei München, Germany
\and
Scuola Superiore Meridionale,
Largo San Marcellino 10,
80138 Napoli, Italy
\and
Dipartimento di Fisica ``Ettore Pancini'',
Università degli Studi di Napoli Federico II,
Via Cinthia 21, Edificio 6,
80126 Napoli, Italy
}

\date{July 21, 2026 / August 27, 2026}

\abstract
{Classical Cepheids are fundamental distance indicators that anchor the cosmic distance ladder through their Period-Luminosity, Period-Luminosity-Color, and Period-Wesenheit relations, while also providing valuable constraints on the properties of young stellar populations. Recent advances in space-based near-infrared observations have highlighted the need for theoretical calibrations of Classical Cepheid observables in modern photometric systems optimized for precision distance-scale applications.}
{This work provides a homogeneous theoretical framework for Classical Cepheids in the JWST/NIRCam and Roman/WFI photometric systems and explores its implications for their use as standard candles in the near-infrared regime.}
{We exploit an extensive grid of nonlinear convective pulsation models transformed into the JWST/NIRCam and Roman/WFI systems. The models span a broad range of masses, effective temperatures, periods, metallicities, and assumptions on the efficiency of superadiabatic convection and on the adopted mass-luminosity relation. From these models, we derive mean magnitudes, colors, pulsation amplitudes, amplitude ratios, and new theoretical Period-Luminosity-Color, Period-Wesenheit, and metal-dependent Period-Wesenheit relations.}
{The theoretical Wesenheit relations are compared with recent JWST observations of Classical Cepheids in six SN Ia host galaxies from the SH0ES collaboration. We find excellent agreement between theoretical predictions and observations, with the largest systematic effects arising from the adopted mass-luminosity relation. Distance moduli derived from the theoretical relations are fully consistent with the empirical JWST determinations when a mildly overluminous mass-luminosity relation is adopted instead of the canonical one.}
{These results establish a self-consistent pulsation-based calibration for JWST and Roman observations and provide a theoretical framework for future precision studies of the local distance scale.}


\maketitle

\section{Introduction}
\label{sec:intro}

Classical Cepheids (CCs) have played a central role in astrophysics for more than a century, providing one of the most robust and widely used primary distance indicators in the local Universe. Their well-established correlations between pulsation period, luminosity, and color make them fundamental anchors of the extragalactic distance scale and a cornerstone of the cosmic distance ladder. In the context of precision cosmology, CCs remain essential for the calibration of secondary distance indicators, most notably Type~Ia supernovae, and therefore for local determinations of the Hubble constant \citep[][]{Cosmoverse2025wp}.

The advent of the \textit{James Webb Space Telescope} (JWST) has opened a new observational window for the study of CCs, particularly in crowded and distant extragalactic environments. Thanks to its unprecedented near-infrared (NIR) sensitivity and angular resolution, JWST significantly mitigates the effects of crowding and blending that affect optical and NIR observations with the \textit{Hubble Space Telescope} (HST). Early JWST observations of CCs in Type~Ia supernova host galaxies have demonstrated a substantial reduction in photometric scatter and have enabled stringent tests of potential systematic biases affecting HST-based distance measurements. In particular, detailed comparisons between JWST and HST CC photometry have shown excellent agreement in distance moduli, providing strong evidence that crowding-related systematics are not responsible for the current tension in local measurements of the Hubble constant \citep{Riess2024,Riess2025}.
At the same time, these results better clarify the present role of JWST in CC-based distance-scale studies. While the superior angular resolution and NIR performance of JWST represent a major advance for mitigating crowding and blending effects, current CC samples observed in Type~Ia supernova host galaxies remain largely limited by the availability of multi-epoch optical data, which are still primarily provided by HST. As a consequence, the number of CCs accessible to JWST observations is, at present, comparable to that of previous HST-based studies. In this sense, JWST observations should not be regarded as a replacement for HST CC surveys, but rather as a powerful and complementary resource. Their primary strength currently lies in providing high-precision NIR photometry and in enabling decisive cross-checks of systematic uncertainties affecting HST-based measurements. As JWST time-domain strategies evolve and additional multi-epoch programs become available, its role in expanding CC samples and improving statistical constraints is expected to grow further \citep{Riess2024,Riess2025}.

Looking ahead, the \textit{Nancy Grace Roman Space Telescope} (Roman) is expected to play a transformative role in time-domain astrophysics and in the study of variable stars. With its wide field of view, stable NIR photometry, and survey-driven observing strategy, Roman will enable the detection and characterization of large samples of CCs across nearby galaxies and stellar systems, substantially increasing the statistical power of CC-based investigations. In particular, Roman surveys are expected to provide homogeneous and well-sampled light curves over wide areas of the sky, opening new opportunities for population-level studies of CCs pulsation properties and distance indicators across a broad range of environments \citep{Roman_report}. However, the scientific return from Roman for CC studies will critically depend on the availability of robust, homogeneous theoretical predictions in the Roman photometric system. Unlike targeted observations, survey data products require direct and quantitative comparisons with theory, spanning instability strip boundaries, full-amplitude light curves, pulsation amplitudes, and distance-scale relations. Establishing such a theoretical framework in advance is therefore essential for the optimal exploitation of forthcoming Roman CC datasets.

In this context, it is timely to provide a comprehensive theoretical reference tailored to current JWST observations and fully compatible with future Roman data. This includes not only predictions for mean magnitudes and colors, but also detailed modeling of the full light-curve morphology and the resulting period-luminosity (PL), period-luminosity-color (PLC), and period-Wesenheit (PW) relations, together with their dependence on chemical composition. These theoretical counterparts are required to interpret the data products released by current and upcoming facilities in a physically consistent and homogeneous manner.

This paper aims to provide such a theoretical framework by presenting a homogeneous set of nonlinear convective pulsation models for CCs \citep{Desomma2020apj, Desomma2022} (hereafter DS20b and DS22, respectively), transformed into the JWT/NIRCam (hereafter JWST) and Roman/WFI (hereafter Roman) photometric systems, and by exploring their implications for the use of CCs as standard candles in the NIR regime. 

The structure of the paper is as follows. In Section~2, we present the adopted pulsation models. The bolometric light curves are transformed into the JWST and Roman photometric systems in Section~3, where details of the conversion procedure, the adopted photometric systems, and the effects on the Hertzsprung progression are discussed. The predicted pulsation amplitudes are analyzed in Section~4. In Section~5, we present the derived PLC and PW relations, together with their dependence on metallicity. An application of the JWST period-Wesenheit-metallicity (PWZ) relations to observational data is discussed in Section~6. Finally, the main results and concluding remarks are summarized in Section~7.

\section{Theoretical Framework and Pulsation Models}

To derive the first theoretical pulsation relations in the JWST and Roman photometric systems, we rely on an extensive set of nonlinear, time-dependent convective pulsation models for CCs. These models were originally presented and discussed in detail in \citetalias{Desomma2020apj} and \citetalias{Desomma2022}, and are based on a well-tested hydrodynamical approach that self-consistently follows the full-amplitude pulsation behavior of stellar envelopes \citep[][]{Stell1982, BonoStell1994}.

The model grid includes both fundamental (F) and first-overtone (FO) pulsators and was constructed by exploring the stability of pulsating models across a wide range of intrinsic stellar parameters. Four different chemical compositions were considered:
(i) $Z = 0.004$, $Y = 0.25$;
(ii) $Z = 0.008$, $Y = 0.25$;
(iii) $Z = 0.02$, $Y = 0.28$;
(iv) $Z = 0.03$, $Y = 0.28$.
These compositions are representative of environments spanning from metal-poor systems to super-solar metallicity regimes relevant for both Galactic and extragalactic CC populations.

For each chemical composition, stellar masses from $3$ to $11\,M_{\odot}$ were investigated, with a step of $1\,M_{\odot}$, for both F and FO pulsation modes. To account for uncertainties in the CC ML relation arising from non-canonical physical processes, such as core convective overshooting, rotation, and mass loss, multiple luminosity levels were explored at fixed mass \citep[][]{BonoTornambe2000}. Specifically, three luminosity prescriptions were adopted: a canonical level (case~A), based on evolutionary models neglecting non-canonical effects, and two brighter non-canonical levels (case~B and case~C), obtained by increasing the canonical luminosity by $\Delta\log(L/L_{\odot}) = 0.2$ and $0.4$~dex, respectively \citep[e.g.][]{Bono_ml}.

For each combination of chemical composition, mass, and luminosity level, model stability was investigated over a wide range of effective temperatures, approximately from $3600$ to $7200$~K, with a step of $100$~K. In addition, three values of the mixing-length parameter were adopted ($\alpha_{\rm ml} = 1.5$, $1.7$, and $1.9$) to explore the impact of the efficiency of superadiabatic convection. The mixing-length parameter enters the mixing-length theory formalism used to close the system of nonlinear hydrodynamical and convective equations \citep{Stellingwerf1982} and plays a crucial role in regulating the coupling between convection and pulsation.

Variations in $\alpha_{\rm ml}$ significantly affect both the topology of the instability strip and the pulsation amplitudes. Increasing the mixing-length parameter results in a systematic narrowing of the instability strip for all investigated chemical compositions, with the blue boundary shifting toward cooler temperatures by approximately $100$-$200$~K and the red boundary moving toward hotter temperatures by about $200$-$300$~K. This behavior reflects the increased efficiency of convective energy transport, which damps the pulsation driving mechanism, particularly in the cooler regions of the instability strip where the superadiabatic layers are more extended (DS20b; \citealt{Dicriscienzo2004, Fiorentino2007}). This behavior reflects the increased efficiency of convective energy transport, which damps the pulsation driving mechanism, particularly in the cooler regions of the instability strip where the superadiabatic layers are more extended \citepalias{Desomma2020apj}, \citep{Dicriscienzo2004, Fiorentino2007}.

The nonlinear equations were integrated until stable limit-cycle solutions were reached for each model. This approach ensures a physically consistent prediction of pulsation observables, including periods, light-curve morphology, amplitudes, and amplitude ratios. Such quantities are known to be sensitive to both the adopted ML relation and the efficiency of superadiabatic convection, and they provide additional constraints on pulsation models beyond mean magnitudes alone. This aspect is particularly relevant for NIR photometric systems, where reduced sensitivity to extinction and crowding shifts the dominant source of scatter in PW relations toward intrinsic physical effects.

Therefore, the resulting theoretical framework provides a homogeneous and robust basis for transforming bolometric light curves into the JWST and Roman photometric systems and for deriving pulsation relations suitable for precision distance-scale applications in the NIR regime.

\section{Transformation of Bolometric Light Curves to JWST and Roman Photometric Systems}

The non-linear hydrodynamical pulsation code adopted in this work provides, as one of the results, bolometric light curves describing the temporal variation of the stellar luminosity along the pulsation cycle. To enable direct comparison with observations and derive pulsation relations in specific passbands, these bolometric light curves are transformed into synthetic photometry in the JWST and Roman photometric systems. The transformation from bolometric to band-limited magnitudes is carried out following the procedure described by \citet{Chen2019}, using bolometric corrections and synthetic spectra based on the PHOENIX stellar atmosphere models, from which the magnitudes in the JWST and Roman filters are computed.
At each pulsation phase, the bolometric luminosity and effective temperature provided by the hydrodynamical models are coupled with static stellar atmosphere models to compute synthetic magnitudes in the selected photometric bands. For each phase, the stellar spectrum—obtained by interpolating within atmosphere grids as a function of $T_{\rm eff}$, $\log g$, and chemical composition—is convolved with the total system transmission function of the filter, including instrumental and detector responses. Repeating this process over the full pulsation cycle produces synthetic multi-band light curves.
This approach preserves the full non-linear behavior of the pulsation models, including phase-dependent temperature variations and asymmetries, and ensures internal consistency between the bolometric and multi-band light curves.

\subsection{JWST Photometric System}

For the JWST, we focus on the Near Infrared Camera (NIRCam) wide-band filters, which are particularly well suited for observations of CCs owing to their high sensitivity and the reduced impact of interstellar extinction. We consider a total of eight JWST wide filters spanning the near-infrared wavelength range from approximately 0.71 to 4.42~$\mu$m, namely F070W ($\lambda_{\rm c}=0.706~\mu$m), F090W ($\lambda_{\rm c}=0.904~\mu$m), F115W ($\lambda_{\rm c}=1.157~\mu$m), F150W ($\lambda_{\rm c}=1.504~\mu$m), F200W ($\lambda_{\rm c}=1.993~\mu$m), F277W ($\lambda_{\rm c}=2.769~\mu$m), F356W ($\lambda_{\rm c}=3.577~\mu$m), and F444W ($\lambda_{\rm c}=4.416~\mu$m). The filter transmission curves, from which the adopted central wavelengths were derived, were taken from the YBC Synthetic Photometry Database \citep{YBC}. These filters provide broad wavelength coverage and are routinely employed in current and planned extragalactic CC programs with the JWST. The use of NIR bands significantly mitigates reddening effects and reduces the sensitivity of pulsation properties to metallicity and temperature variations, making them particularly suitable for the derivation of PLC and PW relations.

Representative theoretical JWST light curves for F-mode CC models are shown in the left panel of Figure~\ref{fig:lc_F}. The corresponding FO-mode models are presented in Appendix~\ref{sec:lc} (Figure~\ref{fig:lc_FO}).

Consistent with previous theoretical investigations based on different photometric systems \citepalias{Desomma2020apj,Desomma2022,Desomma2024}, both the morphology and amplitude of the light curves depend on metallicity, with smoother light-curve shapes and smaller pulsation amplitudes at higher metal abundance. As expected, the pulsation amplitude also decreases systematically towards longer wavelengths.

\subsection{Roman Photometric System}

For the Roman Space Telescope, we adopt the full set of Wide Field Instrument (WFI) filters. The Roman photometric system provides broad NIR wavelength coverage and is specifically designed for wide-field, high-precision photometry of resolved stellar populations. In this work, we consider the seven Roman filters spanning the wavelength range from approximately 0.63 to 2.13~$\mu$m, namely F062 ($\lambda_{\rm c}=0.632~\mu$m), F087 ($\lambda_{\rm c}=0.874~\mu$m), F106 ($\lambda_{\rm c}=1.065~\mu$m), F129 ($\lambda_{\rm c}=1.288~\mu$m), F158 ($\lambda_{\rm c}=1.584~\mu$m), F184 ($\lambda_{\rm c}=1.841~\mu$m), and F213 ($\lambda_{\rm c}=2.133~\mu$m). The availability of multiple NIR bands, combined with Roman's wide-field capabilities, makes the WFI filter set particularly well suited for systematic studies of CC in nearby galaxies, enabling robust constraints on PLC and PW relations over large and homogeneous samples.

\begin{figure*}[t]
\centering
\setlength{\tabcolsep}{1pt}
\begin{tabular}{cc}
\includegraphics[width=0.28\textwidth]{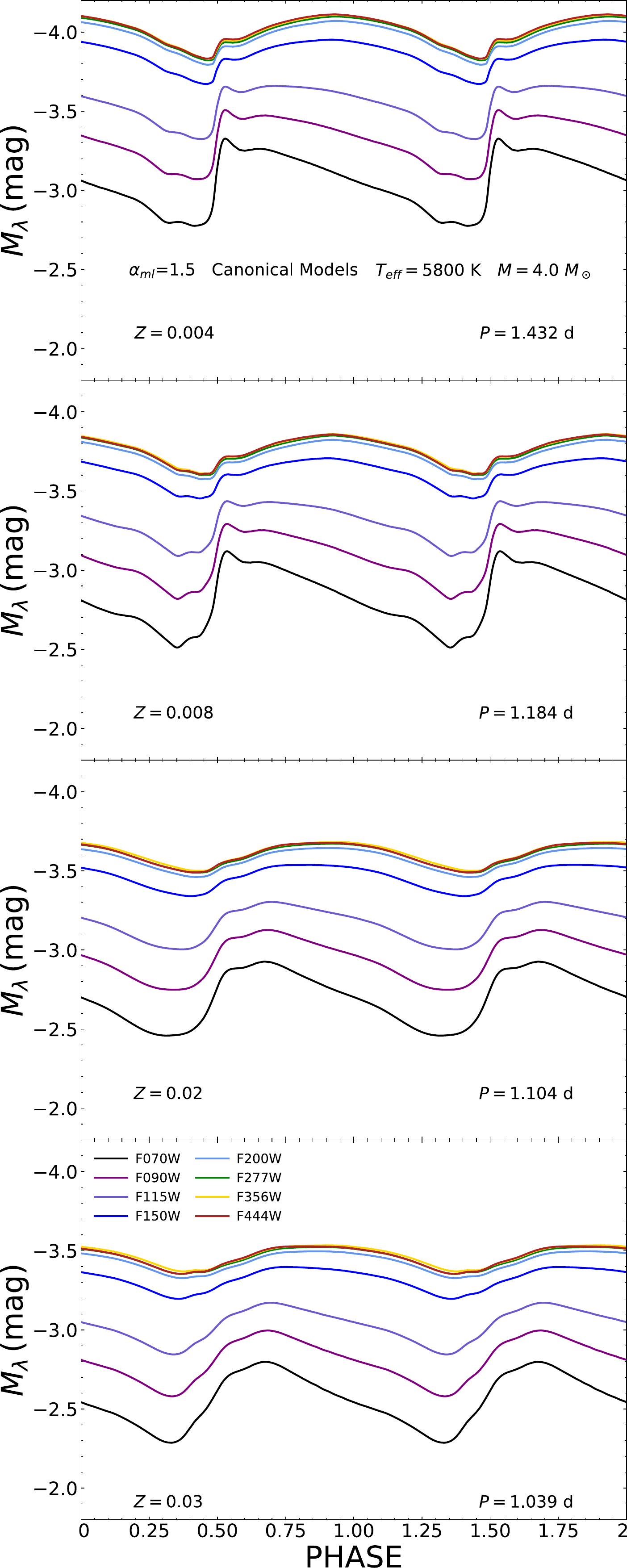} &
\includegraphics[width=0.28\textwidth]{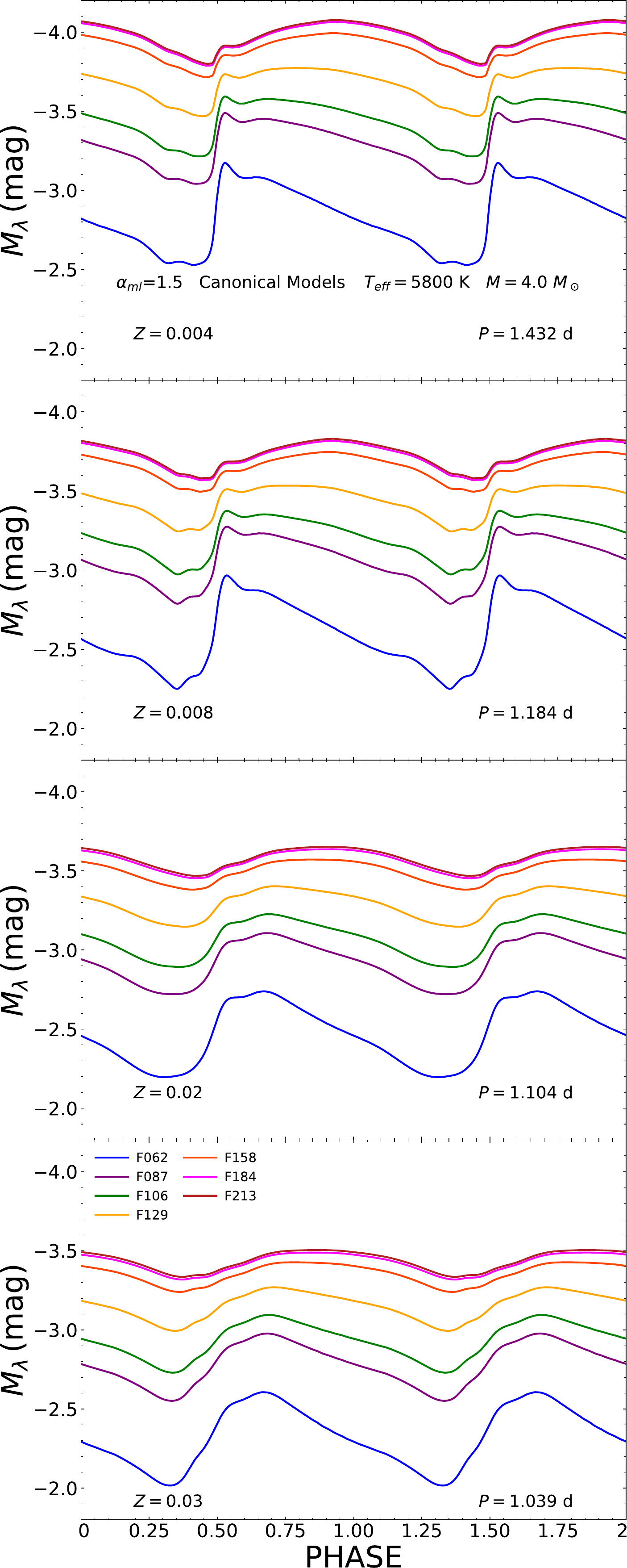}
\end{tabular}
\caption{
Representative theoretical light curves of F-mode CC models transformed into the JWST (left) and Roman (right) photometric systems. The models assume a canonical ML relation for a representative $4\,M_\odot$ model with $\alpha_{\rm ml}=1.5$, located near the center of the instability strip.
For the JWST light curves, the filters are F070W (black), F090W (purple), F115W (lavender), F150W (blue), F200W (light blue), F277W (green), F356W (yellow), and F444W (red). For the Roman light curves, the filters are F062 (blue), F087 (purple), F106 (green), F129 (orange), F158 (red), F184 (magenta), and F213 (brown). Different panels correspond to increasing metallicity from $Z=0.004$ to $Z=0.03$. The adopted chemical composition, effective temperature, stellar mass, and pulsation period are reported in each panel. The complete atlas of light curves in the JWST and Roman photometric systems, for the various assumptions concerning metallicity, ML relation, and super-adiabatic convection efficiency, is available upon request.
}
\label{fig:lc_F}
\end{figure*}

The right panel of Figure~\ref{fig:lc_F} shows the same representative F-mode theoretical models transformed into the Roman photometric system, allowing a direct comparison with the corresponding JWST light curves.

The overall behavior closely mirrors that found in the JWST bands, with pulsation amplitudes decreasing systematically towards longer wavelengths and a clear dependence on metallicity. Minor differences arise from the different wavelength coverage of the Roman filters, which extend to slightly shorter wavelengths than the JWST set.

\subsection{The Hertzsprung Progression in JWST and Roman Filters}
\label{sec:h_prog_text}

The Hertzsprung progression (HP) is a characteristic feature of F-mode light curves, observed over the period range of approximately 6-16 days. It is characterized by the appearance of a secondary bump, giving rise to the so-called bump Cepheids \citep[][]{Bono2002}. The phase of this feature changes systematically with period: it is located along the descending branch of the light curve for periods shorter than $\sim9$ days, shifts toward phases close to maximum light for periods between $\sim9$ and $\sim12$ days, and moves to earlier phases at longer periods. The centre of the progression is commonly defined as the period at which the bump occurs closest to maximum light. This behaviour is generally interpreted as the signature of the resonance between the fundamental mode and the second overtone, i.e. $P_2/P_0\simeq0.5$ \citep[see][]{Marconi2024}.

\begin{figure*}[t]
\centering
\includegraphics[width=0.9\linewidth]{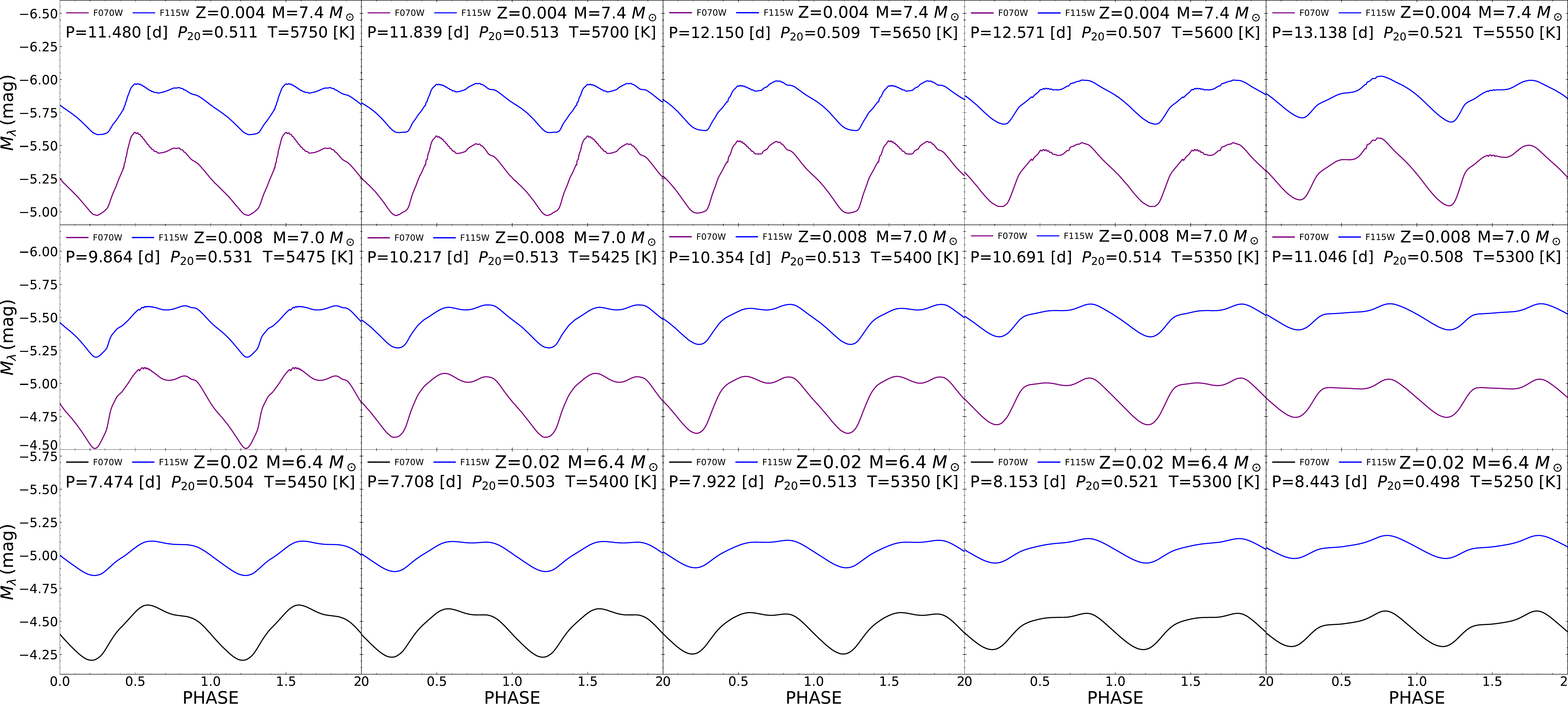} \caption{JWST light curves of representative CC canonical models with  $\alpha_{\rm ml}=1.5$, illustrating the Hertzsprung progression. From top to bottom, the panels correspond to $Z=0.004$, $Z=0.008$, and $Z=0.02$, while the pulsation period increases from left to right. Light curves are shown in the F070W (black) and F115W (blue) filters. The adopted model parameters are indicated in each panel. Although the y-axis limits vary, the amplitude range ($\Delta y$) is kept constant to facilitate comparison.} \label{fig:prog_jwst} 
\end{figure*}

Nonlinear convective pulsation models have long shown that the position of the HP centre depends on several stellar parameters, most notably the stellar mass and chemical composition \citep[e.g.][]{Marconi2024,Desomma2024ApJ}. In particular, the period corresponding to the centre of the progression shifts toward longer values as the metal abundance decreases, whereas variations in the helium abundance produce only minor effects at fixed metallicity. To investigate how the HP is mapped into the NIR regime, we analyse the predicted light curves transformed into the JWST and Roman photometric systems. Figure~\ref{fig:prog_jwst} presents representative JWST light curves, while the corresponding Roman light curves are shown in Appendix~\ref{sec:h_prog} (Figure~\ref{fig:prog_roman}). The models are computed for three representative metallicities ($Z=0.004$, $Z=0.008$, and $Z=0.02$) and five effective temperatures sampling the expected location of the HP within the instability strip.

For the JWST system, the light curves are shown in the F070W and F115W filters (black and blue, respectively). The corresponding Roman light curves are displayed in the F062 and F106 filters (blue and green, respectively). These relatively bluer filters were selected because the pulsation amplitude decreases with increasing wavelength, making the bump progressively more difficult to identify in redder bands. The predicted JWST light curves clearly reproduce the classical behaviour of the HP. The bump migrates from the descending branch toward maximum light as the pulsation period increases, in agreement with results obtained in other photometric systems. Moreover, the position of the HP centre depends strongly on metallicity, occurring at shorter periods for metal-rich models ($Z=0.02$) and shifting toward progressively longer periods for more metal-poor compositions ($Z=0.008$ and $Z=0.004$), fully consistent with previous theoretical predictions based on bolometric light curves.

The Roman light curves exhibit the same overall behaviour, confirming that the transformation into both NIR photometric systems preserves the underlying physical dependence of the HP on stellar parameters. In both systems, metal-rich models display smoother light curves with smaller amplitudes, whereas metal-poor models exhibit more pronounced bumps and larger pulsation amplitudes.

\section{The pulsation amplitudes}

The synthetic light curves obtained in the JWST and Roman photometric systems are used to derive intensity-averaged mean magnitudes and peak-to-peak pulsation amplitudes for both F and FO modes. For each model, amplitudes are computed directly from the full non-linear light curves, ensuring consistency with the underlying hydrodynamical pulsation calculations. The resulting mean magnitudes and amplitudes in all adopted filters are collected in Tables~\ref{tab:jwst_models} and \ref{tab:roman_models} for the JWST and Roman photometric systems, respectively. These tables constitute the basis for the analysis of the Period–Amplitude behavior and Period-Amplitude ratios, which are discussed in the following subsections.

\begin{table*}[ht!]
\caption{\label{tab:jwst_models}
Intensity-averaged mean magnitudes and peak-to-peak amplitudes of the F- and FO-mode Classical Cepheid models in the JWST photometric system.}
\centering
\scriptsize
\setlength{\tabcolsep}{0.8pt}
\begin{tabular}{ccccccccccccccccccccccccc}
\hline\hline
$Z$ &
$Y$ &
Mode &
$P$ &
$M/M_\odot$ &
$\log L/L_\odot$ &
$T_{\rm eff}$ &
$\alpha_{\rm ml}$ &
ML &
$\langle$F070W$\rangle$ &
$A_{\rm F070W}$ &
$\langle$F090W$\rangle$ &
$A_{\rm F090W}$ &
$\langle$F115W$\rangle$ &
$A_{\rm F115W}$ &
$\langle$F150W$\rangle$ &
$A_{\rm F150W}$ &
$\langle$F200W$\rangle$ &
$A_{\rm F200W}$ &
$\langle$F277W$\rangle$ &
$A_{\rm F277W}$ &
$\langle$F356W$\rangle$ &
$A_{\rm F356W}$ &
$\langle$F444W$\rangle$ &
$A_{\rm F444W}$ \\
\hline
(1)&(2)&(3)&(4)&(5)&(6)&(7)&(8)&(9)&
(10)&(11)&(12)&(13)&(14)&(15)&
(16)&(17)&(18)&(19)&(20)&
(21)&(22)&(23)&(24)&(25) \\
\hline
0.004 & 0.25 & F  & 1.460 & 3.0 & 2.49 & 5900 & 1.5 & A & -2.008 & 0.409 & -2.250 & 0.313 & -2.463 & 0.242 & -2.755 & 0.186 & -2.861 & 0.193 & -2.885 & 0.195 & -2.894 & 0.195 & -2.897 & 0.195 \\
0.008 & 0.25 & F  & 1.154 & 3.0 & 2.39 & 6000 & 1.5 & A & -1.772 & 0.563 & -2.000 & 0.430 & -2.202 & 0.313 & -2.474 & 0.190 & -2.574 & 0.202 & -2.598 & 0.203 & -2.606 & 0.203 & -2.606 & 0.198 \\
\hline
\end{tabular}
\tablefoot{
Columns list:
(1) metal abundance $Z$;
(2) helium mass fraction $Y$;
(3) pulsation mode;
(4) pulsation period $P$ (days);
(5) stellar mass ($M_\odot$);
(6) logarithmic luminosity $\log L/L_\odot$;
(7) effective temperature $T_{\rm eff}$ (K);
(8) mixing-length parameter $\alpha_{\rm ml}$;
(9) adopted ML relation;
(10--25) intensity-averaged mean magnitudes $\langle m\rangle$ and corresponding peak-to-peak amplitudes $A$ in the JWST filters F070W, F090W, F115W, F150W, F200W, F277W, F356W, and F444W.
Only the first two rows are shown. The complete table, including all F- and FO-mode models for $Z=0.004$, $0.008$, $0.02$, and $0.03$, is available as supplementary material.
}
\end{table*}

\begin{table*}[ht!]
\caption{\label{tab:roman_models}
Same as Table~\ref{tab:jwst_models}, but for the Roman photometric system.}
\centering
\scriptsize
\setlength{\tabcolsep}{0.9pt}
\begin{tabular}{ccccccccccccccccccccccc}
\hline\hline
$Z$ &
$Y$ &
Mode &
$P$ &
$M/M_\odot$ &
$\log L/L_\odot$ &
$T_{\rm eff}$ &
$\alpha_{\rm ml}$ &
ML &
$\langle$F062$\rangle$ &
$A_{\rm F062}$ &
$\langle$F087$\rangle$ &
$A_{\rm F087}$ &
$\langle$F106$\rangle$ &
$A_{\rm F106}$ &
$\langle$F129$\rangle$ &
$A_{\rm F129}$ &
$\langle$F158$\rangle$ &
$A_{\rm F158}$ &
$\langle$F184$\rangle$ &
$A_{\rm F184}$ &
$\langle$F213$\rangle$ &
$A_{\rm F213}$ \\
\hline
(1)&(2)&(3)&(4)&(5)&(6)&(7)&(8)&(9)&
(10)&(11)&(12)&(13)&(14)&(15)&
(16)&(17)&(18)&(19)&(20)&(21)&(22)&(23) \\
\hline
0.004 & 0.25 & F  & 1.460 & 3.0 & 2.49 & 5900 & 1.5 & A & -1.798 & 0.492 & -2.227 & 0.320 & -2.370 & 0.275 & -2.586 & 0.195 & -2.791 & 0.189 & -2.855 & 0.193 & -2.866 & 0.193 \\
0.008 & 0.25 & F  & 1.154 & 3.0 & 2.39 & 6000 & 1.5 & A & -1.569 & 0.676 & -1.978 & 0.441 & -2.113 & 0.368 & -2.318 & 0.241 & -2.508 & 0.194 & -2.568 & 0.202 & -2.580 & 0.203 \\
\hline
\end{tabular}
\tablefoot{
Same as Table~\ref{tab:jwst_models}, but for the Roman/WFI filters F062, F087, F106, F129, F158, F184, and F213.
Only the first two rows are shown. The complete table, including all F- and FO-mode models for $Z=0.004$, $0.008$, $0.02$, and $0.03$, is available as supplementary material.
}
\end{table*}

\subsection{The Period amplitude diagram}

In Appendix ~\ref{sec:amplitude_appendix}, Figures~\ref{fig:amp_logP_jwst} and \ref{fig:amp_logP_roman}, present the multiband pulsation amplitudes as a function of the logarithmic period for F- and FO-mode models in the JWST and Roman photometric systems, respectively. The models span different stellar masses and chemical compositions. Similarly to what is found in the Rubin-LSST filters (see Figs. 2 and 3 by \citetalias{Desomma2024}), the pulsation amplitudes show a clear dependence on both metallicity and wavelength. In particular, the amplitudes systematically decrease as the metal content increases, reflecting the enhanced opacity in more metal-rich envelopes. Moreover, a prominent effect emerging in the JWST and Roman bands is the overall reduction of pulsation amplitudes compared to optical systems. This behavior is consistent with the well-known damping of the contribution of temperature variations to the resulting light curves in the near- and mid-infrared regime. While at lower luminosity levels the amplitudes display a quasi-linear dependence on period, at higher luminosities the fundamental-mode amplitudes exhibit a more complex, non-linear behavior, progressively approaching the bell-shaped morphology that characterizes first-overtone pulsation at lower luminosities \citep[see also][for details]{Bono2000}. 

The behaviour of the $8 M_{\odot}$ models visible in Figures ~\ref{fig:amp_logP_jwst} and ~\ref{fig:amp_logP_roman} deserves some additional discussion. This feature is not associated with changes in either the adopted ML relation or the convective efficiency, since all models shown in these figures were computed assuming the canonical ML relation (case A) and $\alpha_{\rm ml}=1.5$. For each stellar mass, the plotted sequence samples different effective temperatures across the instability strip. Around $8M_{\odot}$, the nonlinear pulsation amplitudes show a particularly strong variation across the strip, producing the pronounced non-monotonic structure observed in the $A-\log P$ plane. This behaviour is associated with the well-known Hertzsprung progression \citep[see, e.g.,][and references therein]{Marconi2024}, which occurs over a period range of approximately 6-16 days and is characterized by a minimum in the pulsation amplitude near the period at which the secondary maximum moves from the descending to the rising branch of the light and radial-velocity curves (see also Subsection~\ref{sec:h_prog_text}). The occurrence of this feature in both the JWST and Roman filters further indicates that it is intrinsic to the pulsation models rather than being introduced by the photometric transformations.

\subsection{Amplitude Ratios}

Amplitude ratios represent a powerful tool in the analysis of pulsating variable stars, as they relate light-curve amplitudes measured in different photometric bands. In particular, when observations are available in a single filter, a common occurrence for extragalactic CC samples observed with facilities such as JWST and Roman, these ratios can be used to reconstruct the expected amplitudes in other bands, enabling a more complete characterization of the light-curve morphology.

In this context, we computed theoretical amplitude ratios for all the models in our grid by measuring the peak-to-peak amplitude in each band and normalizing it to a reference filter. These predictions represent the first theoretical calibration of amplitude ratios in the JWST and Roman photometric systems. For the JWST system, we adopted the F070W band as reference, while for the Roman system we used the F062 band. This choice is motivated by the fact that shorter wavelengths exhibit larger pulsation amplitudes, thus providing a stable and well-defined normalization baseline.

In Appendix~\ref{sec:amplitude_appendix}, Figure~\ref{fig:amp_ratio_jwst_plot} shows the predicted amplitude ratios as a function of $\log P$ for selected JWST filters and representative chemical compositions (see labels in the figure). A first important result is that the amplitude ratios are, to first order, weakly dependent on the pulsation period. In most cases, the distributions appear approximately flat over the entire period range, especially in the bluer and intermediate bands. This behavior supports the common approximation that CC amplitude ratios can be treated as nearly constant quantities, consistent with what is predicted and observed in other photometric systems.

However, a mild period dependence becomes more evident at longer wavelengths, where both an increase in the intrinsic scatter and a slight systematic trend with period can be observed, particularly for F-mode pulsators. This effect reflects the progressive decrease of pulsation amplitudes toward the NIR. A clear and robust trend is observed as a function of wavelength. The amplitude ratios systematically decrease from the bluer to the redder filters, reaching an approximately constant plateau in the longer wavelength regime. In the JWST system, typical values range from $\sim0.8$ for the F090W/F070W amplitude ratio down to $\sim0.45$--$0.50$ for ratios involving filters redder than F150W. A similar behavior is found for the Roman system, where the ratios decrease from $\sim0.7$ when adopting F087 to $\sim0.35$--$0.40$ in the reddest bands. This trend reflects the well-known decrease in pulsation amplitudes with increasing wavelength.

FO-mode pulsators show amplitude ratios generally consistent with those of F-mode CCs in the bluer bands. However, when redder filters are considered, FO pulsators systematically occupy the lower envelope of the F-mode distribution. For instance, in the JWST system, the average ratio $A_{\mathrm{F150W}}/A_{\mathrm{F070W}}$ ranges from approximately 0.45 to 0.53 for F pulsators and from approximately 0.36 to 0.38 for FO models. 

A similar behavior is observed in the Roman system (Figure~\ref{fig:amp_ratio_roman_plot} in Appendix~\ref{sec:amplitude_appendix}), where $A_{\mathrm{F158}}/A_{\mathrm{F062}}$ decreases from $\sim0.42$ for F pulsators to $\sim0.30$ for FO models, and the difference becomes even more evident in the reddest filters, where FO ratios reach values as low as $\sim0.27$. This difference is a direct consequence of the distinct light-curve morphologies of the two pulsation modes and their different wavelength dependence.

On the other hand, the dependence on metallicity and the adopted ML relation appears to be secondary. Variations in chemical composition and ML assumptions produce changes that are generally smaller than, or comparable to, the intrinsic dispersion of the relations. This result indicates that amplitude ratios are robust quantities that can be safely applied across a wide range of stellar parameters.

Although observational amplitude ratios in the JWST and Roman photometric systems are not yet available, empirical studies in optical and near-infrared bands have shown that amplitude ratios decrease monotonically with increasing wavelength and asymptotically approach nearly constant values in the NIR regime \citep[e.g.,][]{Bhardwaj2015,Inno2015}. Our theoretical predictions reproduce this behaviour in both photometric systems and provide the first calibration of amplitude ratios for the JWST and Roman filters.

To facilitate their practical use, we provide in Tables~\ref{tab:amp_ratio_jwst_values} and \ref{tab:amp_ratio_roman_values} the mean amplitude ratios and their standard deviations for all the considered filters, pulsation modes, and chemical compositions. These calibrations can be directly used to estimate amplitudes in different bands when only single-band observations are available, providing a useful tool for the analysis of CC light curves in the NIR, particularly for datasets with limited photometric coverage.

\begin{table}[ht!]
\caption{\label{tab:amp_ratio_jwst_values}
Mean amplitude ratios $A_{\lambda}/A_{\rm F070W}$ in the JWST filter system.}
\centering
\scriptsize
\setlength{\tabcolsep}{4pt}
\renewcommand{\arraystretch}{0.9}
\begin{tabular}{ccccccc}
\hline\hline
$Z$ &
$Y$ &
Mode &
ML &
Filter &
Mean &
Std. dev. \\
\hline
(1)&(2)&(3)&(4)&(5)&(6)&(7)\\
\hline
0.004 & 0.25 & F & A & F090W & 0.809 & 0.026 \\
0.008 & 0.25 & F & A & F090W & 0.808 & 0.029 \\
\hline
\end{tabular}
\tablefoot{
Columns list:
(1) metal abundance $Z$;
(2) helium mass fraction $Y$;
(3) pulsation mode;
(4) adopted ML relation;
(5) JWST filter;
(6) mean amplitude ratio $A_{\lambda}/A_{\rm F070W}$;
(7) corresponding standard deviation.
Only a portion of the table is shown here. The complete table is available as supplementary material.
}
\end{table}

\begin{table}[ht!]
\caption{\label{tab:amp_ratio_roman_values}
Same as Table~\ref{tab:amp_ratio_jwst_values}, but for the Roman filter system.
}
\centering
\scriptsize
\setlength{\tabcolsep}{4pt}
\renewcommand{\arraystretch}{0.9}
\begin{tabular}{ccccccc}
\hline\hline
$Z$ &
$Y$ &
Mode &
ML &
Filter &
Mean &
Std. dev. \\
\hline
(1)&(2)&(3)&(4)&(5)&(6)&(7)\\
\hline
0.004 & 0.25 & F & A & F087 & 0.713 & 0.034 \\
0.008 & 0.25 & F & A & F087 & 0.715 & 0.032 \\
\hline
\end{tabular}
\tablefoot{
Same as Table~\ref{tab:amp_ratio_jwst_values}, but for amplitude ratios normalized to $A_{\rm F062}$ in the Roman filters.
Only a portion of the table is shown here. The complete table is available as supplementary material.
}
\end{table}

\section{The Pulsation Relations}

\subsection{The PLC relations in JWST and Roman}

The intensity-weighted mean magnitudes and colors derived from the transformed JWST and Roman light curves were used to derive theoretical PLC relations of the form $M_{\lambda}=a+b\log P+c\,CI$, where $CI$ denotes the adopted color index. As already discussed in previous theoretical investigations \citepalias[see e.g.][and references therein]{Desomma2022}, the inclusion of a color term allows us to account for the finite-temperature width of the instability strip, which represents the source of the unavoidable intrinsic dispersion in PL relations.

The coefficients of the theoretical PLC relations are reported in Table~\ref{tab:plc_all}, where only a representative subset of the results is shown. The complete machine-readable tables for the JWST and Roman photometric systems are available in the online supplementary material. The relations were derived for the four adopted chemical compositions, the three assumptions on the ML relation, and the three different assumed values of the mixing-length parameter, separately for F and FO pulsators. A first inspection of the derived coefficients shows that, as expected, the PLC relations are intrinsically very tight in both photometric systems (see, e.g., DS20b, DS22, and \citealt{Caputo2000, Marconi2005}). In particular, fundamental-mode models typically show root-mean-square (rms) dispersions of only a few hundredths of a magnitude. This confirms that the adopted color terms efficiently remove the effect of the finite width of the instability strip in the near- and mid-infrared wavelength range covered by JWST and Roman.

The dependence on the adopted physical assumptions is not negligible. The largest systematic variations are generally associated with the assumed ML relation. Moving from the canonical case A to the brighter moderately non-canonical (B) and full non-canonical (C) cases changes the zero point and, to a lesser extent, the slope of the PLC relations. This behavior is expected, since the adopted ML relation directly affects the luminosity level associated with a given mass and effective temperature \citep[see also][]{Caputo2002}. On the other hand, variations in the mixing-length parameter produce, as previously found \citepalias[see][and references therein]{Desomma2020apj, Desomma2022}, a more limited effect on the PLC coefficients, although the choice of the convective efficiency is important in the computation of pulsation amplitudes and on the location in effective temperature of the instability strip boundaries. 

The chemical composition also affects the inferred coefficients, although the impact of metallicity on the predicted luminosity, at fixed period and color, depends on the adopted filter combination and color term coefficient. For F-mode CCs, the PLC relations remain well defined over the entire explored metallicity range, namely from $Z=0.004$ to $Z=0.03$. Conversely, FO-mode relations become less robust at higher metallicity and for larger values of the mixing-length parameter, due to the progressive reduction of the number of pulsating models. In some cases, such as FO models with $Z=0.03$ and $\alpha_{\rm ml}=1.9$, the limited number of stable pulsation
models does not allow a reliable determination of the PLC coefficients, and the corresponding relations are therefore not reported. The color term coefficients are relatively large for some JWST and Roman NIR color combinations. This is not unexpected because purely infrared colors span a narrow range across the instability strip; therefore, small color changes may correspond to sizeable corrections in absolute magnitude.

\begin{table*}[t]
\caption{\label{tab:plc_all}
Coefficients of the theoretical PLC relations
$M_{\lambda}=a+b\log P+c\,CI$
derived for the JWST and Roman photometric systems.}
\centering
\scriptsize
\setlength{\tabcolsep}{3pt}
\begin{tabular}{ccccccccccccccc}
\hline\hline
$Z$ & $Y$ & Band & CI & Mode & $\alpha_{\rm ml}$ & ML &
$a$ & $b$ & $c$ & $\sigma_a$ & $\sigma_b$ & $\sigma_c$ & $\sigma$ & $R^2$ \\
\hline
\multicolumn{15}{c}{JWST}\\
\hline
0.004 & 0.25 & F150W & F090W$-$F150W & F & 1.5 & A & -3.115 & -3.744 & 1.815 & 0.019 & 0.015 & 0.043 & 0.033 & 0.999 \\
0.004 & 0.25 & F150W & F090W$-$F150W & F & 1.5 & B & -3.015 & -3.654 & 1.754 & 0.026 & 0.019 & 0.055 & 0.050 & 0.999 \\
\hline
\multicolumn{15}{c}{Roman}\\
\hline
0.004 & 0.25 & F158 & F129$-$F158 & F & 1.5 & A & -2.996 & -3.748 & 3.705 & 0.019 & 0.016 & 0.102 & 0.036 & 0.999 \\
0.004 & 0.25 & F158 & F129$-$F158 & F & 1.5 & B & -2.901 & -3.644 & 3.526 & 0.026 & 0.020 & 0.127 & 0.054 & 0.999 \\
\hline
\end{tabular}
\tablefoot{
The complete machine-readable tables are available in the online supplementary material as \texttt{PLC\_JWST.dat} and \texttt{PLC\_Roman.dat}.}
\end{table*}

\subsection{Pure JWST Period-Wesenheit relations}
\label{sec:pw}

For distance-scale applications, it is useful to move from the generic PLC formulation to Wesenheit relations \citep[][]{Madore1982}. In contrast to the PLC relations, where the color coefficient is determined from the fit, the PW formulation fixes this coefficient according to the adopted reddening law. The resulting Wesenheit magnitudes are therefore reddening-free by construction and allow a more direct comparison with observational distance-scale analyses.

The intensity-weighted mean magnitudes and colors were used to derive theoretical PW relations in the JWST photometric system. Following the standard approach, the Wesenheit magnitude was defined as $W(\lambda,CI)=M_{\lambda}-R_{\lambda,CI}\,CI$, where $CI$ denotes the adopted color index and $R_{\lambda,CI}$ is the ratio of total to selective absorption for the selected filter combination.

For JWST, we selected a representative set of Wesenheit combinations spanning different wavelength baselines, including both purely NIR and wider-baseline definitions. This choice allows a direct comparison between different PW formulations while limiting the analysis to a representative subset of filter combinations rather than an exhaustive exploration of all possible Wesenheit definitions.
For JWST, the adopted combinations are:
$W(F150W,F150W-F277W)$, 
$W(F200W,F150W-F200W)$, 
$W(F150W,F090W-F150W)$, 
and $W(F277W,F090W-F150W)$. 
The coefficients of the theoretical PW relations, for the three adopted reddening laws (Cardelli, Fitzpatrick, and Wang), are summarized in Table~\ref{tab:pw_all}, where only a representative subset of the results is shown. The complete machine-readable tables is available in the online supplementary material.

The adopted Wesenheit functions and reddening coefficients are summarized in Table~\ref{tab:wesenheit_coefficients} for various extinction laws, as detailed in the third column.

\begin{table}[ht!]
\caption{\label{tab:wesenheit_coefficients}
Adopted Wesenheit functions and reddening coefficients.}
\centering
\scriptsize
\setlength{\tabcolsep}{1pt}
\renewcommand{\arraystretch}{1.05}
\begin{tabular}{lcc}
\hline\hline
Wesenheit function &
Reddening law &
$R_{\lambda,CI}$ \\
\hline
\multicolumn{3}{c}{\textbf{JWST}}\\
\hline
\multirow{3}{*}{$W(\mathrm{F150W},\mathrm{F150W}-\mathrm{F277W})$}
& Cardelli et al.\ (1989) & 1.59\\
& Fitzpatrick (1999) & 1.73\\
& Wang et al.\ (2024) & 1.43\\
\addlinespace
\multirow{3}{*}{$W(\mathrm{F200W},\mathrm{F150W}-\mathrm{F200W})$}
& Cardelli et al.\ (1989) & 1.78\\
& Fitzpatrick (1999) & 1.88\\
& Wang et al.\ (2024) & 1.33\\
\addlinespace
\multirow{3}{*}{$W(\mathrm{F150W},\mathrm{F090W}-\mathrm{F150W})$}
& Cardelli et al.\ (1989) & 0.80\\
& Fitzpatrick (1999) & 0.72\\
& Wang et al.\ (2024) & 0.58\\
\addlinespace
\multirow{3}{*}{$W(\mathrm{F277W},\mathrm{F090W}-\mathrm{F150W})$}
& Cardelli et al.\ (1989) & 0.30\\
& Fitzpatrick (1999) & 0.30\\
& Wang et al.\ (2024) & 0.17\\
\hline
\multicolumn{3}{c}{\textbf{Roman}}\\
\hline
\multirow{2}{*}{$W(\mathrm{F158},\mathrm{F129}-\mathrm{F158})$}
& Cardelli et al.\ (1989) & 2.64\\
& Fitzpatrick (1999) & 2.62\\
\addlinespace
\multirow{2}{*}{$W(\mathrm{F184},\mathrm{F158}-\mathrm{F184})$}
& Cardelli et al.\ (1989) & 3.39\\
& Fitzpatrick (1999) & 3.92\\
\hline
\end{tabular}
\tablefoot{
The coefficient $R_{\lambda,CI}$ is adopted in the definition of the Wesenheit magnitude
$W(\lambda,CI)=M_\lambda-R_{\lambda,CI}\,CI$. For the Roman filter system, extinction coefficients based on the \citet[][]{Wang2024} extinction law are not currently available.
}
\end{table}

As shown in Fig.~\ref{fig:pw_jwst_card}, the theoretical Wesenheit relations for the two representative JWST band combinations remain very tight over the entire explored metallicity range. The F-mode relations exhibit root-mean-square (rms) dispersions typically ranging from a few hundredths to about one-tenth of a magnitude, with $R^2$ values close to unity. Additional JWST/NIRCam Wesenheit combinations display the same overall behaviour and are presented in Appendix~\ref{sec:pw_appendix} (Figure~\ref{fig:pw_jwst_appendix}). The relations for different chemical compositions remain close to each other, especially for F-mode CCs. The main effect of metallicity is a small shift in the zero point, while the slopes remain broadly similar. This behavior confirms that NIR Wesenheit relations are only mildly affected by metallicity, although the residual dependence is still relevant for precision distance-scale applications. Some FO combinations are not reported because the number of stable FO models is too limited to allow a reliable linear regression. This occurs mainly at high metallicity and for larger values of the mixing-length parameter, where FO pulsation becomes progressively less efficient.

\subsection{Roman Period-Wesenheit relations}

We also derived theoretical PW relations in the Roman photometric system. We adopted two representative NIR Wesenheit combinations $W(F158,F129-F158)$ and $W(F184,F158-F184)$ and computed the corresponding relations using the \citet{Cardelli1989} and \citet{Fitzpatrick1999} extinction laws. The \citet{Wang2024} prescription was not included because extinction coefficients for the Roman filters are not currently available. 

The resulting coefficients are summarized in Table~\ref{tab:pw_all}, where only a representative subset of the results is shown. The complete machine-readable table is available in the online supplementary material.

As for the JWST relations, the Roman PW relations are also very tight for F-mode Cepheids. The corresponding relations are presented in Appendix~\ref{sec:pw_appendix} (Figure~\ref{fig:pw_roman}). Smaller intrinsic dispersions than the corresponding JWST relations generally characterize the Roman PW relations. Typical values range from $\sigma \sim 0.03$ to $0.10$ mag for F-mode Cepheids, compared with $\sigma \sim 0.08$-$0.15$ mag for the JWST combinations. This behavior is likely related to the smaller wavelength separation between the Roman filters entering the Wesenheit definitions, which reduces the sensitivity of the relations to temperature variations across the instability strip. As a consequence, the PW relations appear particularly compact and nearly parallel over the full metallicity range explored.  The effect of chemical composition is again mainly seen as a zero-point shift, while the slopes remain broadly similar. The dependence on the adopted physical assumptions follows the same trends found for JWST filters. The ML relation has the largest impact on the zero point because it changes the luminosity level at fixed mass and temperature. The effect of the mixing-length parameter is generally more limited for F-mode CCs. This confirms that Roman NIR Wesenheit relations are robust distance indicators and will represent a valuable tool for future homogeneous CC studies.

\begin{table*}[t]
\caption{\label{tab:pw_all}
Coefficients of the theoretical PW relations
$W_{\lambda}=a+b\log P$ derived for the JWST and Roman photometric systems. Columns report the adopted extinction law, chemical composition, Wesenheit function, pulsation mode, mixing-length parameter, ML relation, fitted coefficients and uncertainties, rms dispersion ($\sigma$), $R^2$.}
\centering
\scriptsize
\setlength{\tabcolsep}{2pt}
\begin{tabular}{cccccccccccccc}
\hline\hline
Extinction law & $Z$ & $Y$ & Band & CI & Mode & $\alpha_{\rm ml}$ & ML &
$a$ & $b$ & $\sigma_a$ & $\sigma_b$ & $\sigma$ & $R^2$ \\
\hline
\multicolumn{14}{c}{JWST}\\
\hline
Cardelli1989 & 0.004 & 0.25 & F150W & F090W$-$F150W & F & 1.5 & A & -2.756 & -3.467 & 0.033 & 0.025 & 0.092 & 0.996 \\
Cardelli1989 & 0.004 & 0.25 & F150W & F090W$-$F150W & F & 1.5 & B & -2.671 & -3.407 & 0.036 & 0.024 & 0.102 & 0.995 \\
\hline
\multicolumn{14}{c}{Roman}\\
\hline
Cardelli1989 & 0.004 & 0.25 & F158 & F129$-$F158 & F & 1.5 & A & -2.852 & -3.612 & 0.020 & 0.015 & 0.055 & 0.999 \\
Cardelli1989 & 0.004 & 0.25 & F158 & F129$-$F158 & F & 1.5 & B & -2.777 & -3.538 & 0.023 & 0.016 & 0.066 & 0.998 \\
\hline
\end{tabular}
\tablefoot{
The complete machine-readable tables are available in the online supplementary material as \texttt{PW\_JWST.dat} and \texttt{PW\_Roman.dat}.
}
\end{table*}

\subsection{Effect of the adopted reddening law}

Since the definition of a Wesenheit magnitude depends on the adopted extinction law, we computed the theoretical PW relations using different reddening prescriptions. For the JWST filter system, we adopted the extinction laws of \citet{Cardelli1989}, \citet{Fitzpatrick1999}, and \citet{Wang2024}. For the Roman filter system, only the \citet{Cardelli1989} and \citet{Fitzpatrick1999} prescriptions were considered, as extinction coefficients based on the \citet{Wang2024} prescription are not currently available for the Roman filters.

The comparison among the different reddening laws shows that the overall behavior of the PW relations is remarkably stable. In all cases, the relations remain tight, with very similar dispersions and correlation coefficients, and the relative trends as a function of chemical composition, ML relation, and mixing-length parameter are preserved. The resulting slopes and zero points differ only marginally, typically at the level of a few hundredths of a magnitude.

For the JWST Wesenheit relations, the coefficients derived using the Cardelli, Fitzpatrick, and Wang prescriptions differ only marginally, with changes of a few thousandths in the slope and a few hundredths of a magnitude in the zero point. The largest differences are found for Wesenheit definitions involving the widest color baselines; however, even in these cases the variations are significantly smaller than those induced by changes in the adopted ML relation or convective efficiency.

A similar behavior is observed in Roman relations. The PW coefficients derived using the Cardelli and Fitzpatrick extinction laws are nearly indistinguishable, with differences limited to a few hundredths of a magnitude over the full range of chemical compositions explored. The resulting dispersions and correlation coefficients remain unchanged.

This behavior is expected because both the JWST and Roman filters probe the near- and mid-infrared wavelength regime, where the extinction curve is relatively flat and different reddening prescriptions tend to converge. These results indicate that the choice of extinction law is not a dominant source of uncertainty for the theoretical PW relations and confirm their robustness for applications to Classical Cepheid distance determinations and precision cosmology.

\subsection{Hybrid JWST-HST Period-Wesenheit relations for the SH0ES comparison}

In addition to the purely JWST and Roman Wesenheit relations, we derived hybrid PW relations specifically designed to enable a direct comparison with the SH0ES observational framework. Following the formulation adopted by \citet{Riess2025}, we defined
\begin{equation}
W_H = F150W_{\rm JWST} - 0.4(F555W_{\rm HST}-F814W_{\rm HST}) ,
\end{equation}
where the NIR magnitude is taken from JWST, while the color term is based on the optical HST bands.

We first fitted the hybrid relations by leaving the slope as a free parameter, $W_H = a + b(\log P-1)$. This allows us to compare the theoretical slopes with the value adopted in the SH0ES analysis.

Among the F-mode models, the hybrid PW slopes closest to the value adopted by the SH0ES collaboration ($b=-3.25$) are generally obtained for the mildly overluminous ML=B models at low metallicity. In particular, the best agreement is found for the SMC and LMC compositions, which provide slopes of $b=-3.323$ and $b=-3.288$, respectively. At solar metallicity, the ML=B models yield slopes of $b=-3.221$ and $b=-3.231$ for $\alpha_{\rm ml}=1.5$ and $1.7$, respectively, both differing from the SH0ES value by less than $\sim0.03$. At the highest metallicity explored ($Z=0.03$), the closest agreement is instead obtained for the canonical ML=A models, with slopes between $-3.286$ and $-3.326$. Overall, the comparison indicates that the SH0ES slope lies well within the range predicted by the pulsation models, although no single combination of metallicity, ML relation, and mixing-length parameter reproduces the value $b=-3.25$ over the entire metallicity interval explored.

We then repeated the fit by fixing the slope to the SH0ES value, $b=-3.25$, deriving only the corresponding zero point.

The complete coefficients of the theoretical hybrid PW relations derived with a free slope are provided in the machine-readable table \texttt{HybridPW\_free.dat}. The corresponding relations obtained by fixing the slope to the SH0ES value, $b=-3.25$ \citep{Riess2024}, are provided in \texttt{HybridPW\_fixed.dat}. Table~\ref{tab:hybridPW} reports a representative extract of both datasets.

The comparison between the free-slope and fixed-slope relations shows that fixing the slope generally increases the scatter only moderately for F-mode CCs. This suggests that the SH0ES slope is broadly consistent with the theoretical predictions. At the same time, the zero point remains sensitive to the adopted ML relation and, to a lesser extent, to chemical composition and convective efficiency. Therefore, these hybrid relations provide the most appropriate theoretical framework for the direct comparison with JWST CC distances discussed in the next section.

Note that some FO-mode hybrid relations are not reported because the number of available stable FO models is too small to provide a robust fit, especially at high metallicity and high convective efficiency.

\begin{table*}[t]
\caption{\label{tab:hybridPW}
Theoretical hybrid PW relations $W_H=a+b(\log P-1)$, obtained by combining the JWST $F150W$ magnitude with the HST optical color $(F555W-F814W)$. The upper part
reports relations derived with the slope left as a free parameter, whereas the lower part reports relations obtained by fixing the slope to $b=-3.25$, following the SH0ES calibration of \citet{Riess2024}.}
\centering
\scriptsize
\setlength{\tabcolsep}{2pt}
\begin{tabular}{cccccccccccccc}
\hline\hline
Fit & $Z$ & $Y$ & Band & CI & Mode & $\alpha_{\rm ml}$ & ML &
$a$ & $b$ & $\sigma_a$ & $\sigma_b$ & $\sigma$ & $R^2$ \\
\hline
\multicolumn{14}{c}{Free slope}\\
\hline
Free & 0.004 & 0.25 & F150W & F555W$-$F814W & F & 1.5 & A & -6.029 & -3.372 & 0.015 & 0.033 & 0.120 & 0.992 \\
Free & 0.008 & 0.25 & F150W & F555W$-$F814W & F & 1.5 & A & -6.070 & -3.353 & 0.014 & 0.033 & 0.117 & 0.992 \\
\hline
\multicolumn{14}{c}{Fixed slope}\\
\hline
Fixed & 0.004 & 0.25 & F150W & F555W$-$F814W & F & 1.5 & A & -6.059 & -3.250 & 0.014 & --- & 0.129 & 0.991 \\
Fixed & 0.008 & 0.25 & F150W & F555W$-$F814W & F & 1.5 & A & -6.089 & -3.250 & 0.014 & --- & 0.123 & 0.991 \\
\hline
\end{tabular}
\tablefoot{
The complete machine-readable tables are available in the online supplementary
material as \texttt{HybridPW\_free.dat} and \texttt{HybridPW\_fixed.dat}.
}
\end{table*}

\begin{figure}[ht]
\centering
\includegraphics[width=0.75\columnwidth]{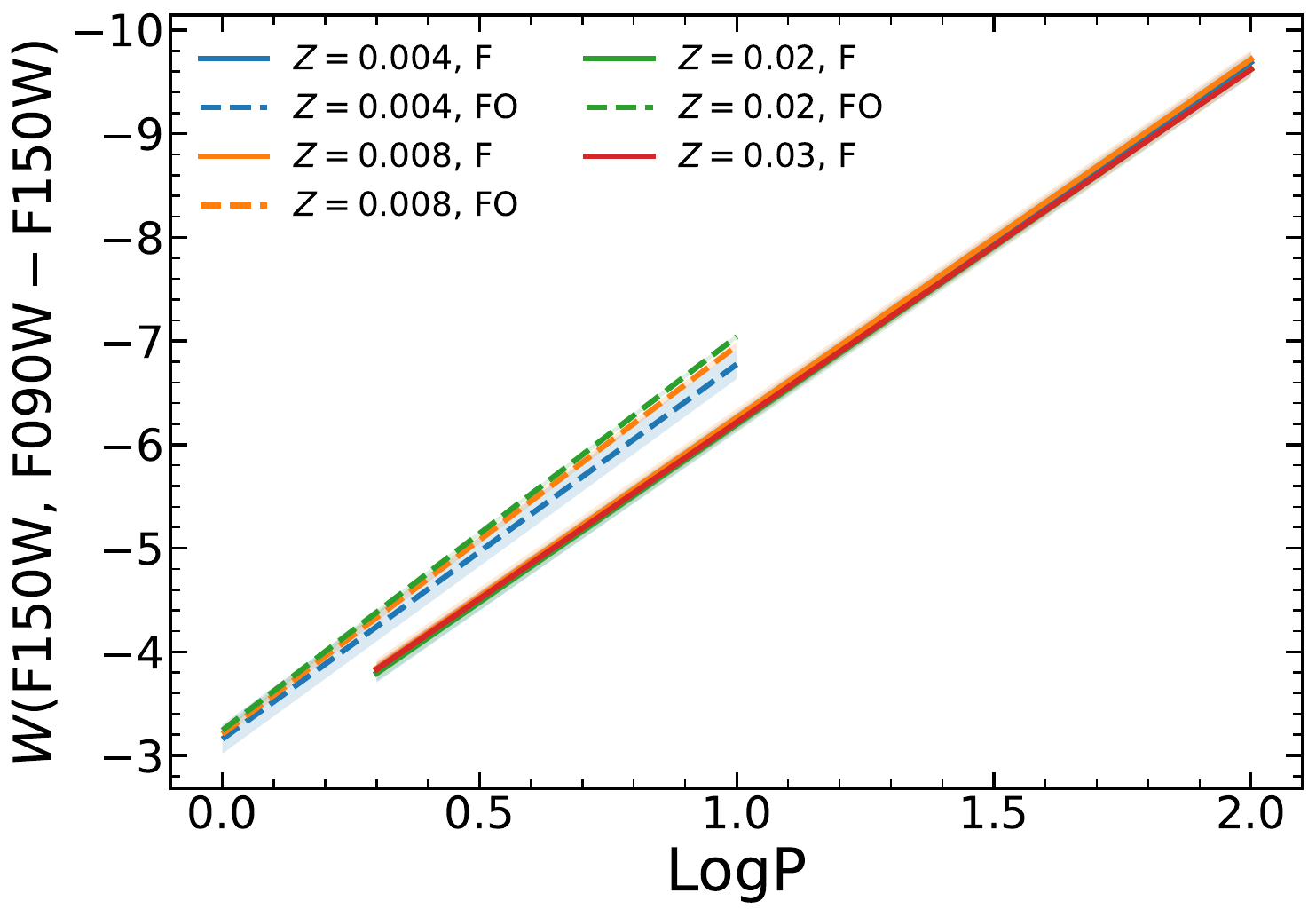}
\vspace{0.2cm}
\includegraphics[width=0.75\columnwidth]{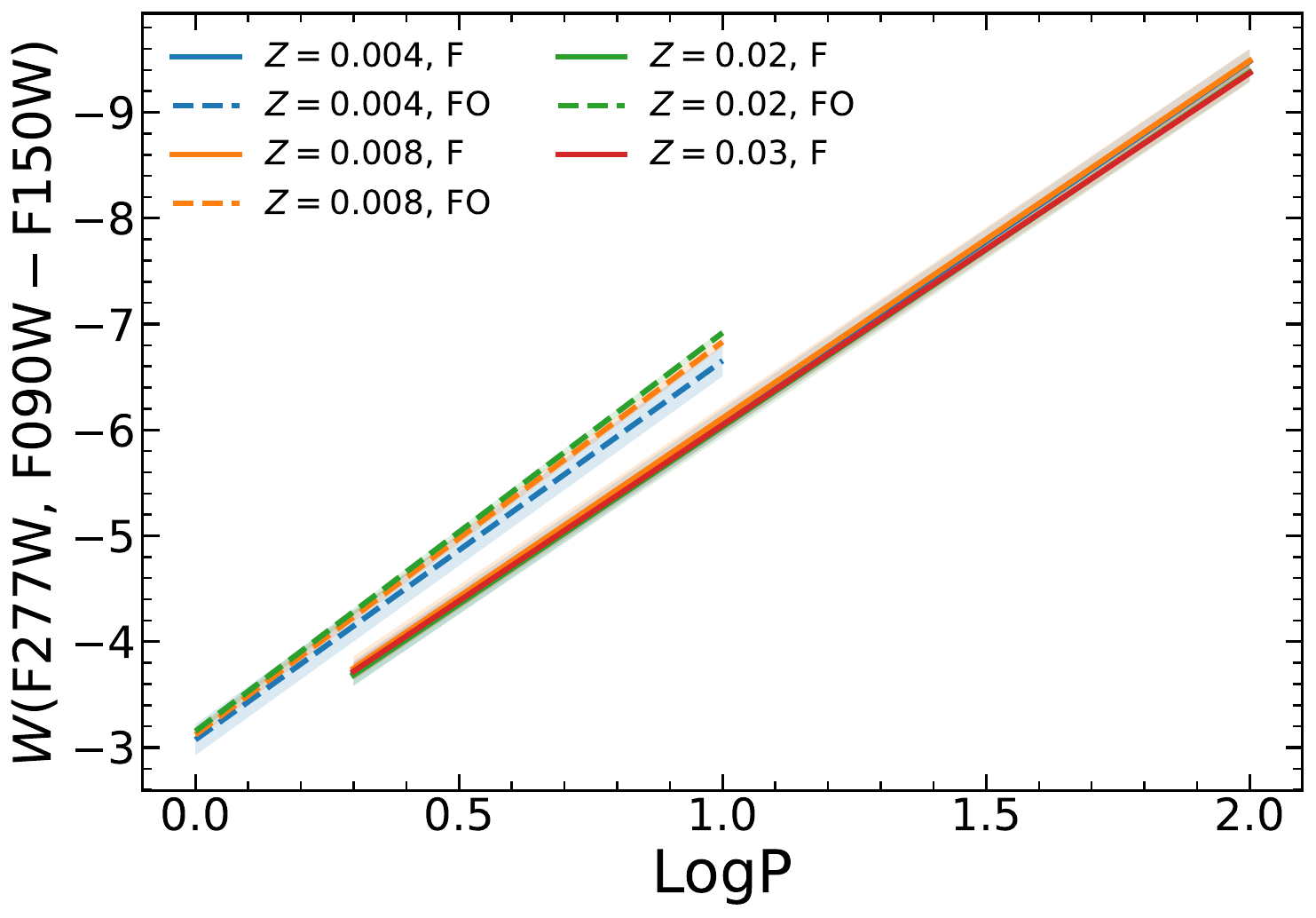}
\caption{
Theoretical PW relations in the JWST photometric system for the two representative Wesenheit combinations
$W(F150W,\,F090W-F150W)$ (top) and
$W(F277W,\,F090W-F150W)$ (bottom),
assuming $\alpha_{\rm ml}=1.5$ and the canonical ML relation.
The relations are shown for different metallicities:
$Z=0.004$ (blue),
$Z=0.008$ (orange),
$Z=0.02$ (green), and
$Z=0.03$ (red).
Solid lines represent F-mode CCs, while dashed lines denote FO pulsators.
The shaded regions indicate the intrinsic dispersion of each relation.
}
\label{fig:pw_jwst_card}
\end{figure}

\subsection{Metal-dependent Period-Wesenheit relations}
\label{sec:PWZ}

The PW relations presented in the previous section already suggest a metallicity effect on the coefficients. This behaviour is evident in both the JWST(Figure~\ref{fig:pw_jwst_card}) and (Appendix~\ref{sec:pw_appendix}, Figures ~\ref{fig:pw_jwst_appendix} and ~\ref{fig:pw_roman}) PW relations, where the Wesenheit relations corresponding to different chemical compositions do not perfectly overlap, even in the NIR regime.

Although the metallicity dependence is reduced with respect to optical bands, small systematic shifts in both slope and zero point remain visible, particularly when comparing metal-poor and metal-rich models. This behavior motivates the introduction of metal-dependent Period-Wesenheit (PWZ) relations, especially in the context of the ongoing debate concerning the size and sign of the metallicity correction to the CC distance scale. Most recent empirical and theoretical determinations converge toward a negative metallicity dependence of the order of $\gamma \sim -0.2$~mag~dex$^{-1}$, although a non-negligible scatter among different methodologies remains (e.g., \citealt{Romaniello2008}, Figure~1; \citealt{Trentin2024}, Figure~19; \citealt{Breuval2025}, Figure~1; see also \citealt{Bhardwaj2024, Breuval2022, Bhardwaj2023, Desomma2022, Riess2021, Ripepi2021, Ripepi2022, Ripepi2026}).
For this reason, we derived PWZ relations of the form $W(\lambda,CI)=\alpha+\beta\log P+\gamma[{\rm Fe/H}]$, where the metallicity term is defined as $[{\rm Fe/H}] = \log(Z/Z_{\odot})$, assuming $Z_{\odot}=0.02$.

The resulting PWZ coefficients for the JWST and Roman filter systems are summarized in Table~\ref{tab:pwz_all}. The complete machine-readable tables is available in the online supplementary material.

\begin{figure}[t]
\centering
\includegraphics[width=\columnwidth]{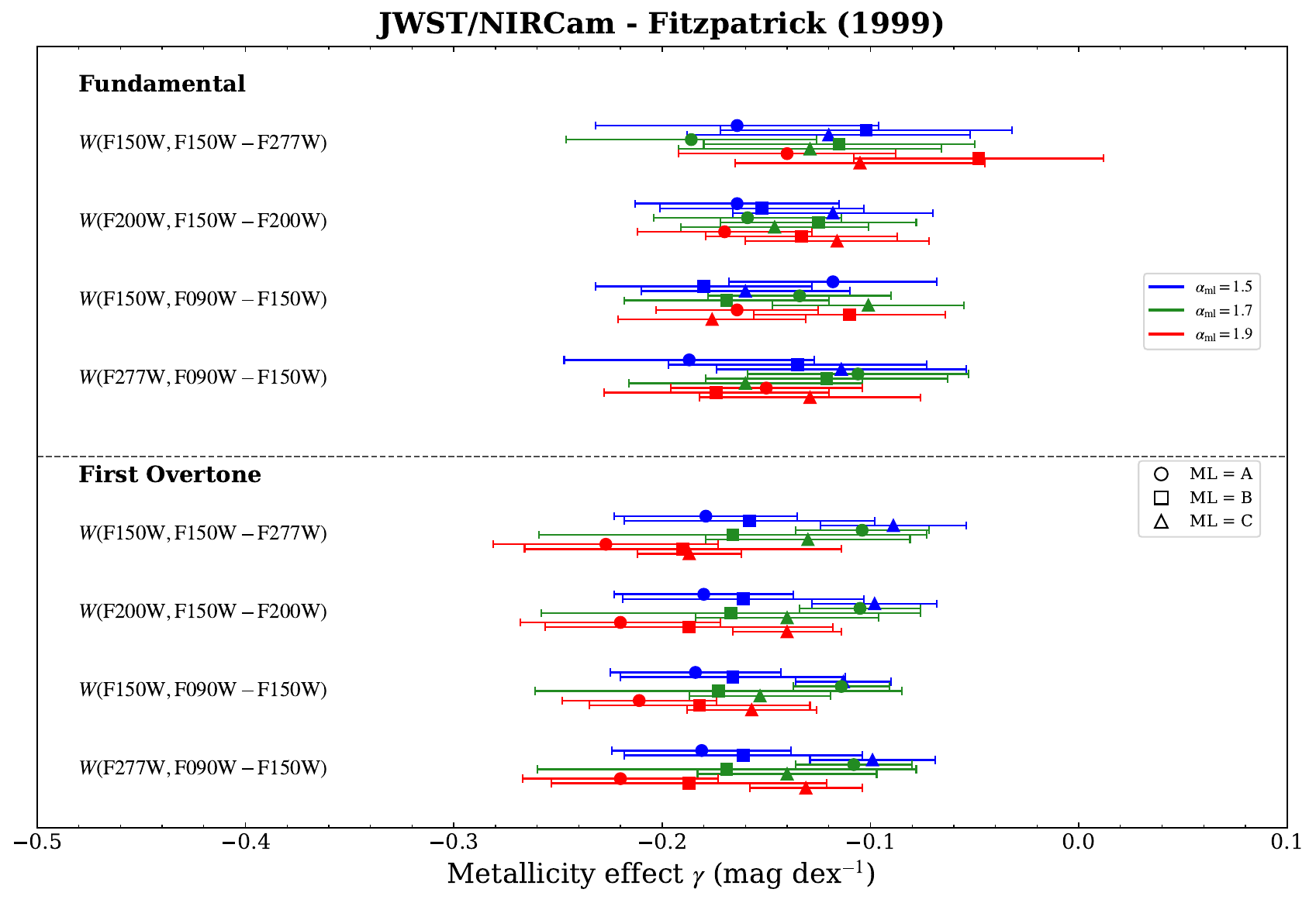}
\caption{
Metallicity coefficient $\gamma$ derived from the theoretical
PWZ relations in the JWST photometric system using the
\citet{Fitzpatrick1999} extinction law, adopted throughout the
comparison with the SH0ES observations.
Colors denote different values of $\alpha_{\rm ml}$, while symbols
identify the adopted ML relations.
Error bars represent the formal uncertainties on $\gamma$.
}
\label{fig:pwz_gamma_comparison_JWST}
\end{figure}

Figure~\ref{fig:pwz_gamma_comparison_JWST} summarizes the behaviour of the metallicity coefficient $\gamma$ for the explored JWST PWZ relations computed using the \citet{Fitzpatrick1999} extinction law, adopted throughout the comparison with the SH0ES observations. The corresponding JWST results based on the \citet{Cardelli1989} and \citet{Wang2024} extinction laws are presented in Appendix~\ref{sec:pwz_appendix} (Figures~\ref{fig:pwz_gamma_appendix}). In all cases, the inferred metallicity coefficients exhibit nearly identical behaviour.

The metallicity effect depends on the adopted Wesenheit combination. In the JWST system, all four Wesenheit definitions investigated, namely $W({\rm F150W},{\rm F150W-F277W})$, $W({\rm F200W},{\rm F150W-F200W})$, $W({\rm F150W},{\rm F090W-F150W})$, and $W({\rm F277W},{\rm F090W-F150W})$, yield metallicity coefficients that generally lie in the range $-0.10 \lesssim \gamma \lesssim -0.23$ mag dex$^{-1}$. 

The Roman PWZ relations exhibit a very similar behaviour (Appendix~\ref{sec:pwz_appendix}, Figure~\ref{fig:pwz_gamma_comparison_roman}).
The two Wesenheit definitions investigated yield metallicity coefficients that generally lie in the range $-0.10\lesssim\gamma\lesssim-0.22$\,mag\,dex$^{-1}$. Among them, the $W(\mathrm{F184},\mathrm{F158}-\mathrm{F184})$ relation generally exhibits the strongest metallicity dependence, whereas $W(\mathrm{F158},\mathrm{F129}-\mathrm{F158})$ yields slightly smaller absolute values of $\gamma$.

The inferred metallicity coefficients also retain a non-negligible sensitivity to the adopted physical assumptions. Variations in the mixing-length parameter and ML relation can introduce differences of several hundredths of a magnitude per dex, particularly for FO pulsators. Overall, most of the predicted coefficients fall within the interval currently discussed in the recent empirical and theoretical literature \citep[e.g.,][and references therein]{Anderson2016, Breuval2022, Breuval2024, Bhardwaj2023, Bhardwaj2024, Cruz2023, Desomma2022, Khan2025, Molinaro2023, Ripepi2021, Ripepi2026, Trentin2024}.

Moreover, and as expected on the basis of what was already found for the PW relations, the predicted PWZ relations show only a weak dependence on the adopted reddening law. For the JWST system, the metallicity coefficients derived using the Cardelli, Wang, and Fitzpatrick prescriptions remain fully consistent within the quoted uncertainties.

Similarly, the Roman PWZ relations computed using the Cardelli and Fitzpatrick extinction laws show only marginal differences. The variations introduced by the adopted reddening prescription are generally much smaller than those associated with changes in the stellar parameters and pulsation assumptions, indicating that the inferred metallicity dependence is largely insensitive to the specific extinction law adopted in the NIR regime.

These results confirm that the metallicity dependence is generally reduced in near-infrared Wesenheit relations. While several filter combinations exhibit metallicity coefficients of about $-0.1$\,mag\,dex$^{-1}$, others reach values of up to about $-0.2$\,mag\,dex$^{-1}$, indicating that the metallicity dependence is not completely removed. The derived PWZ relations therefore provide an appropriate theoretical framework for inferring CC distances in galaxies spanning a broad range of chemical compositions. 

\begin{table*}[t]
\caption{\label{tab:pwz_all}
Coefficients of the theoretical PWZ relations
$W_{\lambda}=\alpha+\beta\log P+\gamma[\mathrm{Fe/H}]$
derived for the JWST and Roman photometric systems.
Columns report the adopted extinction law, Wesenheit function, pulsation mode,
mixing-length parameter, ML relation, fitted coefficients and uncertainties,
rms dispersion ($\sigma$), and coefficient of determination ($R^2$).
The metallicity term is defined as
$[\mathrm{Fe/H}]=\log(Z/Z_{\odot})$, with $Z_{\odot}=0.02$.}
\centering
\scriptsize
\setlength{\tabcolsep}{2pt}
\begin{tabular}{cccccccccccccc}
\hline\hline
Extinction law & Band & CI & Mode & $\alpha_{\rm ml}$ & ML &
$\alpha$ & $\beta$ & $\gamma$ &
$\sigma_{\alpha}$ & $\sigma_{\beta}$ & $\sigma_{\gamma}$ &
$\sigma$ & $R^2$ \\
\hline
\multicolumn{14}{c}{JWST}\\
\hline
Cardelli1989 & F150W & F090W$-$F150W & F & 1.5 & A & -2.849 & -3.463 & -0.125 & 0.033 & 0.018 & 0.047 & 0.089 & 0.996 \\
Cardelli1989 & F150W & F090W$-$F150W & F & 1.5 & B & -2.744 & -3.400 & -0.191 & 0.035 & 0.017 & 0.049 & 0.102 & 0.995 \\
\hline
\multicolumn{14}{c}{Roman}\\
\hline
Cardelli1989 & F158 & F129W$-$F158W & F & 1.5 & A & -2.948 & -3.608 & -0.129 & 0.020 & 0.011 & 0.028 & 0.053 & 0.999 \\
Cardelli1989 & F158 & F129W$-$F158W & F & 1.5 & B & -2.856 & -3.537 & -0.111 & 0.023 & 0.011 & 0.032 & 0.067 & 0.998 \\
\hline
\end{tabular}
\tablefoot{
The table reports a representative extract of the complete datasets.
The full machine-readable tables, including all adopted extinction laws,
pulsation modes, mixing-length parameters, and ML relations, are available
in the online supplementary material as \texttt{PWZ\_JWST.dat} and
\texttt{PWZ\_Roman.dat}.
}
\end{table*}

\section{Application to JWST CC observations}

\subsection{Comparison with observed Period-Wesenheit relations}

To assess the predictive capability of the new JWST PW relations, we compared the theoretical relations derived in this work with the CC observations presented by \citet{Riess20248sigma}. Their sample includes more than one thousand CCs observed with JWST in the geometric distance anchor NGC~4258 and in five Type Ia supernova host galaxies (NGC~5584, NGC~5643, NGC~1559, NGC~5468, and NGC~1448), providing one of the largest homogeneous JWST CC datasets currently available.

Following the definitions adopted by \citetalias{Riess20248sigma}, we considered three Wesenheit formulations:

\begin{equation}
W_H = F150W - 0.4(F555W-F814W),
\end{equation}

\begin{equation}
W_{F150W} = F150W - 0.72(F090W-F150W),
\end{equation}

\begin{equation}
W_{F277W} = F277W - 0.30(F090W-F150W),
\end{equation}

corresponding to the hybrid JWST+HST NIR relation, the JWST-only NIR relation, and the JWST-only mid-infrared (MIR) relation, respectively. These Wesenheit magnitudes were defined by \citet{Riess2024} using the \citet[][]{Fitzpatrick1999} reddening law.

For each galaxy, the apparent Wesenheit magnitudes provided by \citetalias{Riess20248sigma} were transformed into absolute magnitudes using the published distance moduli and metallicity corrections adopted in their analysis (see Table 3 of \citetalias[][]{Riess20248sigma}). We then superimposed the observed CCs on the corresponding theoretical PW relations computed using the Fitzpatrick reddening law

The comparison was initially carried out using the full set of F-mode models computed for $\alpha_{\rm ml}=1.5$ and $1.7$ and for both canonical (ML=A) and mildly overluminous (ML=B) ML assumptions. The predicted PW relations show only a very weak dependence on the adopted mixing-length parameter. For this reason, and to improve the readability of the figures, only the $\alpha_{\rm ml}=1.5$ models are displayed in the final comparison plots, for both ML=A and ML=B.

\begin{figure*}
\centering
\includegraphics[width=0.34\textwidth]{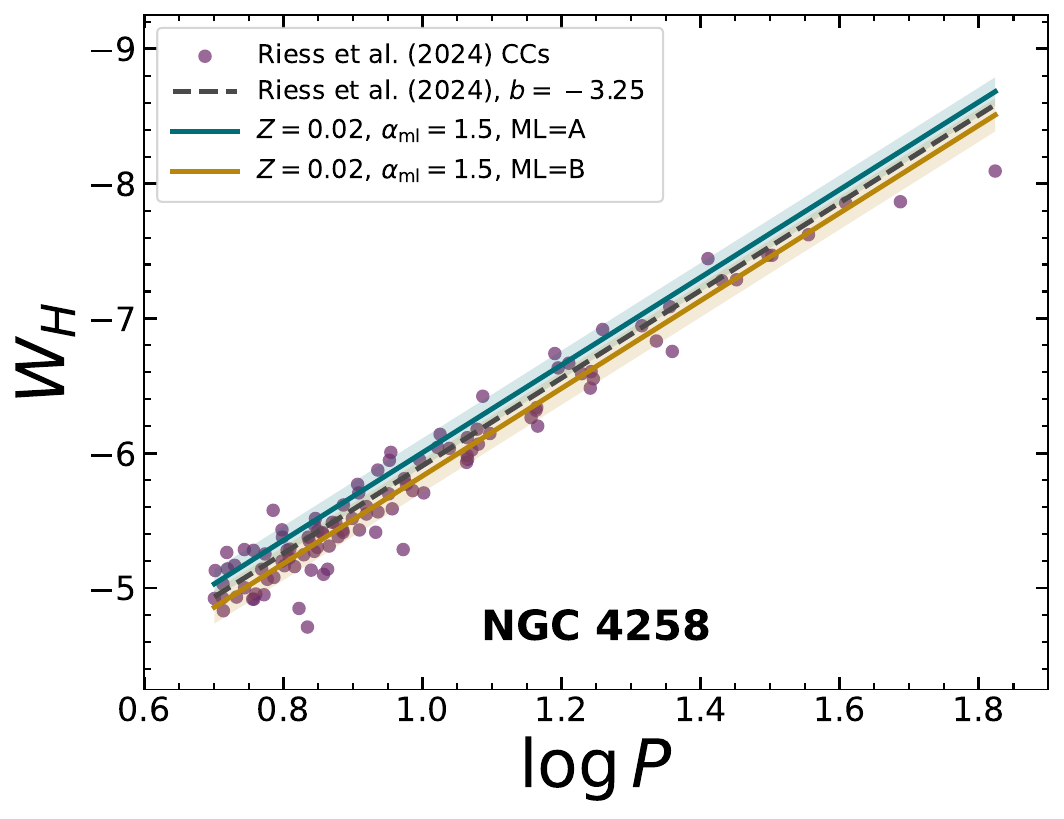}
\vspace{0.2cm}
\includegraphics[width=0.34\textwidth]{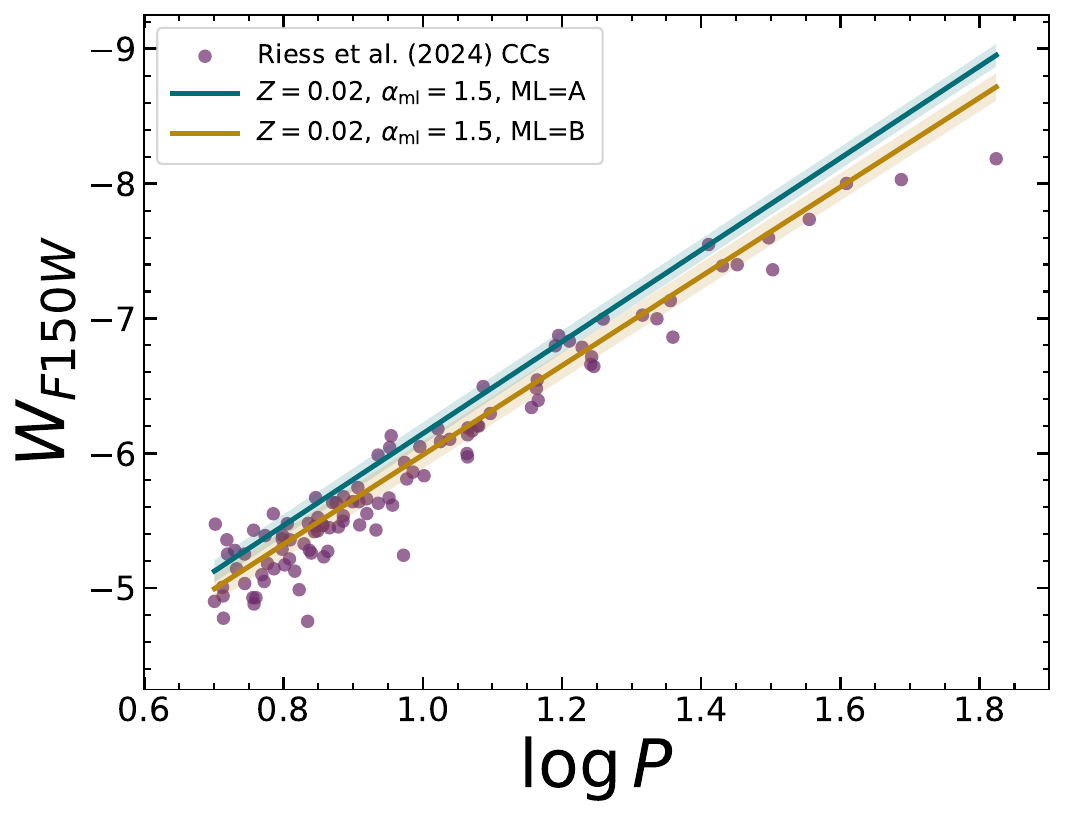}
\vspace{0.2cm}
\includegraphics[width=0.34\textwidth]{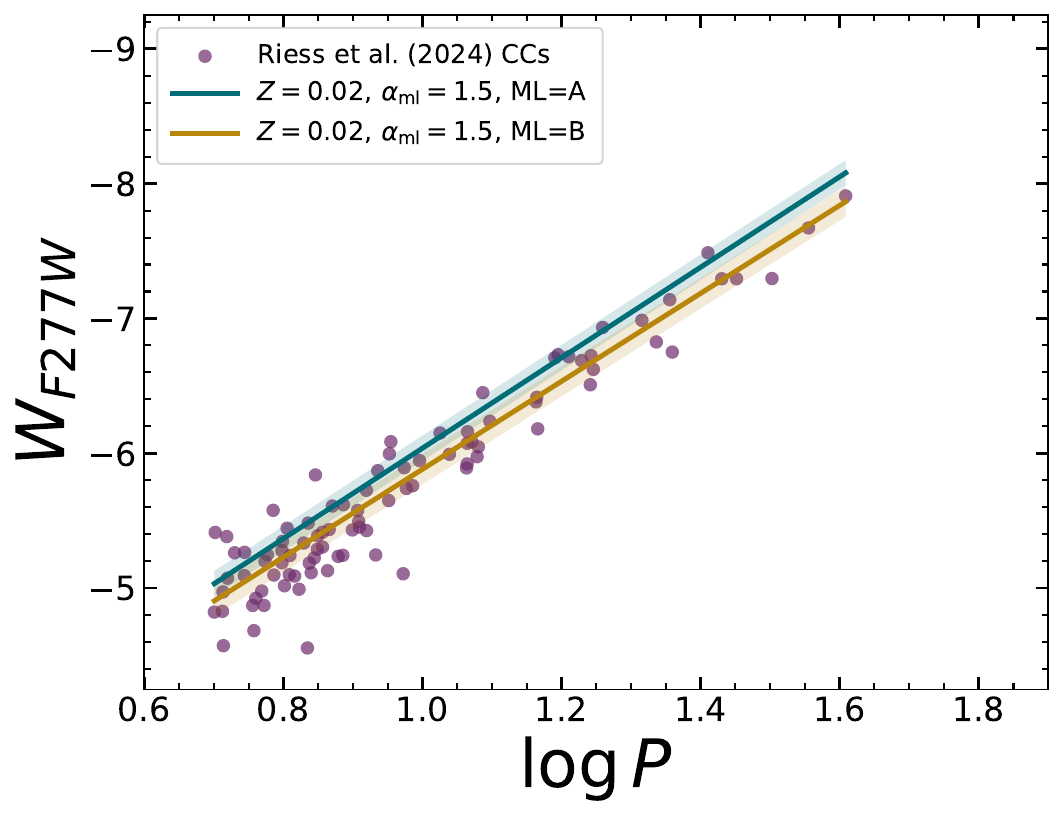}
\caption{
Comparison between the observed JWST CC Wesenheit relations in NGC~4258 and the theoretical predictions based on pulsation models with $Z=0.02$ and $\alpha_{\rm ml}=1.5$. The panels show the hybrid JWST+HST Wesenheit relation (upper left), the JWST NIR Wesenheit relation based on $W(F150W,F090W-F150W)$ (upper right), and the JWST MIR Wesenheit relation based on $W(F277W,F090W-F150W)$ (bottom). The filled circles represent the observed Cepheids from \citet{Riess2024}. Solid lines correspond to the theoretical PW relations for the canonical (ML=A) and noncanonical (ML=B) ML assumptions, while the shaded regions indicate the intrinsic dispersion of the theoretical relations. In the top panel, the dashed line shows the empirical SH0ES calibration derived only from the hybrid JWST+HST Wesenheit relation in \citetalias{Riess20248sigma}, adopting a fixed slope of $-3.25$.
}
\label{fig:ngc4258_pw}
\end{figure*}

Figure \ref{fig:ngc4258_pw} shows the comparison between the observed JWST CCs in NGC 4258 and the theoretical PW relations. Analogous figures for the remaining host galaxies are provided in Appendix~\ref{sec:w_comp}.

The observed CCs closely follow the theoretical PW relations over the entire period range covered by the observations. The agreement is particularly remarkable for the hybrid relation $W_H$, in the case ML=B, with the theoretical sequences reproducing both the observed slope and zero-point of the Cepheid distribution. Case ML=B is more consistent with the observations than ML=A, also in the other two Wasenheit formulations. The agreement confirms that the theoretical framework developed in this work provides a consistent description of the observed JWST CC population.

This result is consistent with previous empirical tests of nonlinear pulsation models. In particular, \citetalias{Desomma2022} showed that PW relations based on mildly overluminous models provide distance moduli for the Large Magellanic Cloud (LMC) that are in better agreement with the most accurate geometric distance determinations, yielding values close to $\mu = 18.477 \pm 0.026$ mag \citep{Pietr2019}.

The preference for the ML=B scenario found here therefore reinforces the evidence that a moderate overluminosity relative to canonical evolutionary predictions may offer a more realistic description of CCs.

The good agreement between theory and observations supports the reliability of the newly derived JWST PW relations and motivates their use as distance indicators, which is explored in the following section through a direct comparison of theoretical and observed distance moduli.

\subsection{Comparison between theoretical and observed distance moduli}

As a further validation of the theoretical PW relations, distance moduli were derived for the six SN~Ia host galaxies analyzed by \citetalias{Riess20248sigma}. The comparison was performed using the three adopted Wesenheit definitions, named $W_{H}$, $W_{F150W}$, and $W_{F277W}$ and the \citet{Fitzpatrick1999} reddening laws. For each Cepheid, an individual theoretical distance modulus was derived as the difference between the observed Wesenheit magnitude and the absolute Wesenheit magnitude predicted by the corresponding theoretical PW relation at the same pulsation period. The distance modulus of each host galaxy was then estimated as the mean value of the individual CC moduli. Distance moduli were computed using the theoretical PW relations corresponding to the Galactic chemical composition ($Z=0.02$, $Y=0.28$), since the CC populations in the SH0ES host galaxies considered are characterized by metallicities close to the solar value,
considering the four different model assumptions, namely $\alpha_{\rm ml}=1.5$ and $1.7$ combined with both ML=A and ML=B relations. This should be regarded as a first-order approximation, since the observed CC span a significant range of galactocentric distances and are therefore expected to sample different metallicities within their host galaxies. The resulting distance moduli are listed in Table~\ref{tab:mu_theory_riess}, together with the values reported by \citetalias{Riess20248sigma}. As already suggested by the comparison of the PW relations, the inferred distance moduli are only marginally affected by the adopted mixing-length parameter. Differences between the $\alpha_{\rm ml}=1.5$ and $\alpha_{\rm ml}=1.7$ solutions are typically below a few hundredths of a magnitude, confirming the weak sensitivity of NIR Wesenheit relations to the efficiency of superadiabatic convection. Conversely, the choice of the ML relation produces a systematic effect, with the noncanonical models generally yielding distance moduli in closer agreement with the empirical determinations.

To show this behavior, Figure~\ref{fig:mu_comparison_hybrid} compares the theoretical and observed distance moduli obtained from the hybrid Wesenheit relation. Since the dependence on $\alpha_{\rm ml}$ is negligible, only the $\alpha_{\rm ml}=1.5$ solutions are shown. The left panel displays the results obtained assuming the canonical ML relation (ML=A), while the right panel corresponds to the noncanonical case (ML=B). The upper panels show the direct comparison between the theoretical distance moduli and those derived by \citetalias{Riess20248sigma}, whereas the lower panels present the residuals. 

The ML=A models systematically overestimate the observed distance moduli by approximately $0.1$-$0.2$ mag, while the ML=B models produce significantly smaller residuals and a distribution more closely centered around $\Delta\mu = 0$ mag. This result is fully consistent with the empirical tests discussed in the previous subsection.

Overall, the comparison demonstrates that the theoretical JWST Wesenheit relations reproduce the observed CC distance scale with a high degree of accuracy and further confirms that mildly overluminous pulsation models provide the most consistent description of the available observations.

\begin{table*}[ht!]
\caption{\label{tab:mu_theory_riess}
Comparison between the distance moduli derived by \citet{Riess20248sigma} and those inferred from the theoretical JWST PW relations.}
\centering
\scriptsize
\setlength{\tabcolsep}{4pt}
\renewcommand{\arraystretch}{1.05}
\begin{tabular}{lccccccccccccc}
\hline\hline
Galaxy &
$\mu_{\rm Riess}$ &
\multicolumn{4}{c}{$\mu(W_H)$} &
\multicolumn{4}{c}{$\mu(W_{\rm F150W})$} &
\multicolumn{4}{c}{$\mu(W_{\rm F277W})$} \\
\cline{3-6}
\cline{7-10}
\cline{11-14}
&
(mag) &
1.5A & \cellcolor{gray!15}1.5B & 1.7A & 1.7B &
1.5A & \cellcolor{gray!15}1.5B & 1.7A & 1.7B &
1.5A & \cellcolor{gray!15}1.5B & 1.7A & 1.7B \\
\hline
NGC~1448 & 31.289 & 31.438 & \cellcolor{gray!15}31.265 & 31.463 & 31.276 &
31.510 & \cellcolor{gray!15}31.308 & 31.545 & 31.310 &
31.452 & \cellcolor{gray!15}31.253 & 31.490 & 31.255 \\
NGC~1559 & 31.371 & 31.473 & \cellcolor{gray!15}31.300 & 31.498 & 31.311 &
31.660 & \cellcolor{gray!15}31.454 & 31.697 & 31.457 &
31.617 & \cellcolor{gray!15}31.413 & 31.657 & 31.415 \\
NGC~4258 & 29.397 & 29.550 & \cellcolor{gray!15}29.377 & 29.575 & 29.388 &
29.608 & \cellcolor{gray!15}29.452 & 29.626 & 29.450 &
29.578 & \cellcolor{gray!15}29.425 & 29.599 & 29.425 \\
NGC~5468 & 32.975 & 33.156 & \cellcolor{gray!15}32.983 & 33.181 & 32.994 &
33.343 & \cellcolor{gray!15}33.117 & 33.387 & 33.121 &
33.236 & \cellcolor{gray!15}33.013 & 33.283 & 33.016 \\
NGC~5584 & 31.838 & 32.010 & \cellcolor{gray!15}31.837 & 32.035 & 31.848 &
32.141 & \cellcolor{gray!15}31.939 & 32.176 & 31.941 &
32.034 & \cellcolor{gray!15}31.834 & 32.072 & 31.836 \\
NGC~5643 & 30.520 & 30.592 & \cellcolor{gray!15}30.419 & 30.617 & 30.430 &
30.705 & \cellcolor{gray!15}30.504 & 30.739 & 30.506 &
30.666 & \cellcolor{gray!15}30.467 & 30.703 & 30.469 \\
\hline
\end{tabular}
\tablefoot{
The empirical distance moduli are taken from \citet{Riess20248sigma}. The theoretical moduli are derived using the hybrid Wesenheit relation
$W_H=F150W-0.40(V-I)$,
the JWST near-infrared relation
$W_{\rm F150W}=F150W-0.72(F090W-F150W)$,
and the JWST mid-infrared relation
$W_{\rm F277W}=F277W-0.30(F090W-F150W)$.
The shaded columns correspond to the $\alpha_{\rm ml}=1.5$, ML relation B case, namely the mildly overluminous models, which provide the best overall agreement with the empirical distance moduli.
}
\end{table*}

\begin{figure*}[t]
\centering
\setlength{\tabcolsep}{1pt}
\begin{tabular}{cc}
\includegraphics[width=0.4\textwidth]{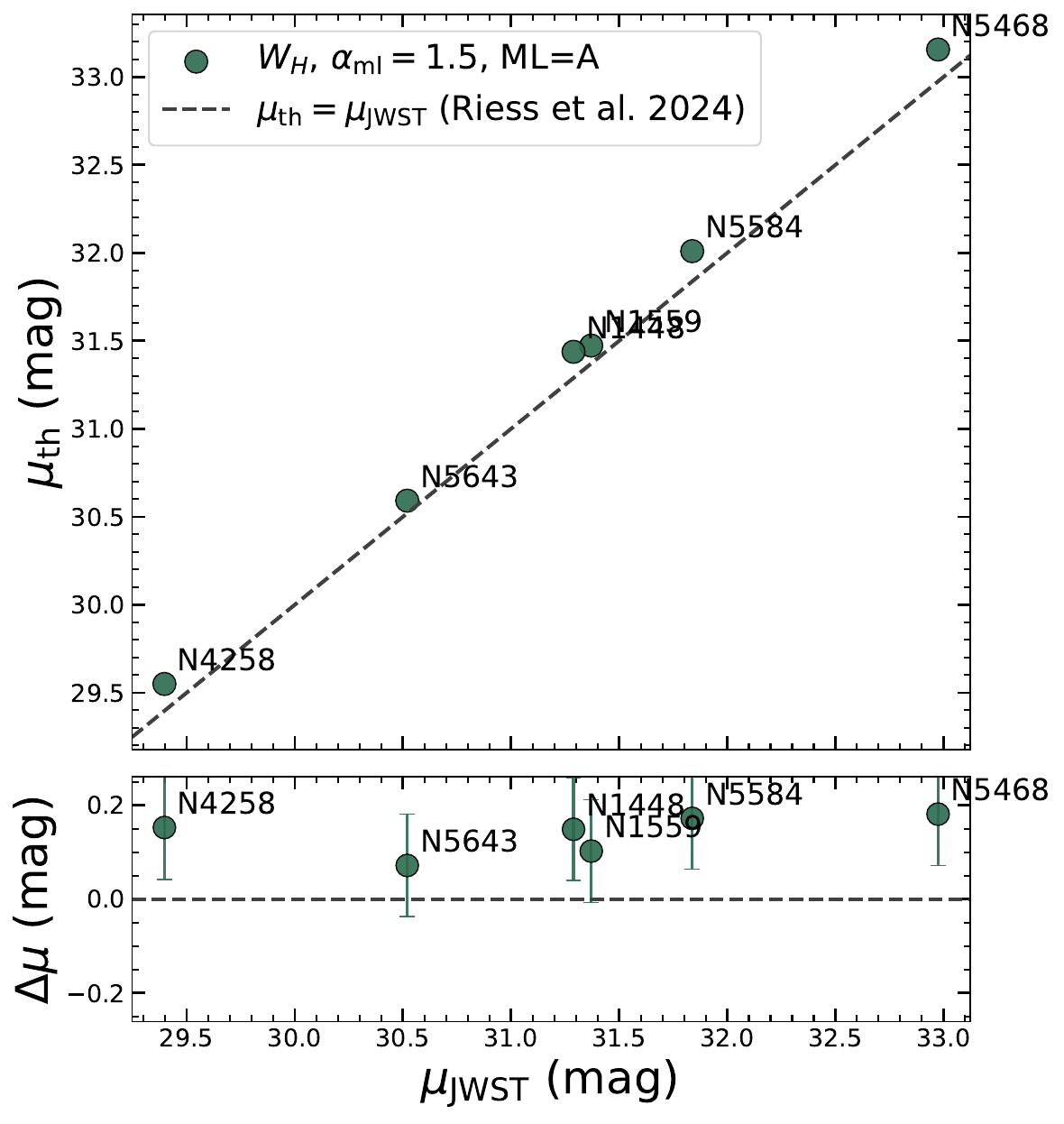} &
\includegraphics[width=0.4\textwidth]{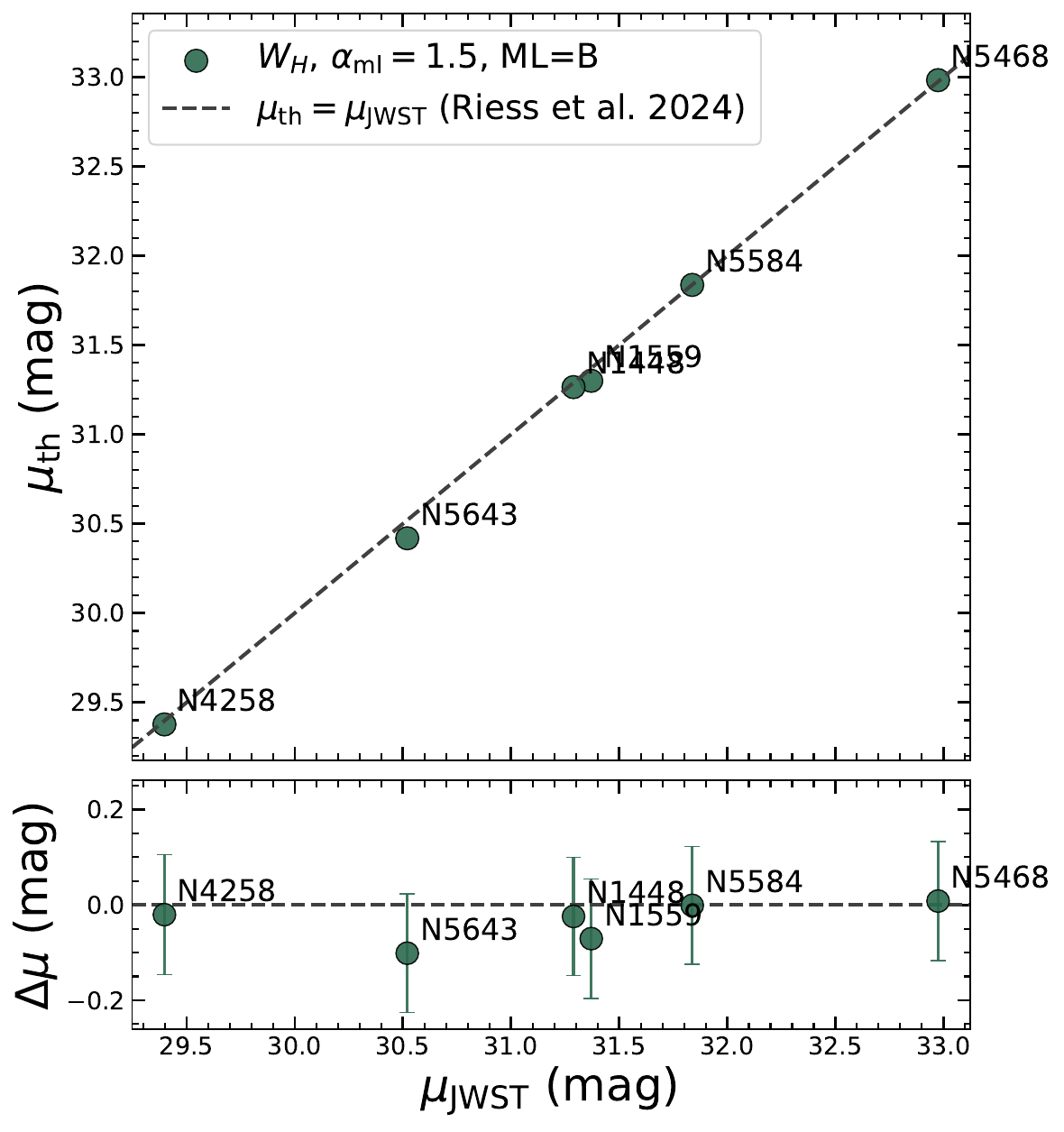}
\end{tabular}
\caption{
Comparison between the distance moduli derived from the theoretical hybrid Period--Wesenheit relation,
$W_H = F150W - 0.4(F555W-F814W)$,
and the observed JWST distance moduli from \citetalias{Riess20248sigma}. The left and right panels show the results obtained assuming the canonical (ML=A) and mildly overluminous (ML=B) mass--luminosity relations, respectively, both computed with $\alpha_{\rm ml}=1.5$. In each panel, the upper sub-panel compares the theoretical and observed distance moduli, while the lower sub-panel shows the residuals,
$\Delta\mu=\mu_{\rm th}-\mu_{\rm JWST}$.
The dashed line in the upper sub-panels indicates the one-to-one relation, whereas the horizontal dashed line in the residual panels marks $\Delta\mu=0$.
}
\label{fig:mu_comparison_hybrid}
\end{figure*}

\section{Conclusions}

In this paper, we have presented the first homogeneous theoretical framework for CCs in the JWST and Roman photometric systems, based on an extensive grid of nonlinear convective pulsation models spanning a wide range of metallicities, ML assumptions, and superadiabatic convective efficiencies. The theoretical bolometric light curves were transformed into the JWST and Roman passbands, allowing a detailed characterization of CC pulsation properties in these passbands.

The main results of this work can be summarized as follows:

\begin{itemize}

\item We derived, for both F and FO pulsators, new theoretical PLC, PW, and metal-dependent PW relations in the JWST and Roman filter systems. The resulting relations exhibit very small intrinsic dispersions, confirming the excellent distance-scale potential of NIR observations.

\item The transformed light curves allowed us to investigate the wavelength dependence of pulsation amplitudes and the morphology of the HP in the JWST and Roman bands. As expected, pulsation amplitudes decrease toward longer wavelengths, while the characteristic signatures of the HP remain detectable across the explored filters.

\item The derived PW relations show only a weak dependence on the adopted reddening law. The coefficients obtained using the \citet[][]{Cardelli1989}, \citet[][]{Fitzpatrick1999}, and \citet[][]{Wang2024} extinction prescriptions remain consistent within the uncertainties, indicating that extinction-law systematics are strongly reduced in the near- and mid-infrared regime.

\item The metallicity dependence is generally weaker than in optical Wesenheit relations, but it is not completely removed. The derived PWZ relations predict metallicity coefficients typically in the range $-0.10 \lesssim \gamma \lesssim -0.23$\,mag\,dex$^{-1}$, depending on the adopted Wesenheit definition and pulsation assumptions. These values are broadly consistent with recent empirical and theoretical determinations reported in the literature.

\item We derived, for the first time, purely JWST-based and Roman-based PWZ relations, providing a theoretical framework reliable for distance determinations in galaxies spanning a broad range of chemical compositions.

\item To enable a direct comparison with the SH0ES framework, we also constructed hybrid JWST-HST PW relations following the formulation adopted by \citetalias[][]{Riess20248sigma}. The predicted slopes are generally consistent with the value adopted by the SH0ES collaboration. On the other hand, the zero points show a clear dependence on both metallicity and the ML relation. 

\item We compared the theoretical PW relations with the recent JWST CC observations obtained by the SH0ES collaboration in NGC 4258 and five Type Ia supernova host galaxies. The agreement between theoretical predictions and observed Wesenheit relations is generally very good. The largest systematic differences are associated with the adopted ML relation, with mildly overluminous models typically providing the best overall agreement.

\item Distances derived from the theoretical hybrid PW relations are consistent with the empirical JWST determinations within the quoted uncertainties, supporting the reliability of the pulsation models and their applicability to precision distance-scale studies.

\end{itemize}

\begin{acknowledgements}
We sincerely thank the anonymous Referee for the very positive feedback on our work and for the insightful suggestion, which helped us further improve the manuscript and highlight an interesting aspect of our results.
The authors acknowledge funding from Gaia DPAC through INAF and ASI (2025-10-HH.0; PI: M.G. Lattanzi).
This project has received funding from the PRIN MUR 2022 project (code 2022ARWP9C) ``Early Formation and Evolution of Bulge and Halo (EFEBHO)'', PI: M. Marconi, funded by the European Union -- Next Generation EU, and from the Large Grant INAF 2023 MOVIE, PI: M. Marconi.
We acknowledge the INAF GO-GTO grant 2023 ``C-MetaLL -- Cepheid Metallicities in the Leavitt Law'' (PI: V. Ripepi).
G.D.S. and T.S. thank INFN, Naples section, for support via QGSKY initiatives, with additional INFN support for Moonlight2 (G.D.S.).
M.M., V.R., and M.D. acknowledge ISSI for funding the project ``EXPANDING'' (ISSI--ISSI Beijing Team).
This research also benefited from COST Action CA21136 (CosmoVerse), addressing cosmological tensions through systematics and fundamental physics, funded by COST (European Cooperation in Science and Technology).
\end{acknowledgements}

\bibliographystyle{aa}
\bibliography{desomma_main_apj}

\appendix

\section{FO-mode JWST and Roman theoretical light curves}
\label{sec:lc}

\begin{figure*}[t]
\centering
\setlength{\tabcolsep}{1pt}
\begin{tabular}{cc}
\includegraphics[width=0.28\textwidth]{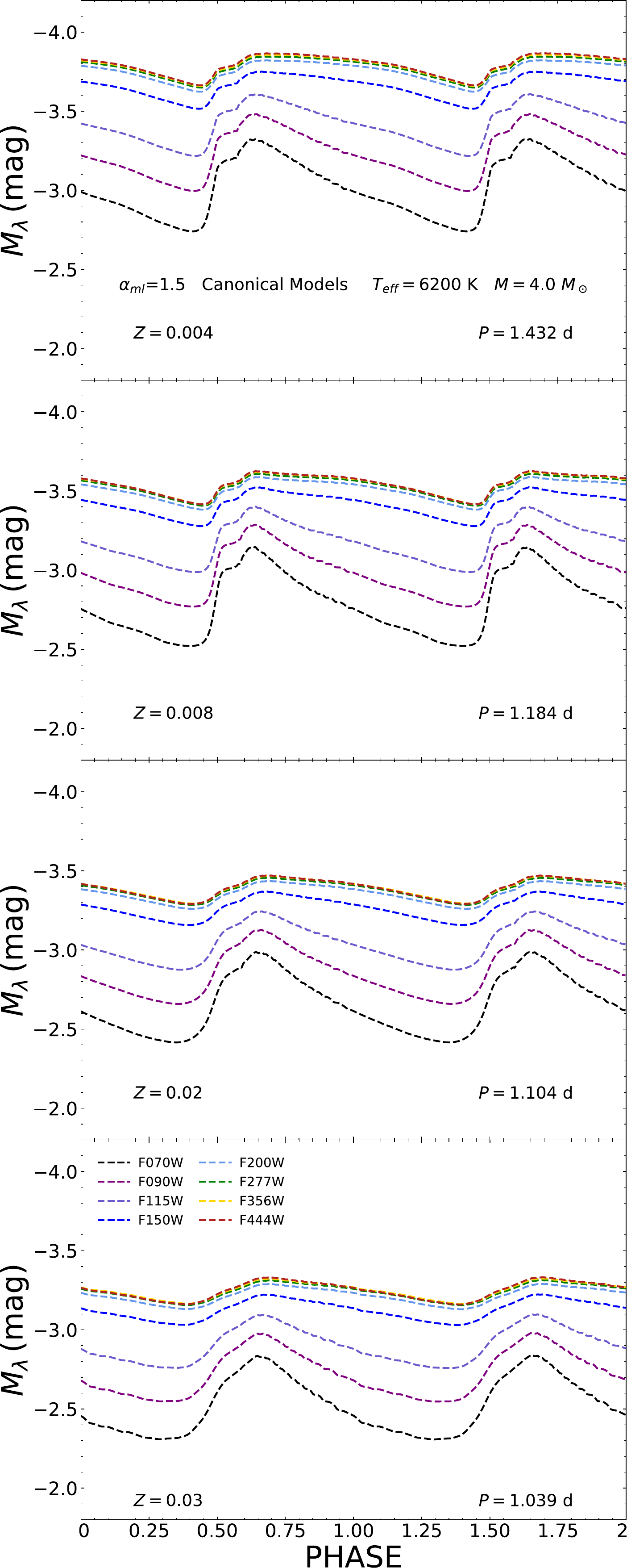} &
\includegraphics[width=0.28\textwidth]{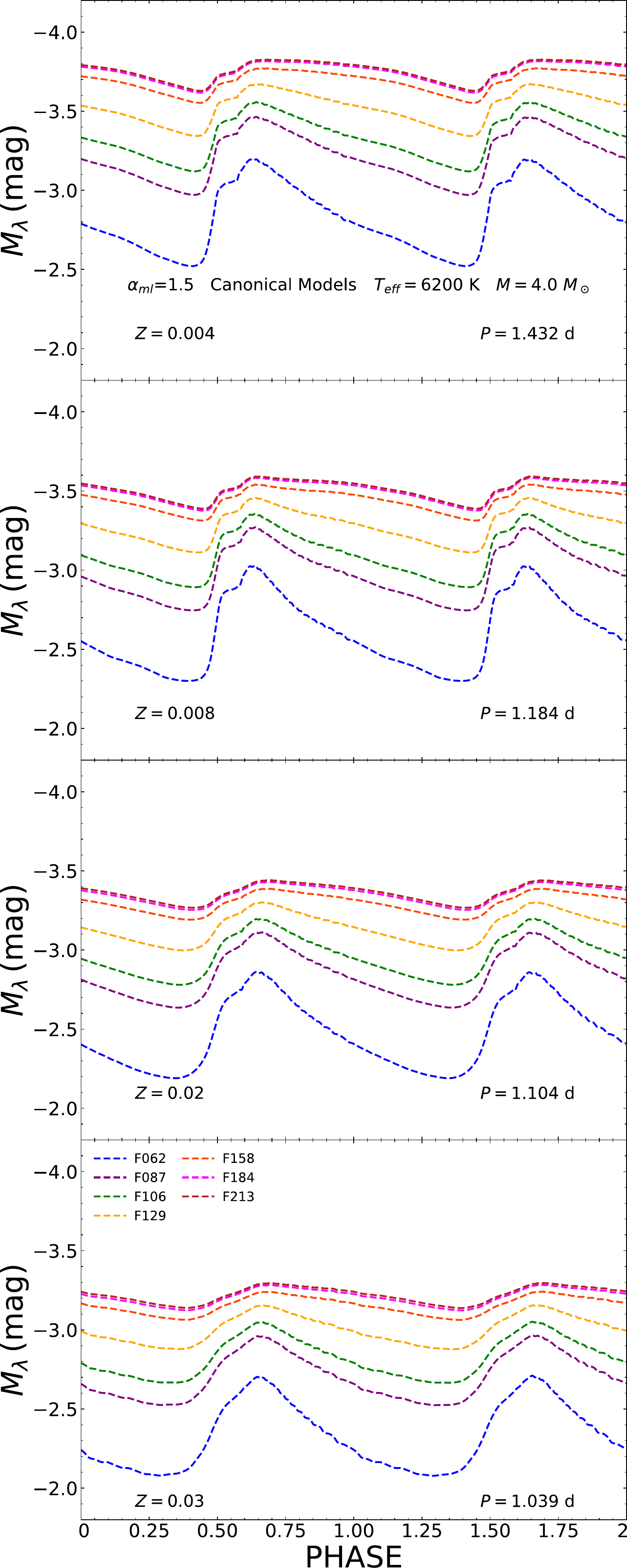}
\end{tabular}
\caption{
Same as Figure~\ref{fig:lc_F}, but for representative FO-mode CC models.}
\label{fig:lc_FO}
\end{figure*}

Figure~\ref{fig:lc_FO} presents the corresponding representative theoretical light curves for FO-mode CC models transformed into the JWST (left panel) and Roman (right panel) photometric systems. The adopted stellar mass, mixing-length parameter, and position within the instability strip are the same as those assumed for the F-mode models shown in Figure~\ref{fig:lc_F}. As in the main text, the panels are ordered according to increasing metallicity, while the colors identify filters from shorter to longer wavelengths.

\section{The Hertzsprung Progression in the Roman Photometric System}
\label{sec:h_prog}

\begin{figure*}[t]
\centering
\includegraphics[width=0.9\linewidth]{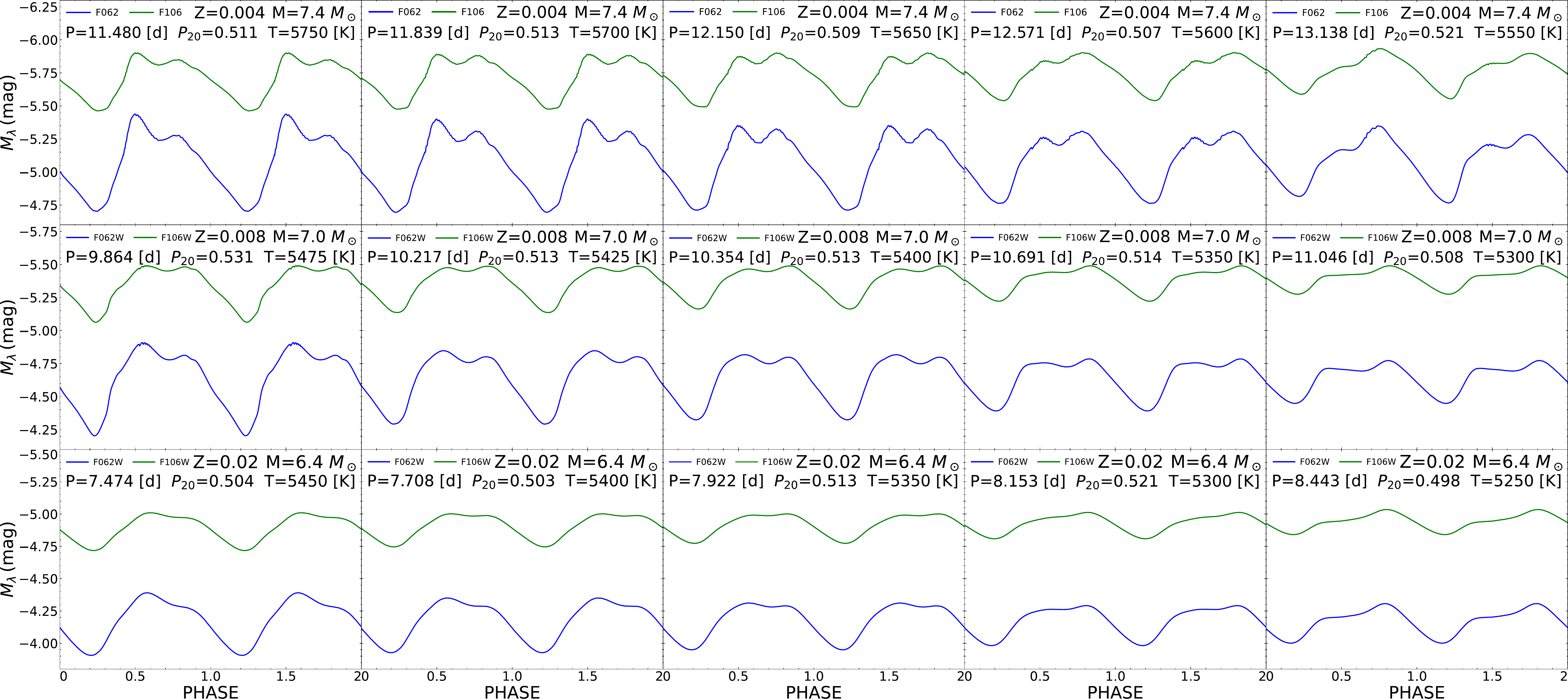}
\caption{
Same as Figure~\ref{fig:prog_jwst}, but for the Roman filters F062 (blue) and F106 (green).
}
\label{fig:prog_roman}
\end{figure*}

Figure~\ref{fig:prog_roman} presents the corresponding theoretical Roman light curves illustrating the HP for the same representative F-mode CC models discussed in subsection~\ref{sec:h_prog_text}. The adopted stellar parameters are identical to those used for the JWST models shown in Figure~\ref{fig:prog_jwst}. The light curves are displayed in the F062 (blue) and F106 (green) filters, chosen to maximize the visibility of the bump feature. The same metallicity dependence observed in the JWST light curves is recovered, with the centre of the HP shifting toward longer periods as the metal abundance decreases.

\section{Amplitude properties}
\label{sec:amplitude_appendix}

\subsection{Period-amplitude diagrams}
\label{sec:amp_diag}

\begin{figure*}
\centering
\setlength{\tabcolsep}{1pt}
\renewcommand{\arraystretch}{1.0}
\begin{tabular}{cc}
\includegraphics[width=0.33\textwidth]{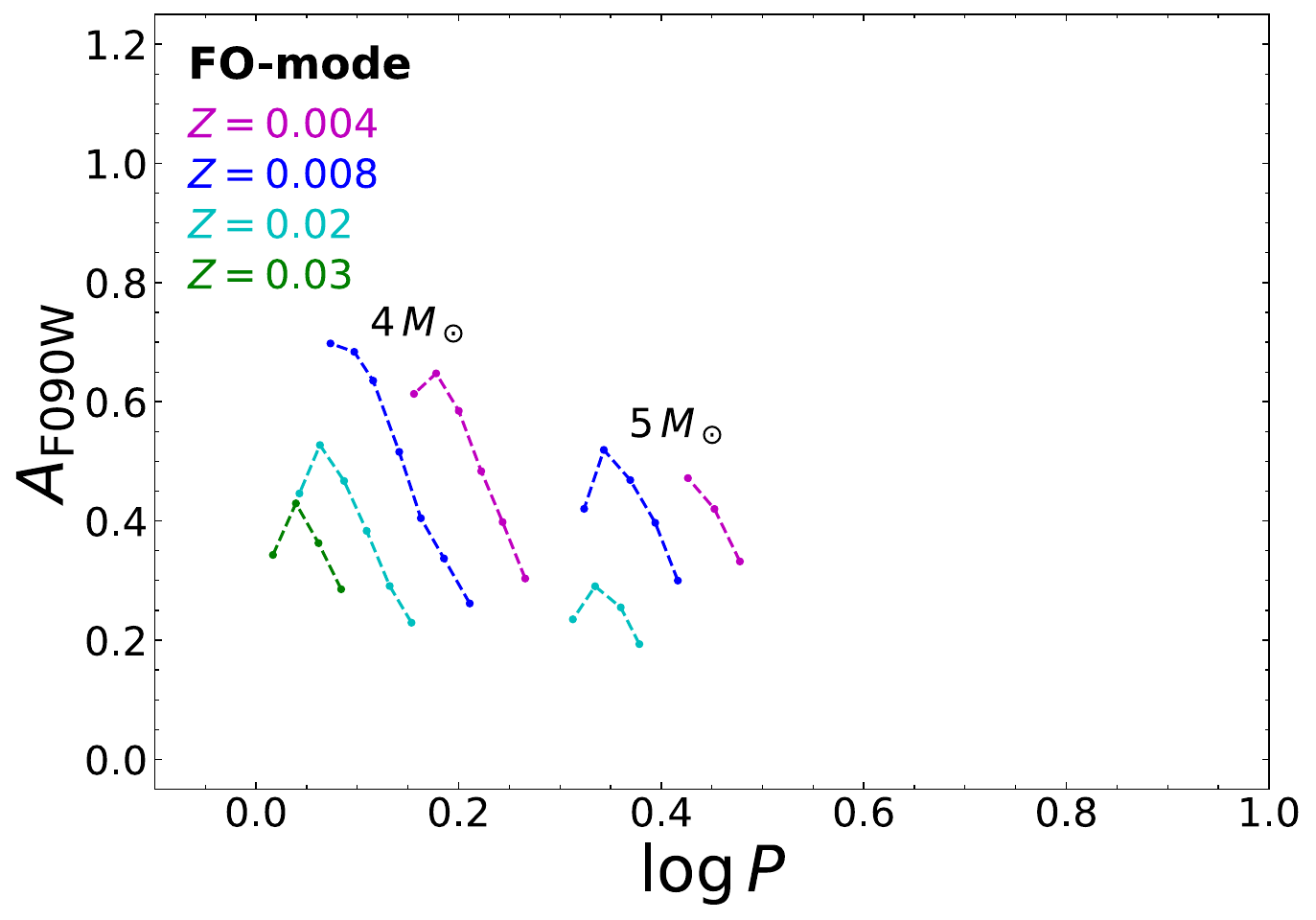} &
\includegraphics[width=0.33\textwidth]{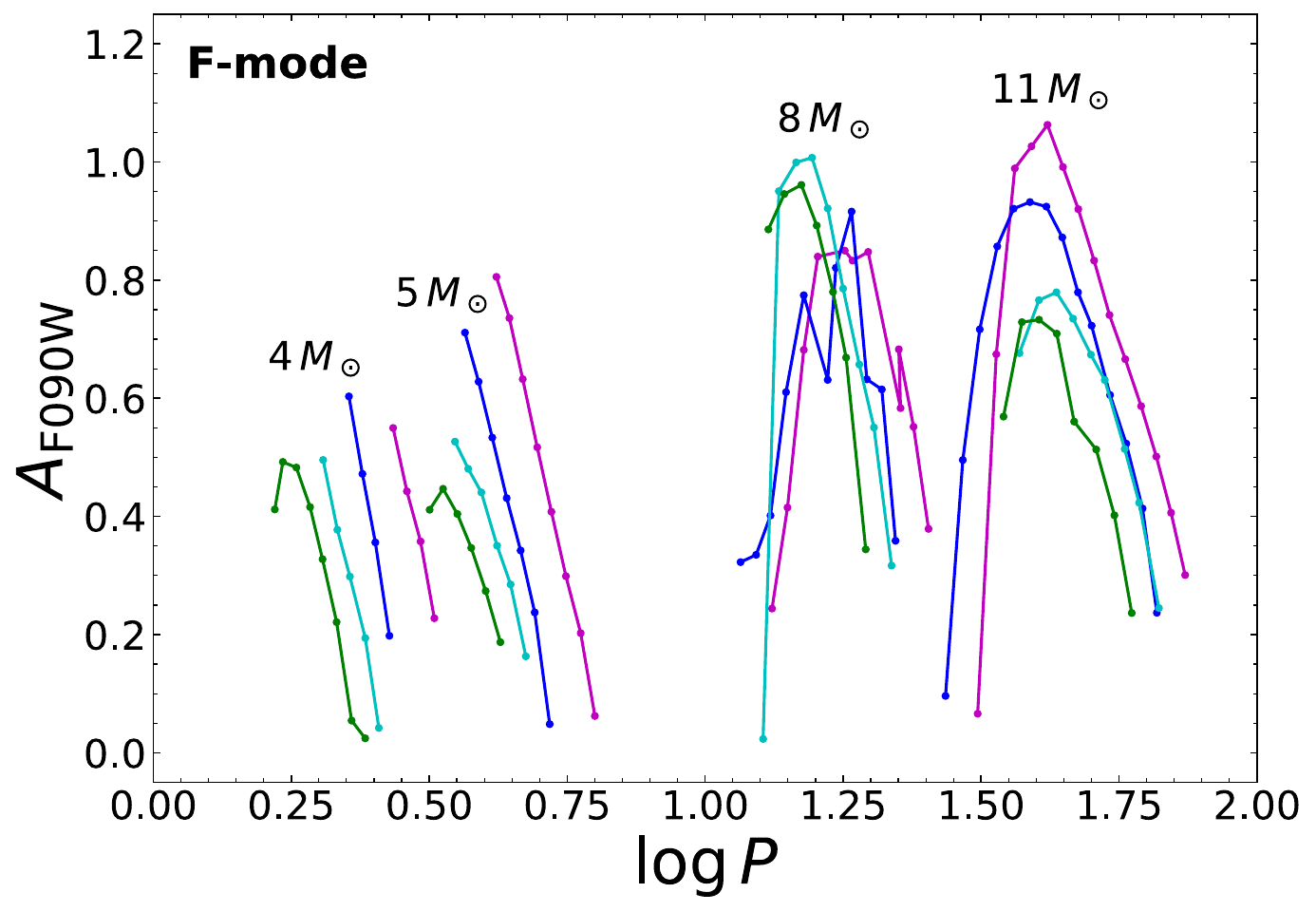} \\
\includegraphics[width=0.33\textwidth]{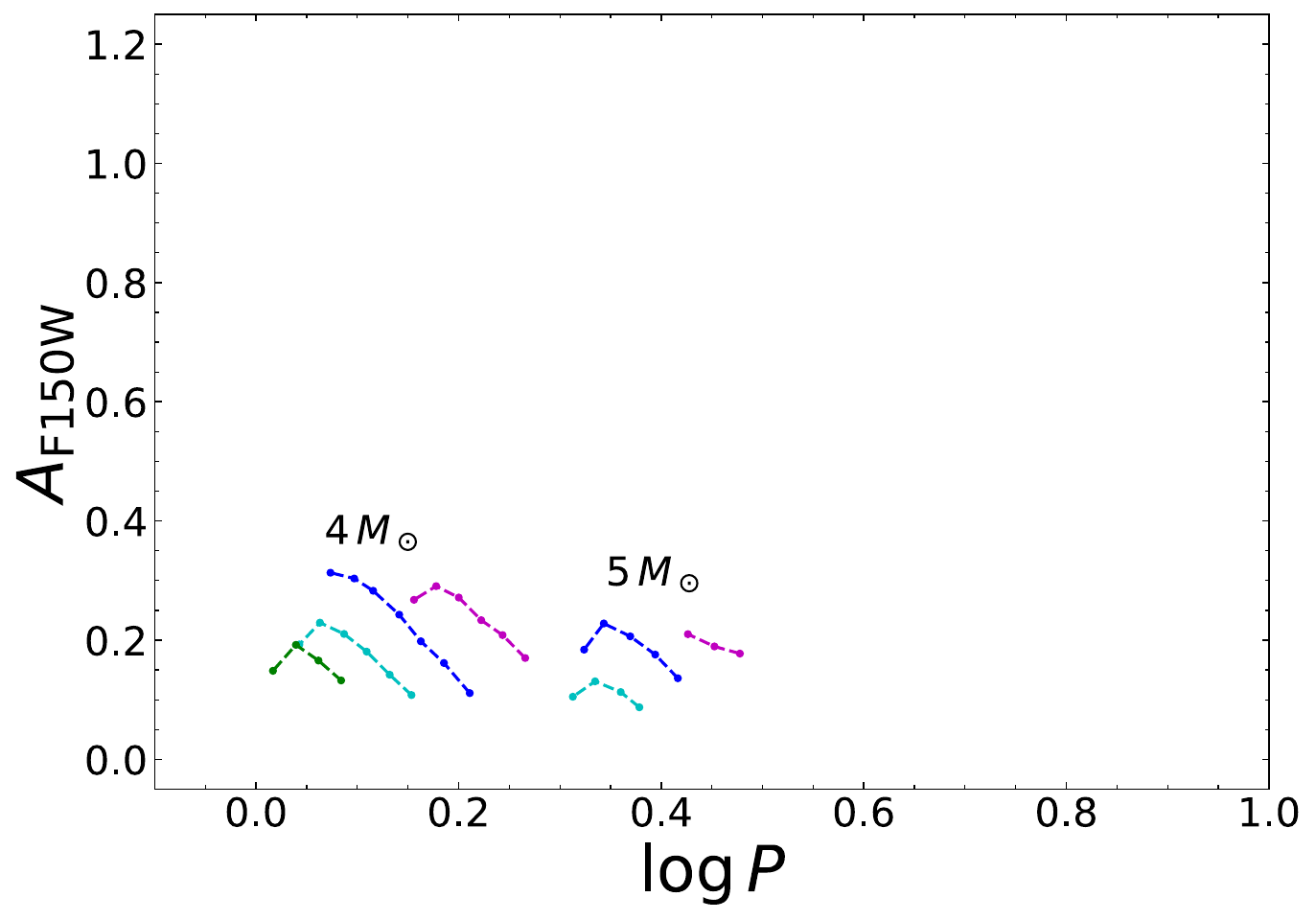} &
\includegraphics[width=0.33\textwidth]{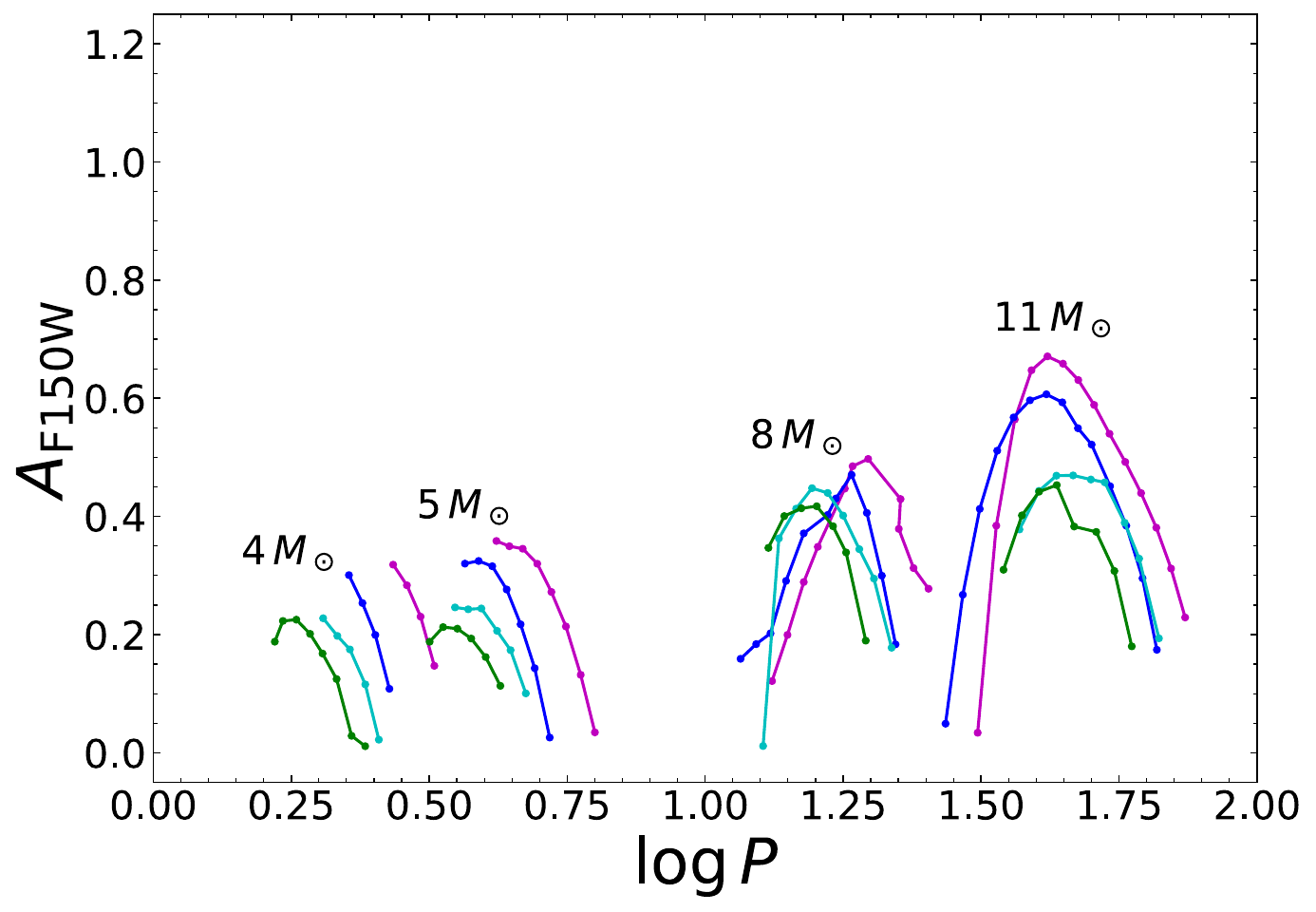} \\
\end{tabular}
\caption{
Period-amplitude diagrams for CC models in selected JWST filters, computed assuming the canonical ML relation (case A) and $\alpha_{\rm ml}=1.5$. The models span different stellar masses, effective temperatures, and chemical compositions. The left panels show FO-mode models, while the right panels show F-mode models. From top to bottom, the results refer to the F090W and F150W filters. The period ranges differ between the left and right panels because FO-mode models cover systematically shorter periods than F-mode models.
}
\label{fig:amp_logP_jwst}
\end{figure*}

Figure~\ref{fig:amp_logP_jwst} presents the period-amplitude diagrams
for F- and FO-mode CC models in selected JWST filters. The models
span different stellar masses, luminosity levels, chemical compositions,
mixing-length parameters, and ML assumptions. As discussed in the main
text, the pulsation amplitudes depend on both metallicity and wavelength
and systematically decrease toward the longer-wavelength filters.

\begin{figure*}
\centering
\setlength{\tabcolsep}{1pt}
\renewcommand{\arraystretch}{1.0}
\begin{tabular}{cc}
\includegraphics[width=0.33\textwidth]{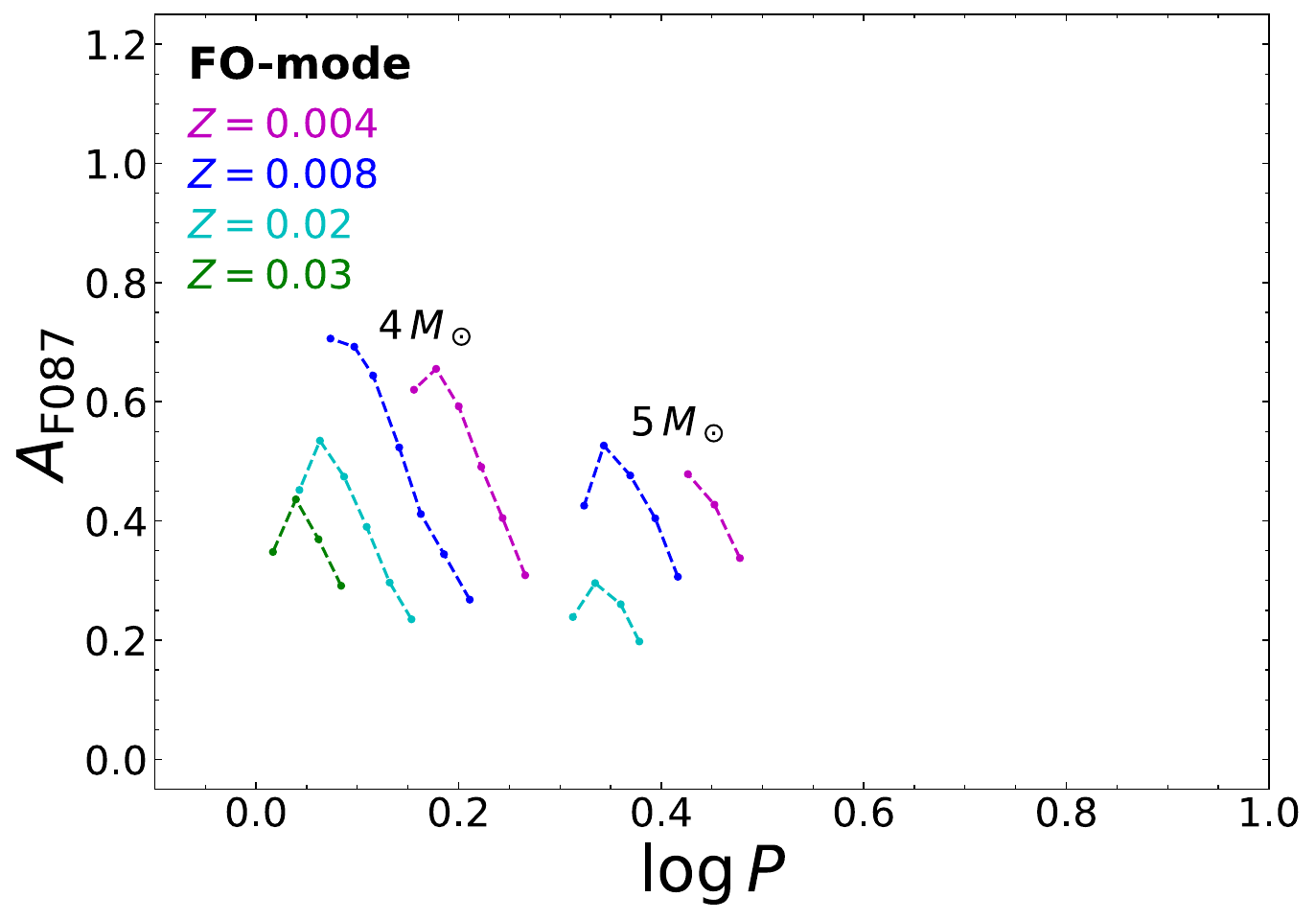} &
\includegraphics[width=0.33\textwidth]{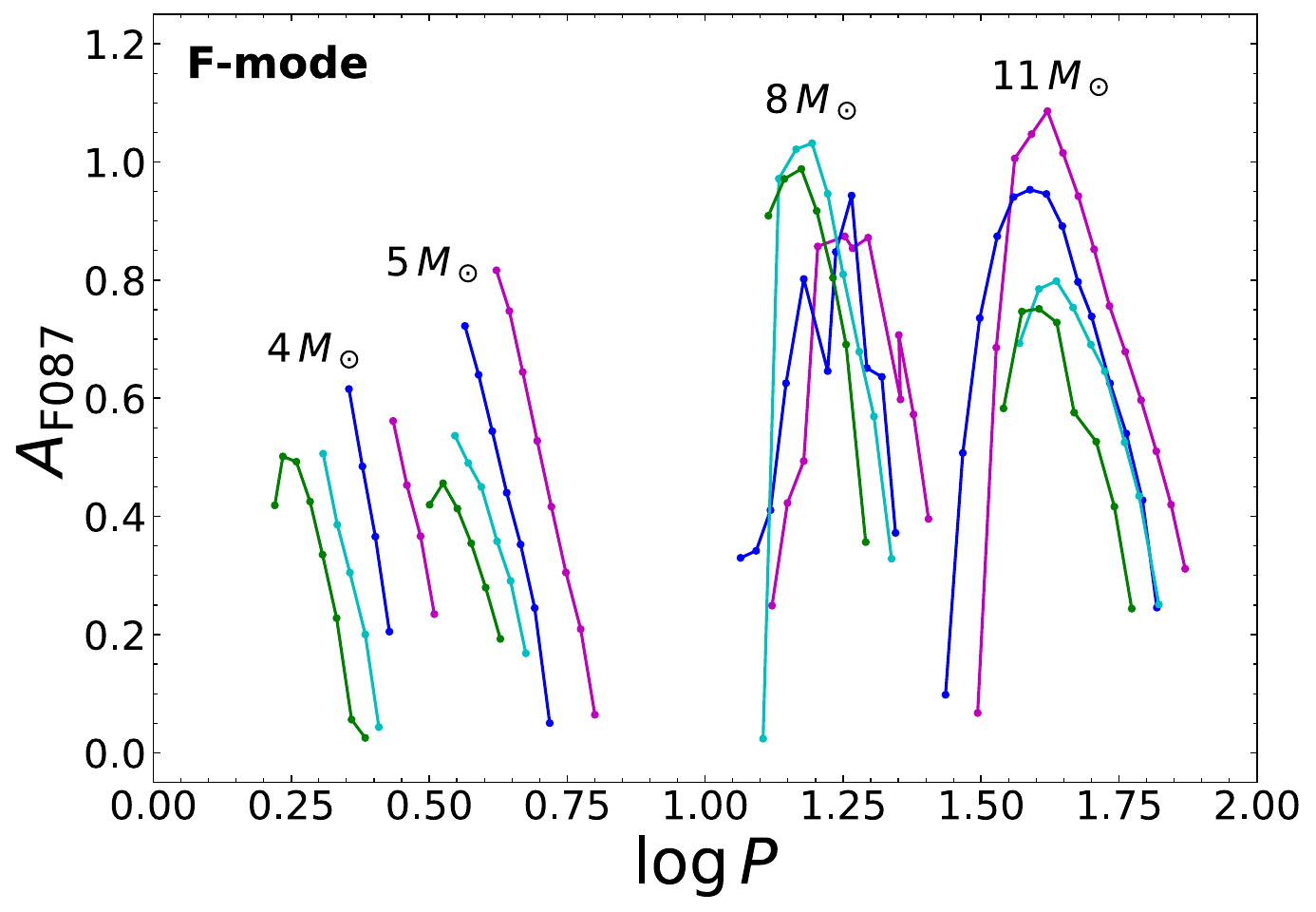} \\
\includegraphics[width=0.33\textwidth]{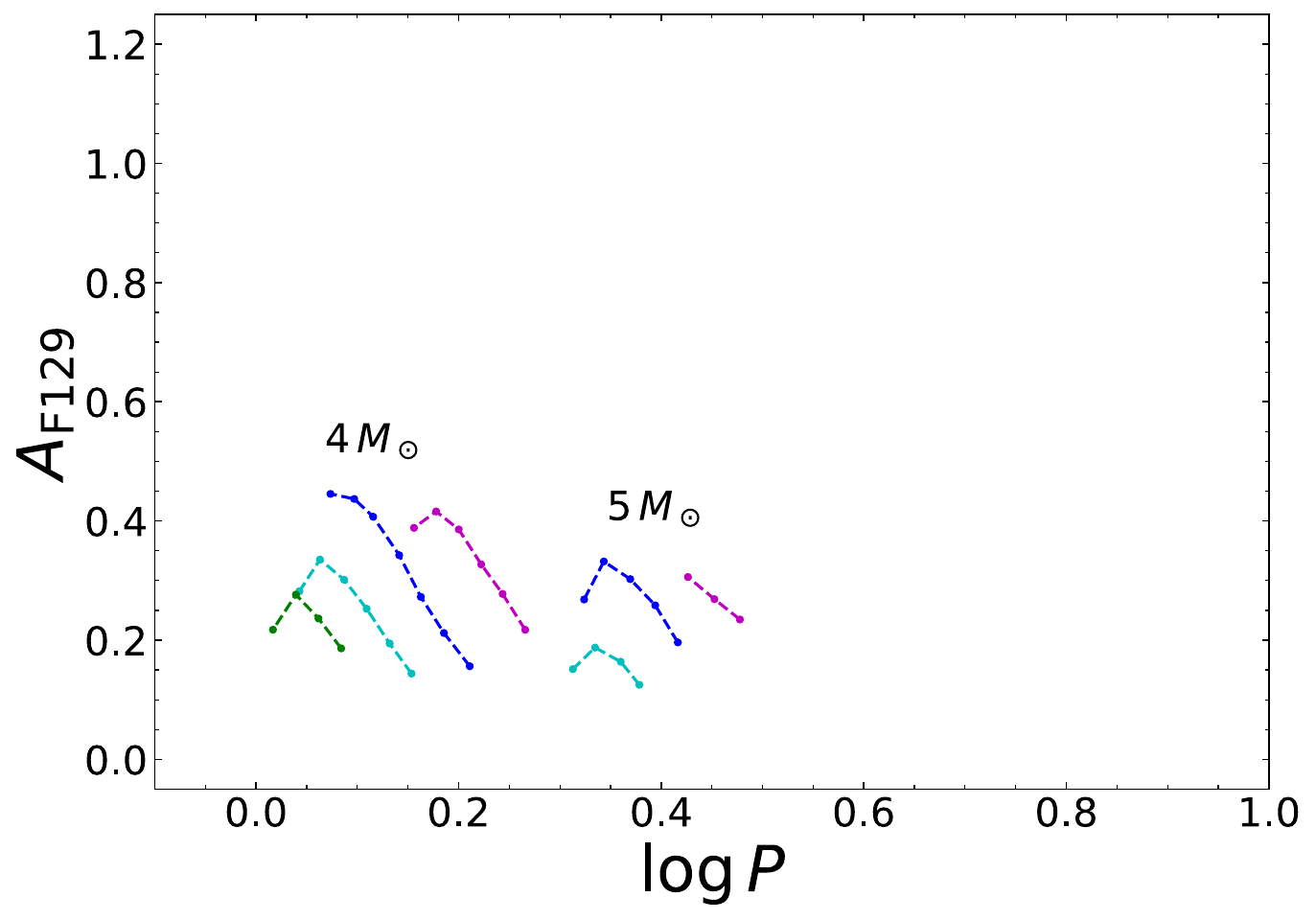} &
\includegraphics[width=0.33\textwidth]{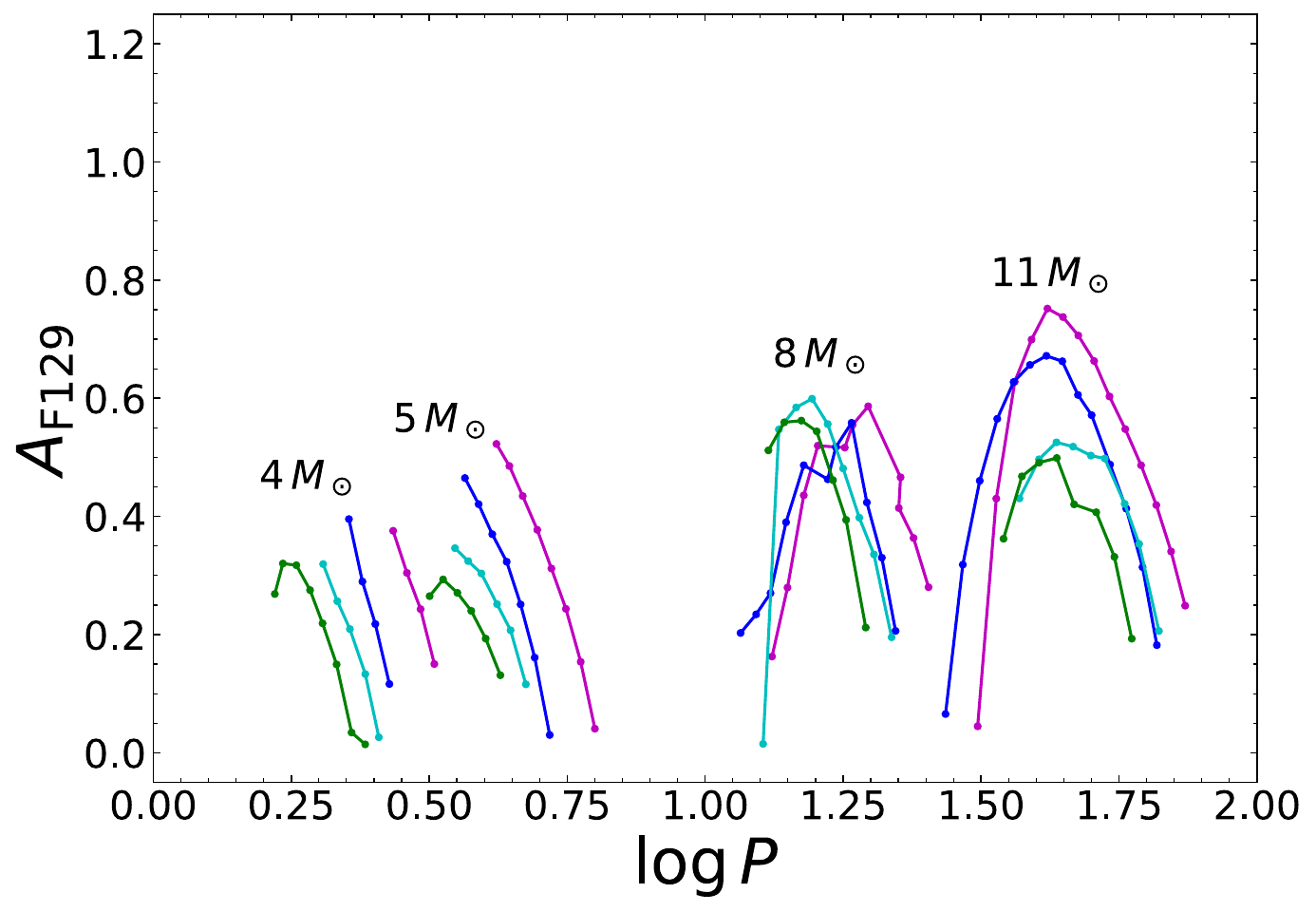} \\
\end{tabular}
\caption{
Same as Figure~\ref{fig:amp_logP_jwst}, but for the Roman/WFI filters
F087 (top panels) and F129 (bottom panels).
}
\label{fig:amp_logP_roman}
\end{figure*}

Figure~\ref{fig:amp_logP_roman} presents the corresponding
period-amplitude diagrams in selected Roman/WFI filters. The same stellar
masses, luminosity levels, chemical compositions, mixing-length parameters,
and ML assumptions adopted for the JWST diagrams in
Figure~\ref{fig:amp_logP_jwst} are considered. The Roman amplitudes
display the same general dependence on metallicity and pulsation mode,
together with the characteristic decrease in pulsation amplitude toward
longer wavelengths.

\subsection{Amplitude ratios}
\label{sec:amp_ratio}

\begin{figure*}
\centering
\setlength{\tabcolsep}{0pt}
\renewcommand{\arraystretch}{0.0}
\begin{tabular}{cc}
\includegraphics[width=0.33\textwidth]{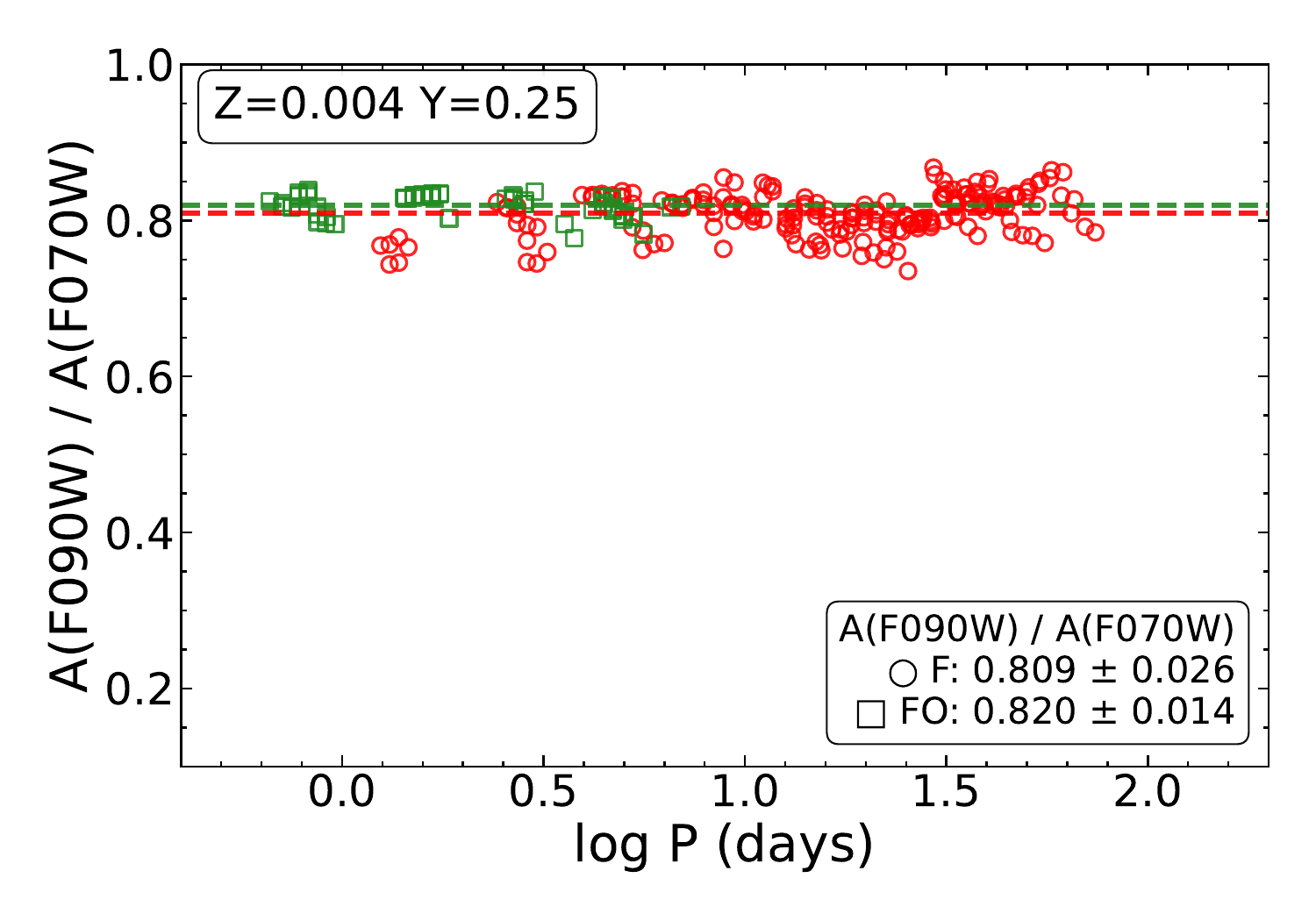} &
\includegraphics[width=0.33\textwidth]{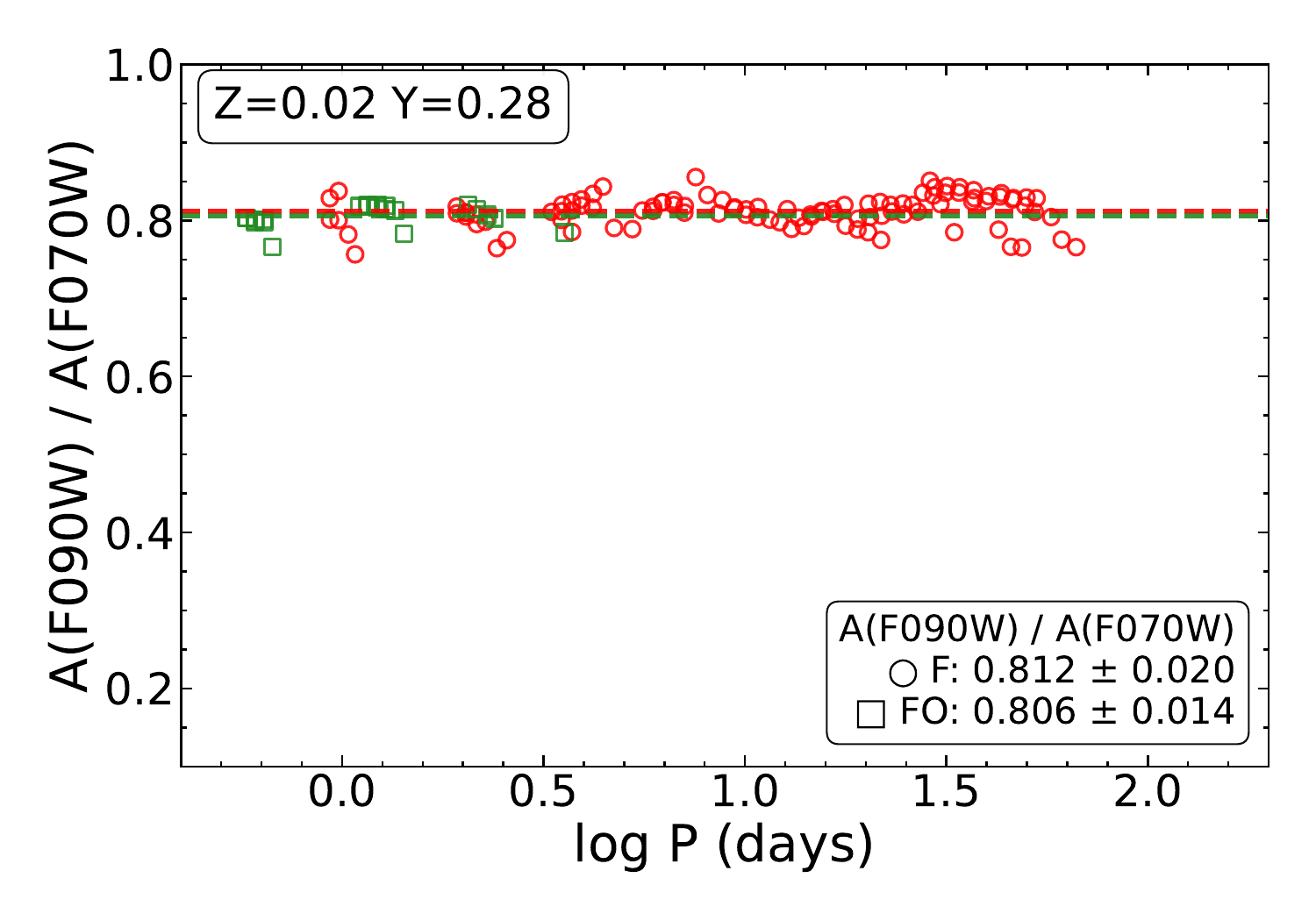} \\
\includegraphics[width=0.33\textwidth]{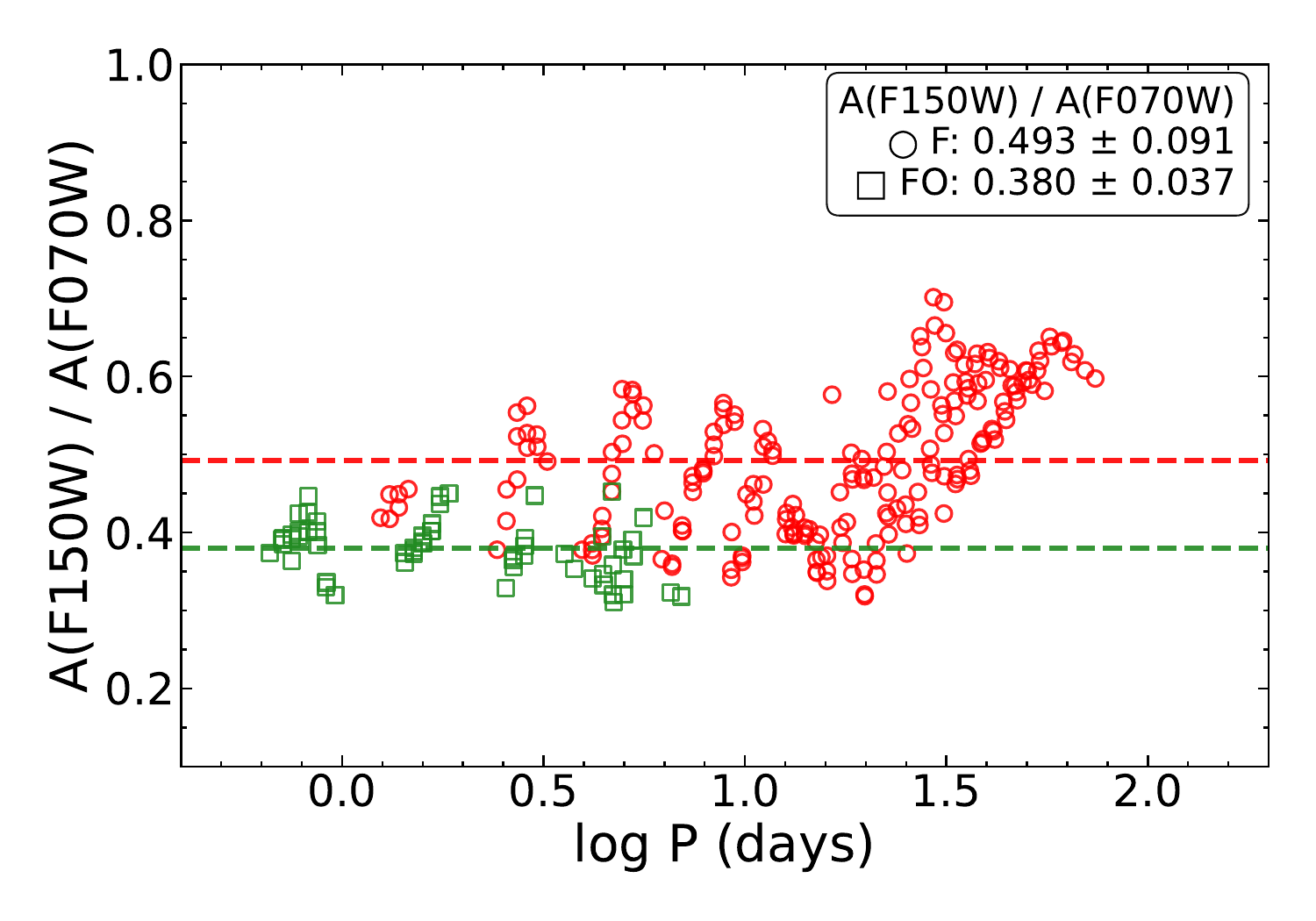} &
\includegraphics[width=0.33\textwidth]{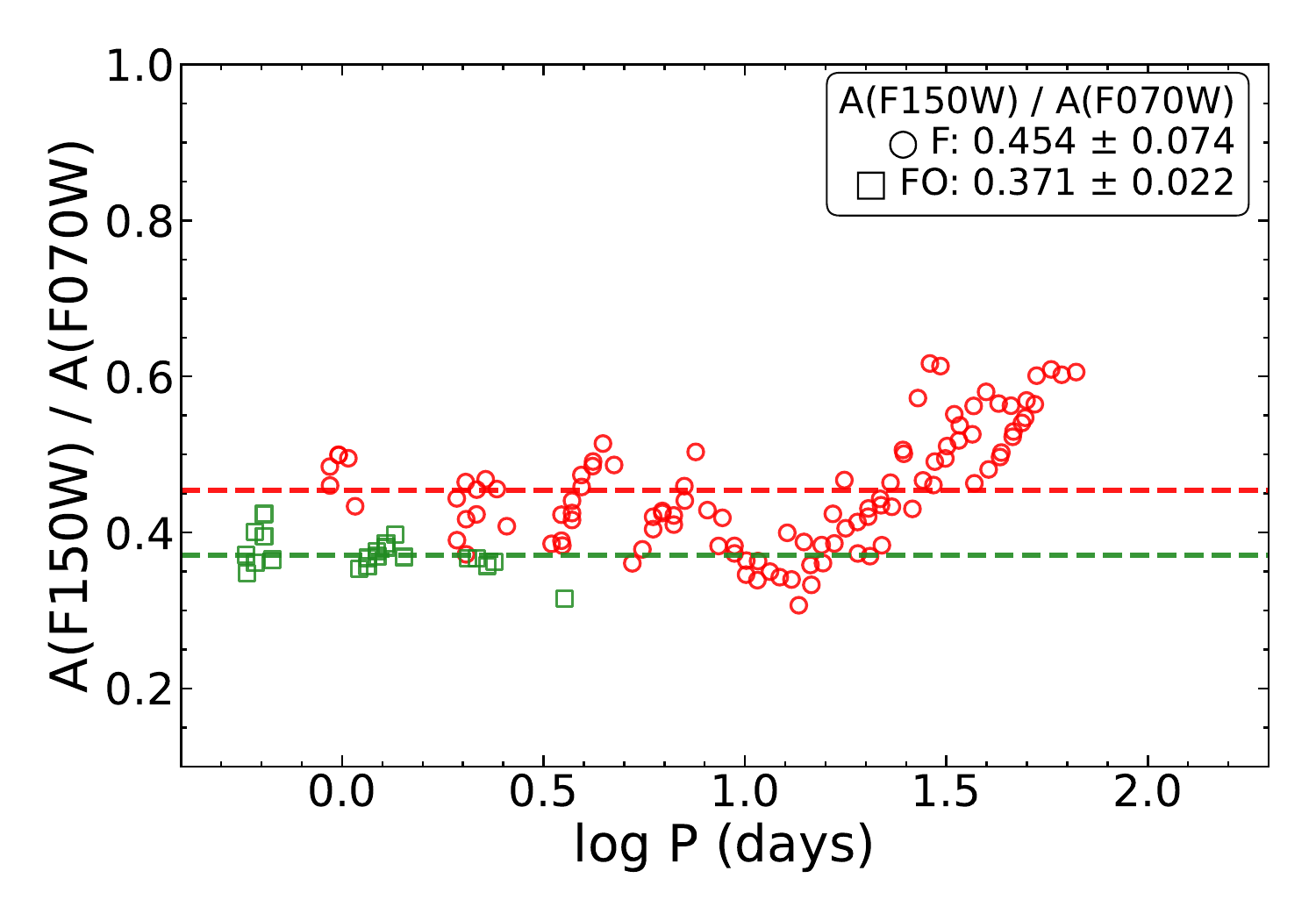} \\
\end{tabular}
\caption{
Predicted amplitude ratios as a function of $\log P$ for CC models in
selected JWST/NIRCam filters. The ratios are defined as
$A_{\lambda}/A_{\rm F070W}$, where $A_{\lambda}$ is the peak-to-peak
amplitude in a given filter. The left panels show results for $Z=0.004$,
while the right panels show results for $Z=0.02$. Open circles and squares
denote F- and FO-mode pulsators, respectively. The mean amplitude ratio and
corresponding standard deviation are reported in each panel. For clarity,
only two representative amplitude ratios are shown, while the complete set
of theoretical JWST/NIRCam amplitude ratios is provided in
Table~\ref{tab:amp_ratio_jwst_values}.
}
\label{fig:amp_ratio_jwst_plot}
\end{figure*}

Figure~\ref{fig:amp_ratio_jwst_plot} presents the predicted amplitude
ratios as a function of $\log P$ for selected JWST filters and
representative chemical compositions. The ratios are normalized to the
F070W amplitude. As discussed in the main text, they show only a weak
dependence on pulsation period and systematically decrease toward longer
wavelengths.

\begin{figure*}
\centering
\setlength{\tabcolsep}{0pt}
\renewcommand{\arraystretch}{0.0}
\begin{tabular}{cc}
\includegraphics[width=0.33\textwidth]{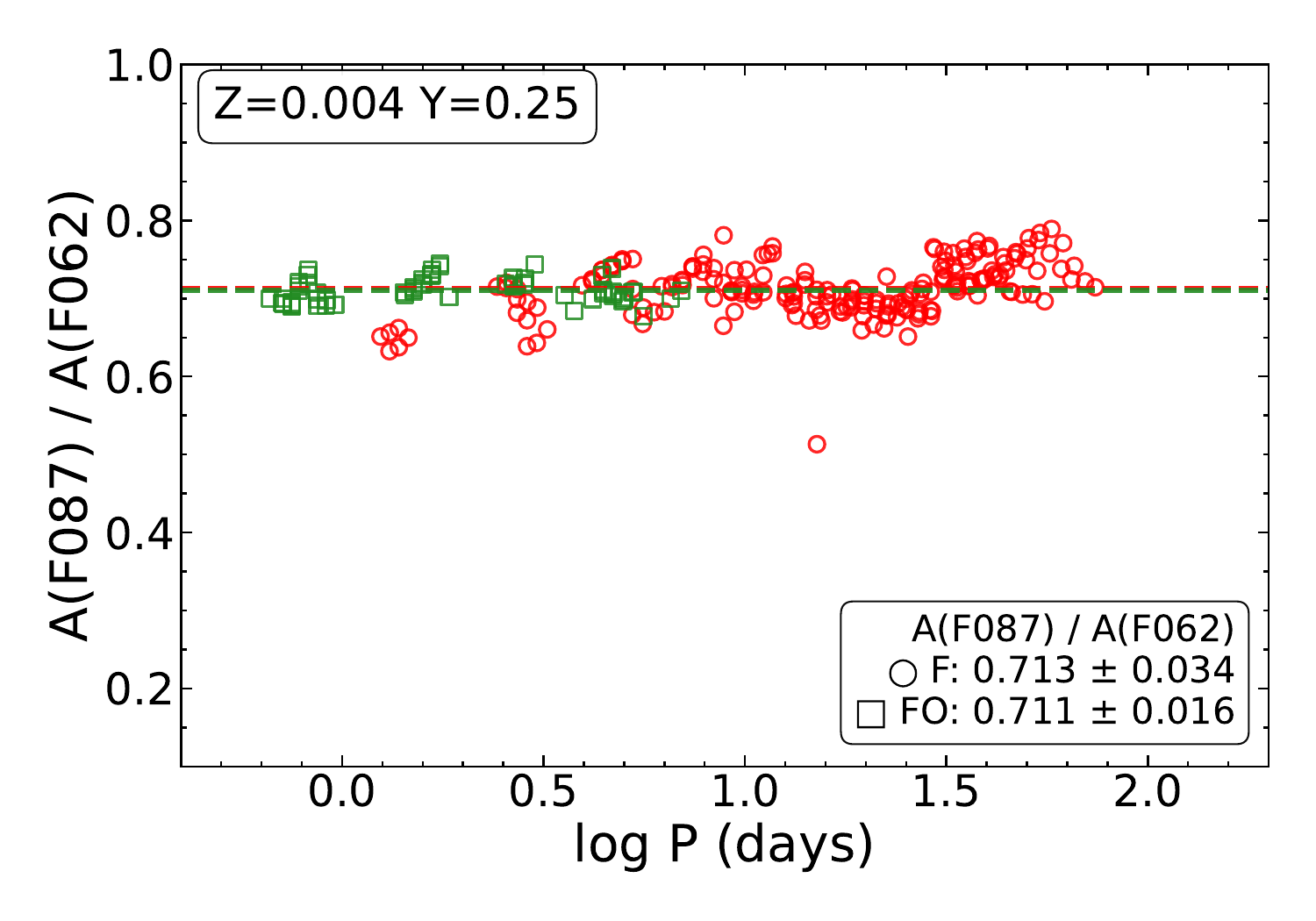} &
\includegraphics[width=0.33\textwidth]{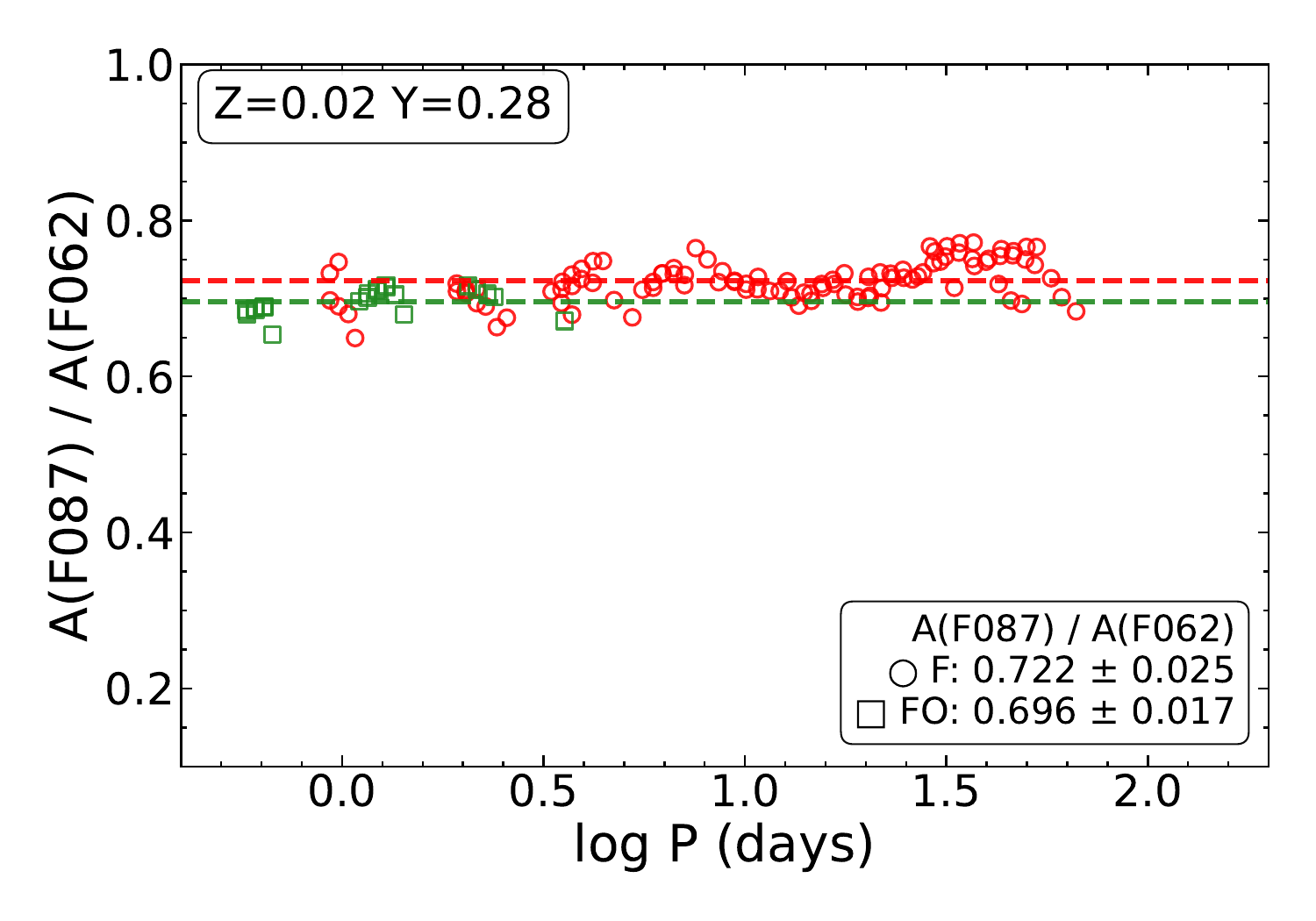} \\
\includegraphics[width=0.33\textwidth]{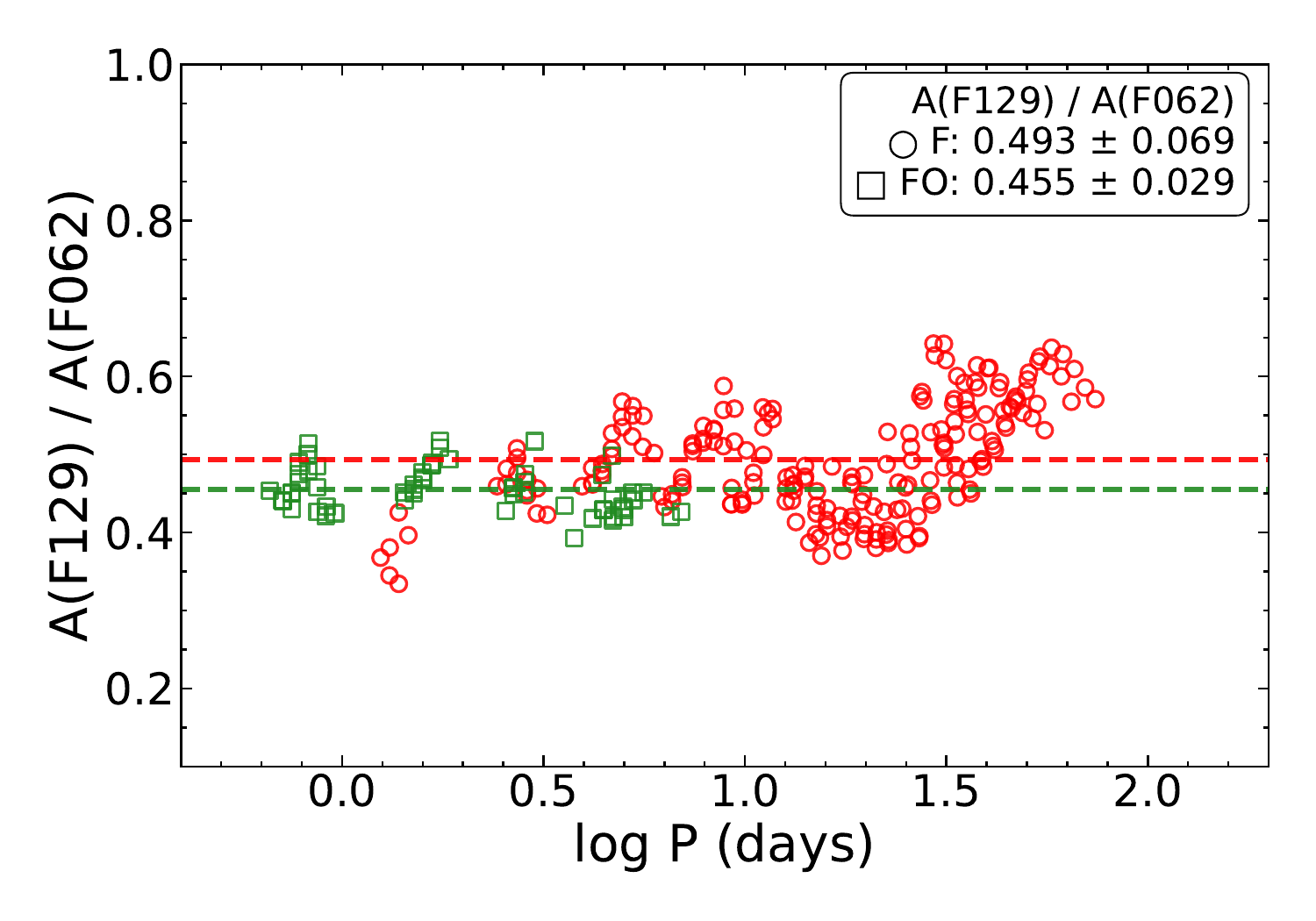} &
\includegraphics[width=0.33\textwidth]{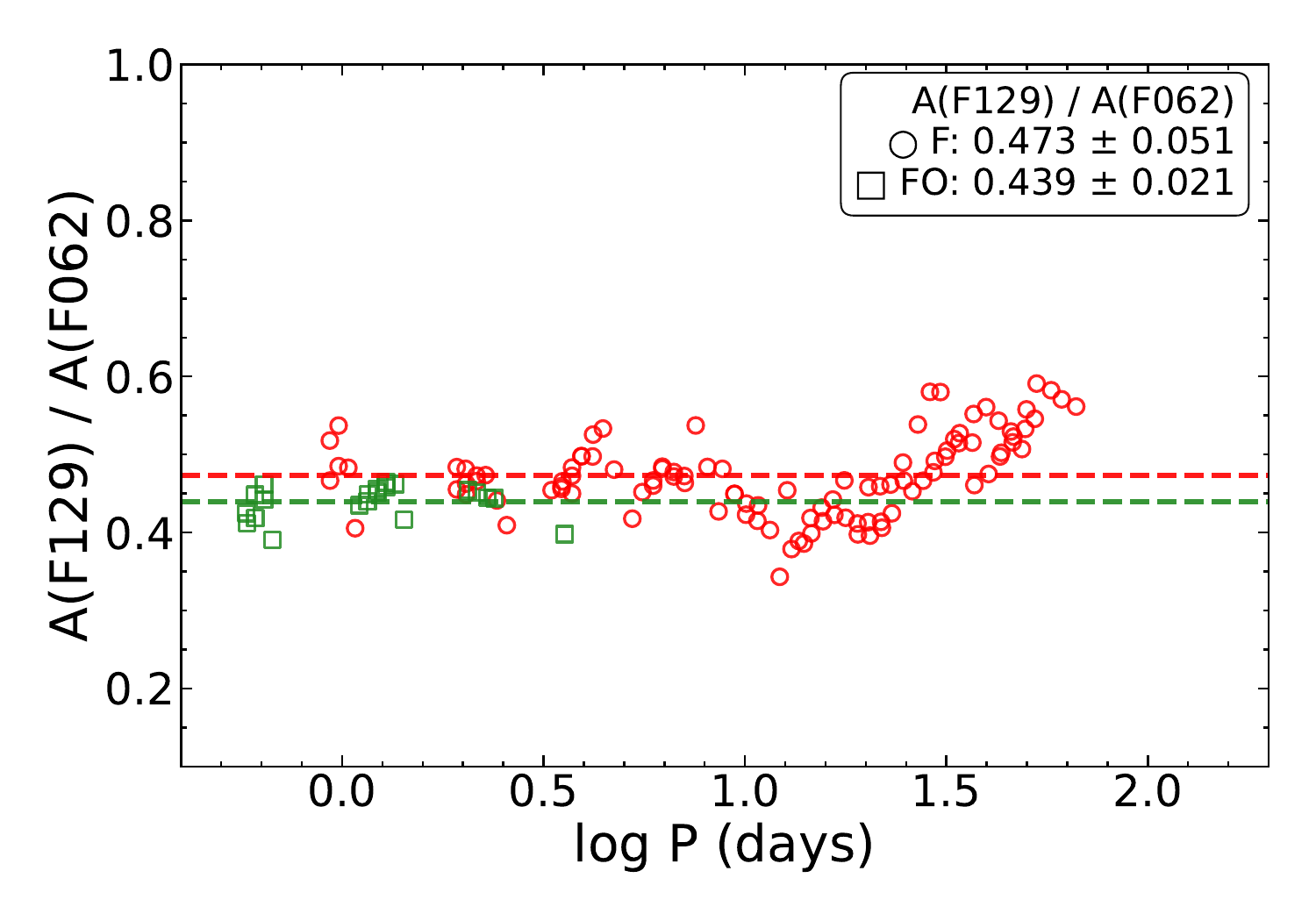} \\
\end{tabular}
\caption{
Same as Figure~\ref{fig:amp_ratio_jwst_plot}, but for the Roman/WFI
filters, adopting F062 as the reference band. The complete set of
Roman/WFI amplitude ratios is provided in
Table~\ref{tab:amp_ratio_roman_values}.
}
\label{fig:amp_ratio_roman_plot}
\end{figure*}

Figure~\ref{fig:amp_ratio_roman_plot} presents the corresponding
theoretical amplitude ratios in the Roman/WFI photometric system for
representative chemical compositions. The amplitudes are normalized to
the F062 band. The Roman/WFI ratios display the same overall behaviour as
the JWST/NIRCam predictions, with only a weak dependence on pulsation
period and a systematic decrease toward longer wavelengths. F- and
FO-mode models also follow the same general mode-dependent trends
discussed in the main text.

\section{Period-Wesenheit Relations}
\label{sec:pw_appendix}

\subsection{Additional JWST/NIRCam PW relations}

In this subsection we present two additional theoretical
PW relations in the JWST photometric system that complement the representative relations discussed in ~\ref{sec:pw}. These additional
Wesenheit combinations sample different wavelength
baselines and confirm the same overall behaviour described
in the main text.

\begin{figure}[ht]
\centering
\includegraphics[width=0.75\columnwidth]{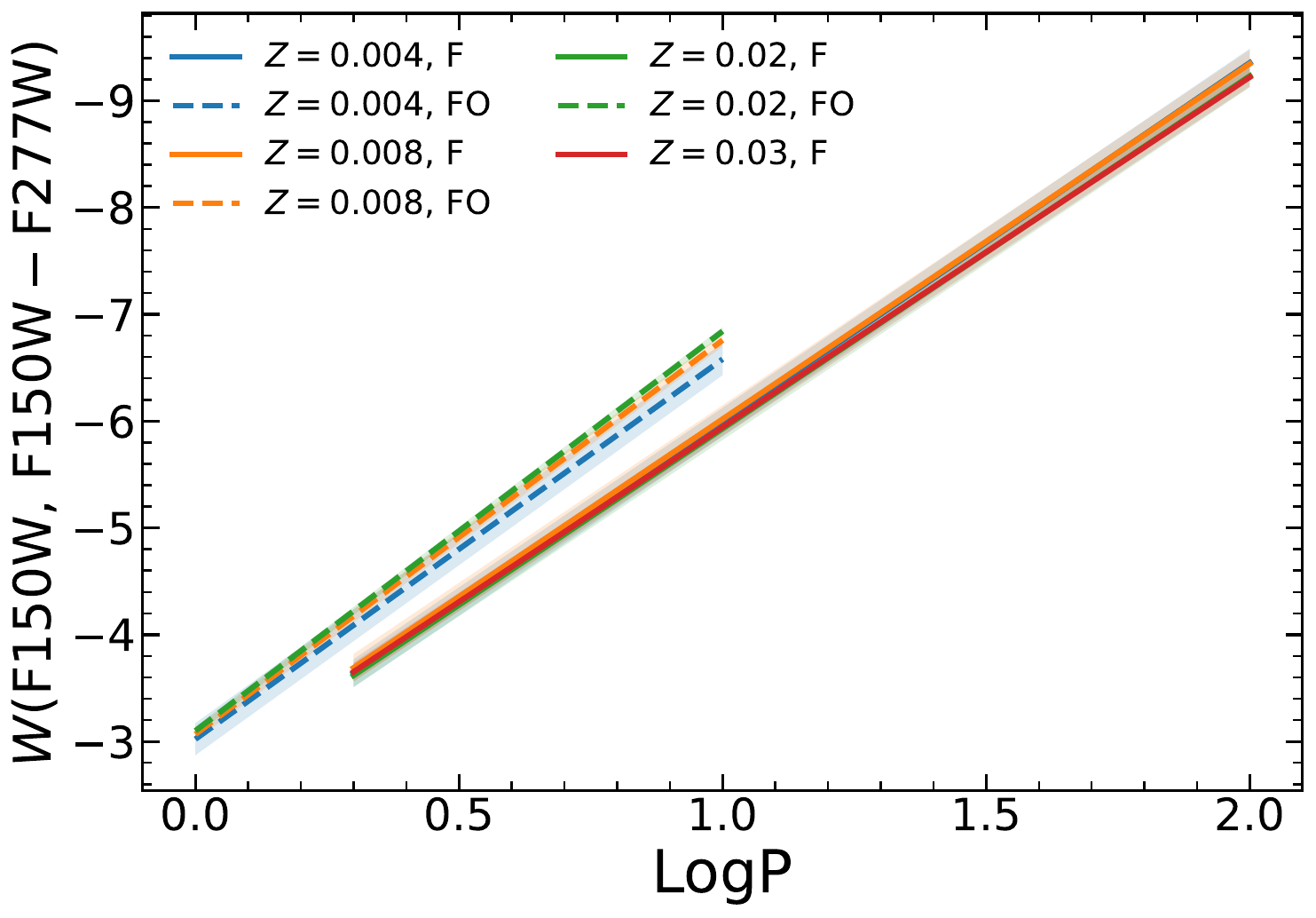}
\vspace{0.2cm}
\includegraphics[width=0.75\columnwidth]{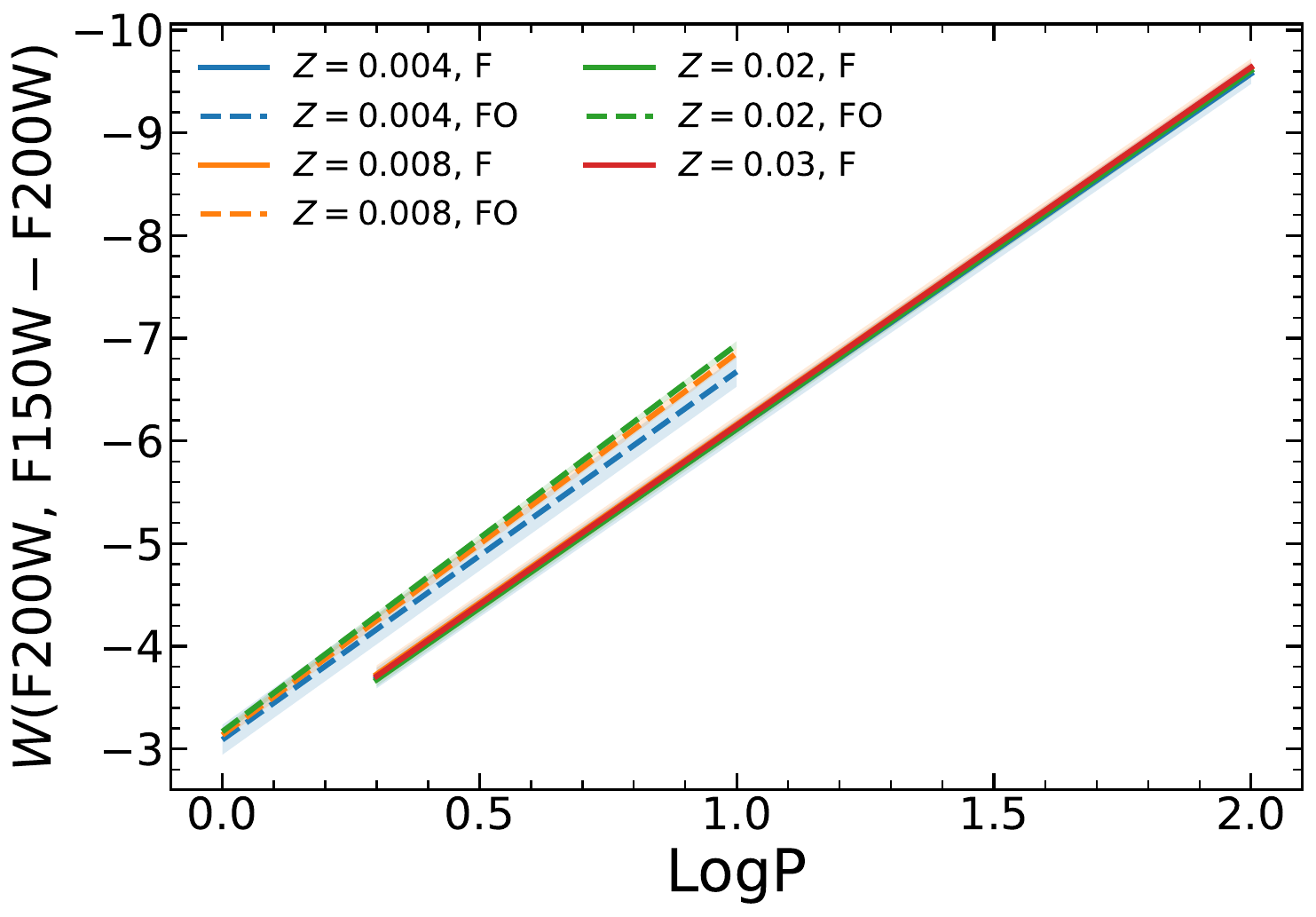}
\caption{ Same as Figure~\ref{fig:pw_jwst_card}, but for the additional
JWST Wesenheit combinations
$W(F150W,\,F150W-F277W)$ (left) and
$W(F200W,\,F150W-F200W)$ (right),
assuming $\alpha_{\rm ml}=1.5$ and the canonical ML relation. As in Figure~\ref{fig:pw_jwst_card},
the relations are shown for
$Z=0.004$ (blue),
$Z=0.008$ (orange),
$Z=0.02$ (green), and
$Z=0.03$ (red).
Solid and dashed lines denote F- and FO-mode
CCs, respectively, while the shaded regions indicate the
intrinsic dispersion of each relation.
}
\label{fig:pw_jwst_appendix}
\end{figure}

\subsection{Roman PW relations}
\label{sec:pw_roman}
\begin{figure}[ht]
\centering
\includegraphics[width=0.75\columnwidth]{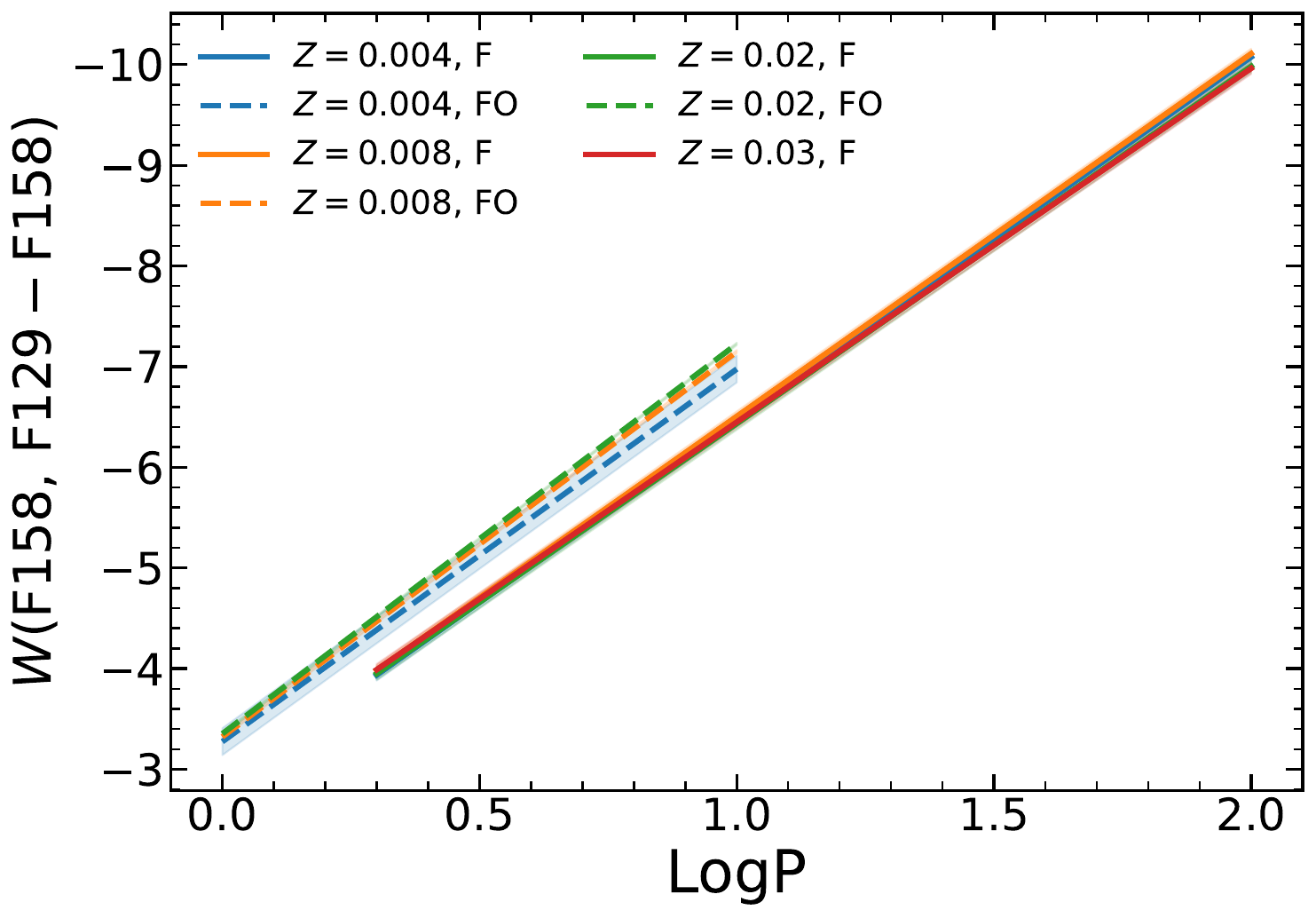}
\vspace{0.2cm}
\includegraphics[width=0.75\columnwidth]{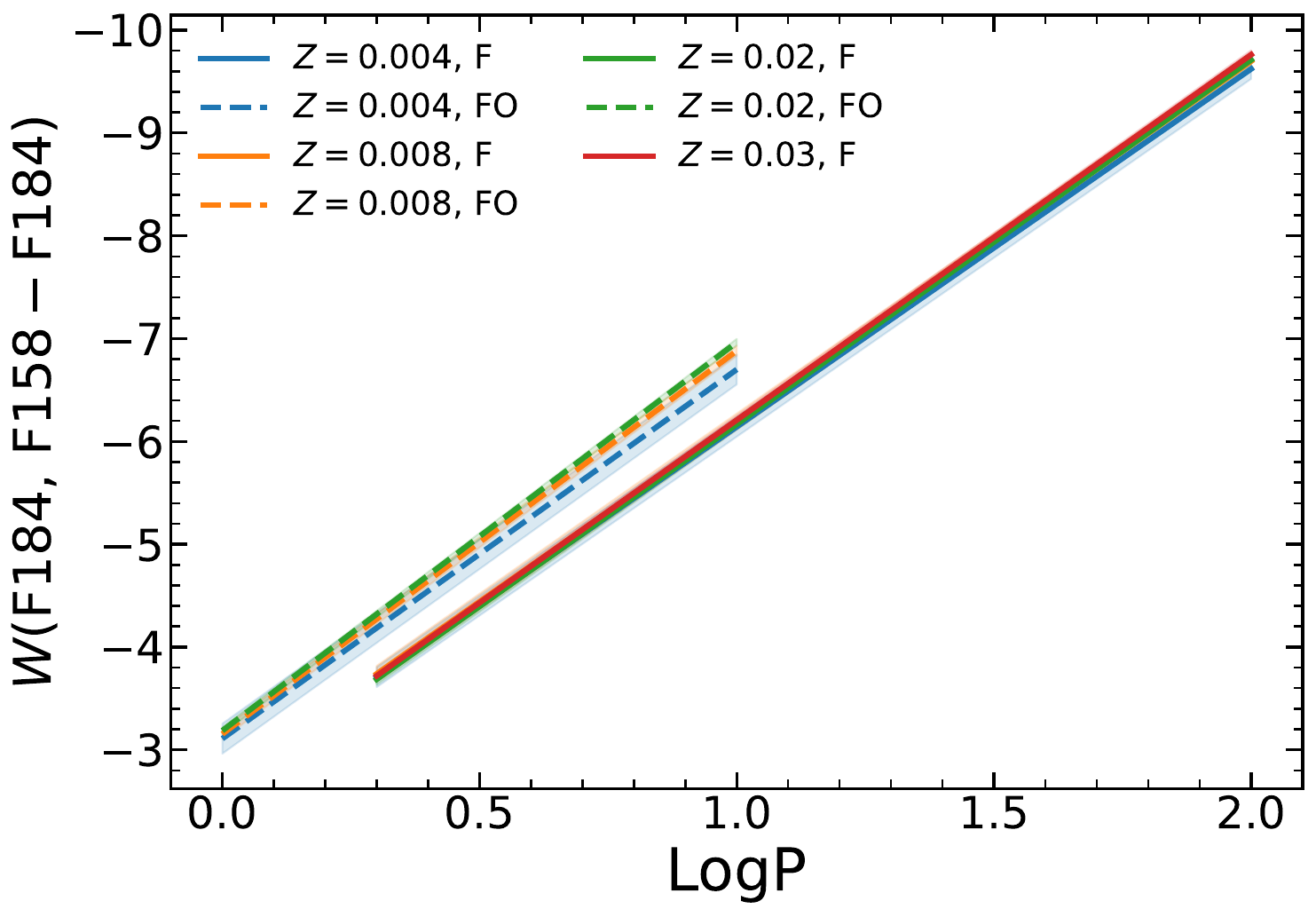}
\caption{Same as Figure~\ref{fig:pw_jwst_card}, but for the selected Roman Wesenheit combinations.}
\label{fig:pw_roman}
\end{figure}

Figure~\ref{fig:pw_roman} presents theoretical PW relations in the Roman photometric system for the $W(F158,F129-F158)$ and $W(F184,F158-F184)$ Wesenheit combinations. As for the JWST relations discussed in Section~\ref{sec:pw}, the different panels correspond to increasing metallicity and illustrate that the effect of chemical composition is mainly reflected in the zero point, while the slopes remain nearly unchanged. The Roman relations preserve the same overall behaviour discussed in the main text, exhibiting small intrinsic dispersions and a limited sensitivity to the adopted physical assumptions.

\section{PWZ Metallicity Coefficients}
\label{sec:pwz_appendix}

\subsection{Additional JWST/NIRCam results}

\begin{figure}[t]
\centering
\setlength{\tabcolsep}{0pt}
\renewcommand{\arraystretch}{1.0}
\begin{tabular}{c}
\includegraphics[width=\columnwidth]{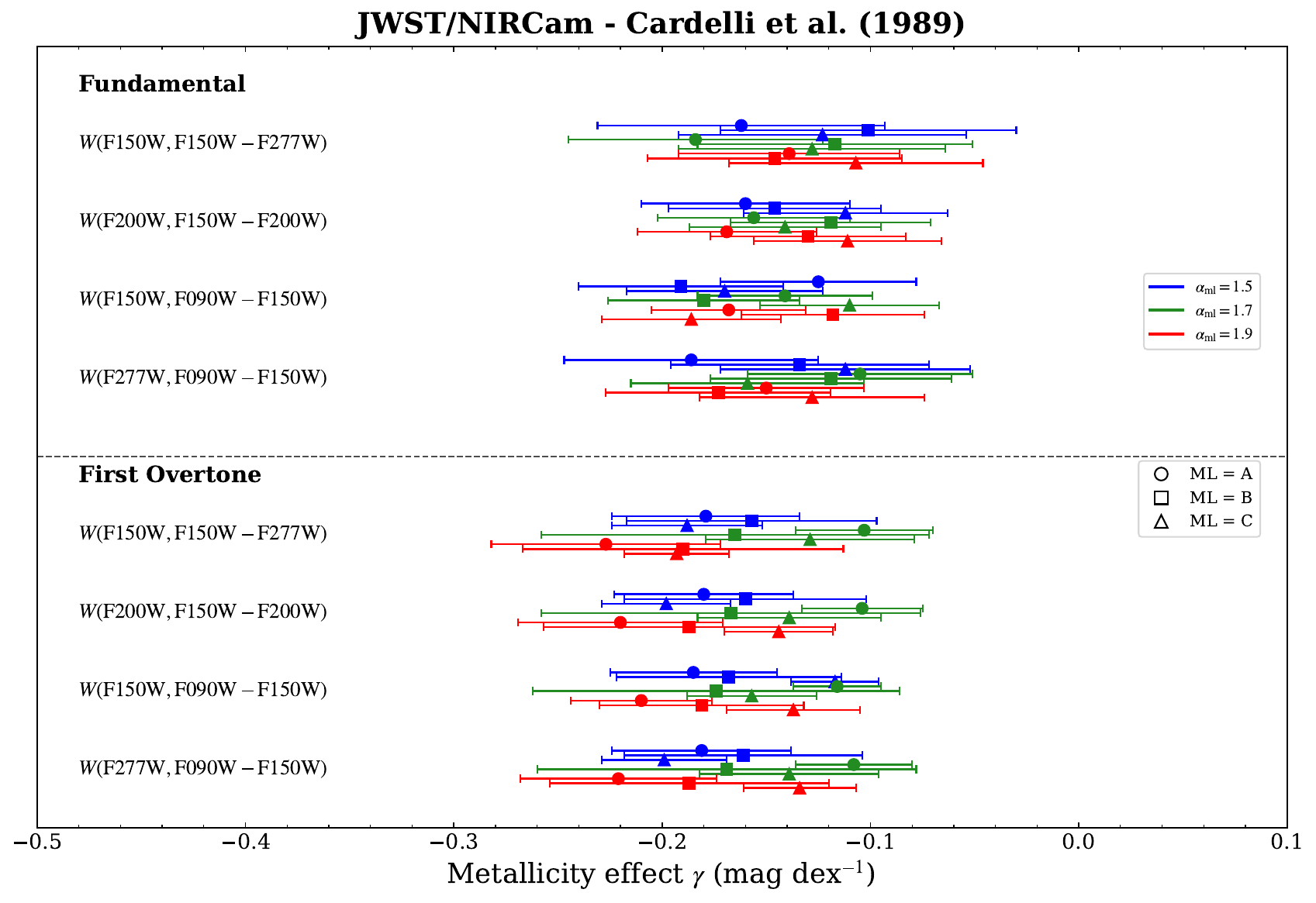} \\
\vspace{2mm}
\includegraphics[width=\columnwidth]{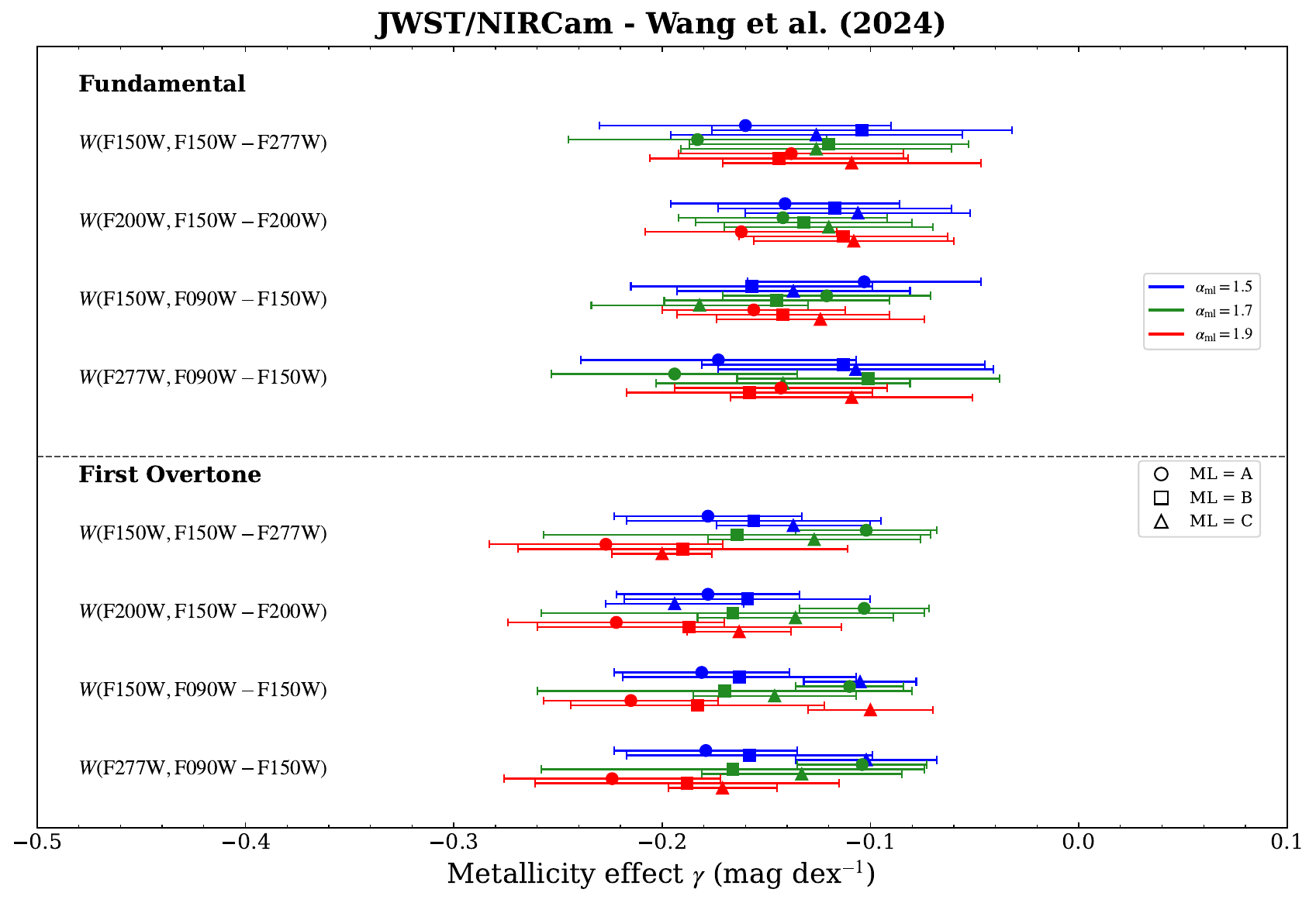}
\end{tabular}
\caption{
Same as Figure~\ref{fig:pwz_gamma_comparison_JWST}, but using
the extinction laws of \citet{Cardelli1989} (top) and
\citet{Wang2024} (bottom).
}
\label{fig:pwz_gamma_appendix}
\end{figure}

Figure~\ref{fig:pwz_gamma_appendix} presents the metallicity
coefficient $\gamma$ derived from the theoretical JWST
PWZ relations using the extinction laws of
\citet{Cardelli1989} and \citet{Wang2024}. As discussed in
Section~5.6, the inferred metallicity coefficients closely
resemble those obtained using the \citet{Fitzpatrick1999}
prescription adopted in the main text, confirming that the
predicted metallicity dependence is largely insensitive to the
choice of extinction law in the near-infrared regime.

\subsection{Roman PWZ metallicity coefficients}
\label{sec:pwz_roman}

\begin{figure}[t]
\centering
\setlength{\tabcolsep}{0pt}
\renewcommand{\arraystretch}{1.0}
\begin{tabular}{c}
\includegraphics[width=\columnwidth]{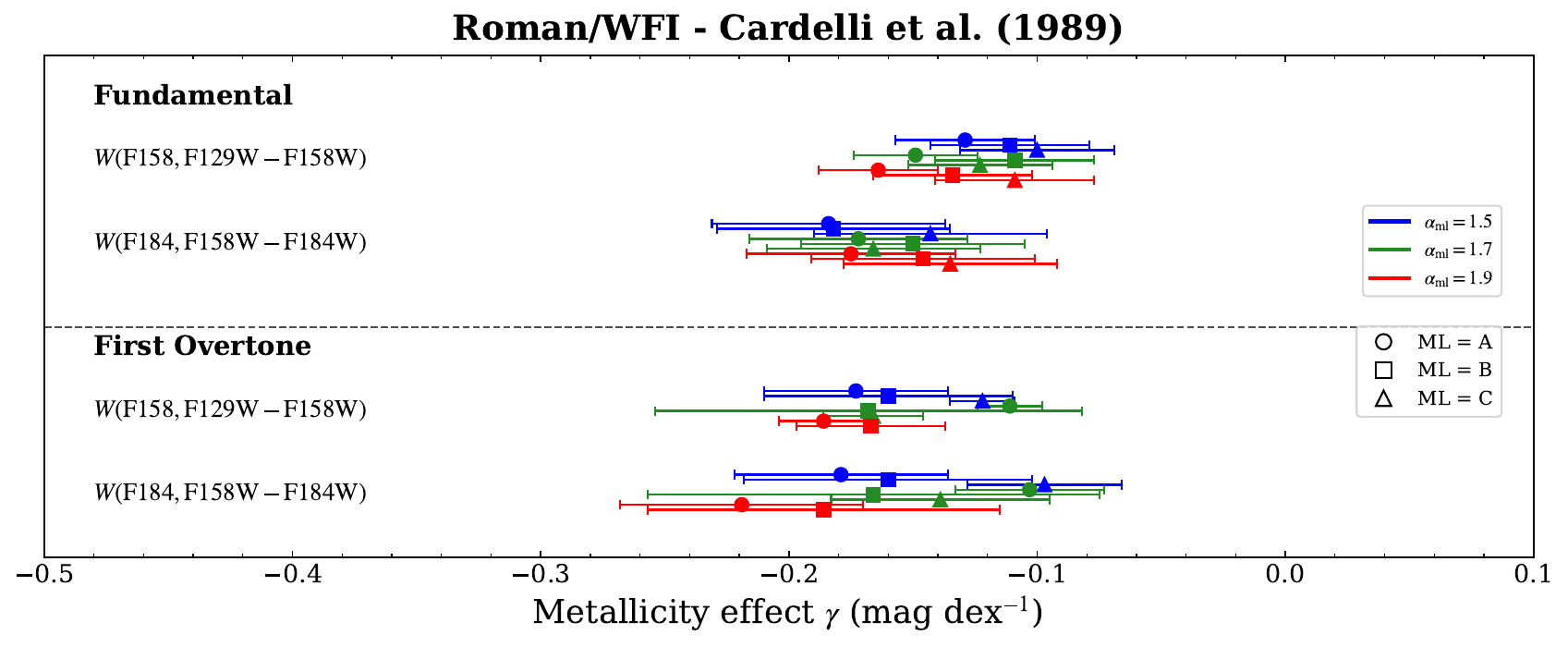} \\[0.1cm]
\includegraphics[width=\columnwidth]{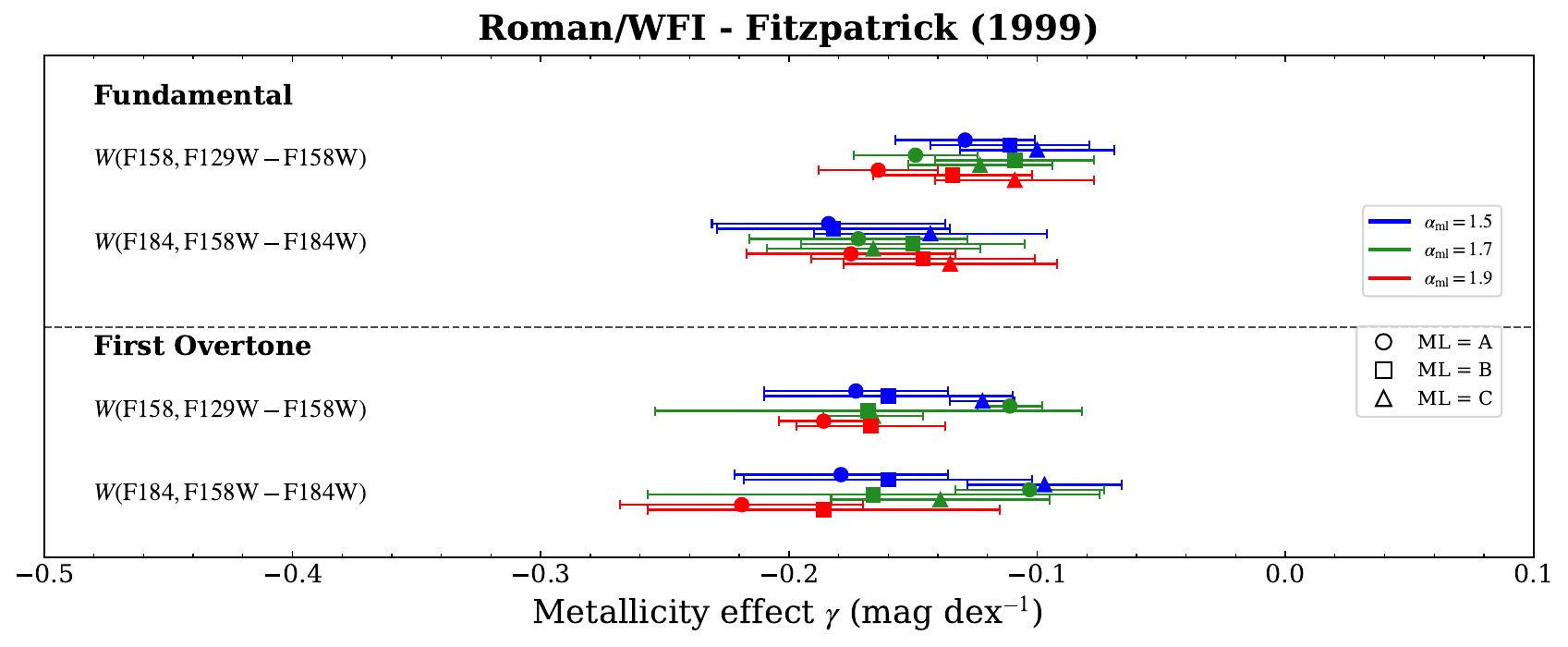}
\end{tabular}
\caption{
Same as Figure~\ref{fig:pwz_gamma_comparison_JWST}, but for the Roman photometric system. The top and bottom panels correspond to the extinction laws of \citet{Cardelli1989} and \citet{Fitzpatrick1999}, respectively. The \citet{Wang2024} extinction law is not shown because extinction coefficients for the Roman filters are not currently available.
}
\label{fig:pwz_gamma_comparison_roman}
\end{figure}

Figure~\ref{fig:pwz_gamma_comparison_roman} summarizes the behaviour of the metallicity coefficient $\gamma$ for the explored Roman PWZ relations. The upper and lower panels show the results obtained using the extinction laws of \citet{Cardelli1989} and \citet{Fitzpatrick1999}, respectively. As discussed in subsection~\ref{sec:PWZ}, the inferred metallicity coefficients closely resemble those obtained for the JWST photometric system, with only minor differences among the adopted reddening prescriptions and a metallicity dependence that remains primarily controlled by the adopted Wesenheit combination and the pulsation model assumptions.

\section{Comparison between theoretical and observed PW relations for individual host galaxies}
\label{sec:w_comp}

In this Section we present the comparison between the theoretical PW relations and the observed JWST CC samples from \citetalias[][]{Riess20248sigma} for all host galaxies considered in this work. The figures are analogous to Figure~\ref{fig:ngc4258_pw}, shown in the main text for NGC~4258. For each galaxy, the three panels show, from top to bottom, the hybrid JWST+HST Wesenheit relation, the JWST NIR Wesenheit relation based on $W(F150W,F090W-F150W)$, and the JWST MIR Wesenheit relation based on $W(F277W,F090W-F150W)$.

\begin{figure}[t]
\centering
\includegraphics[width=0.8\columnwidth]{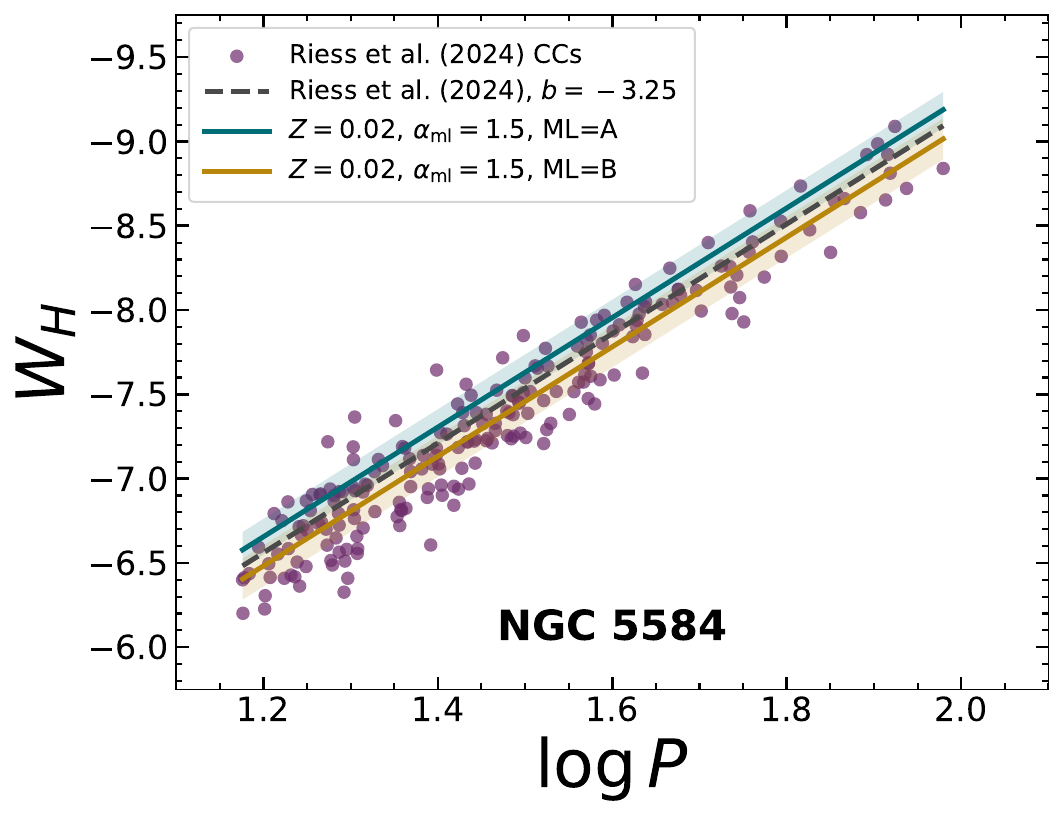}
\vspace{0.03cm}
\includegraphics[width=0.8\columnwidth]{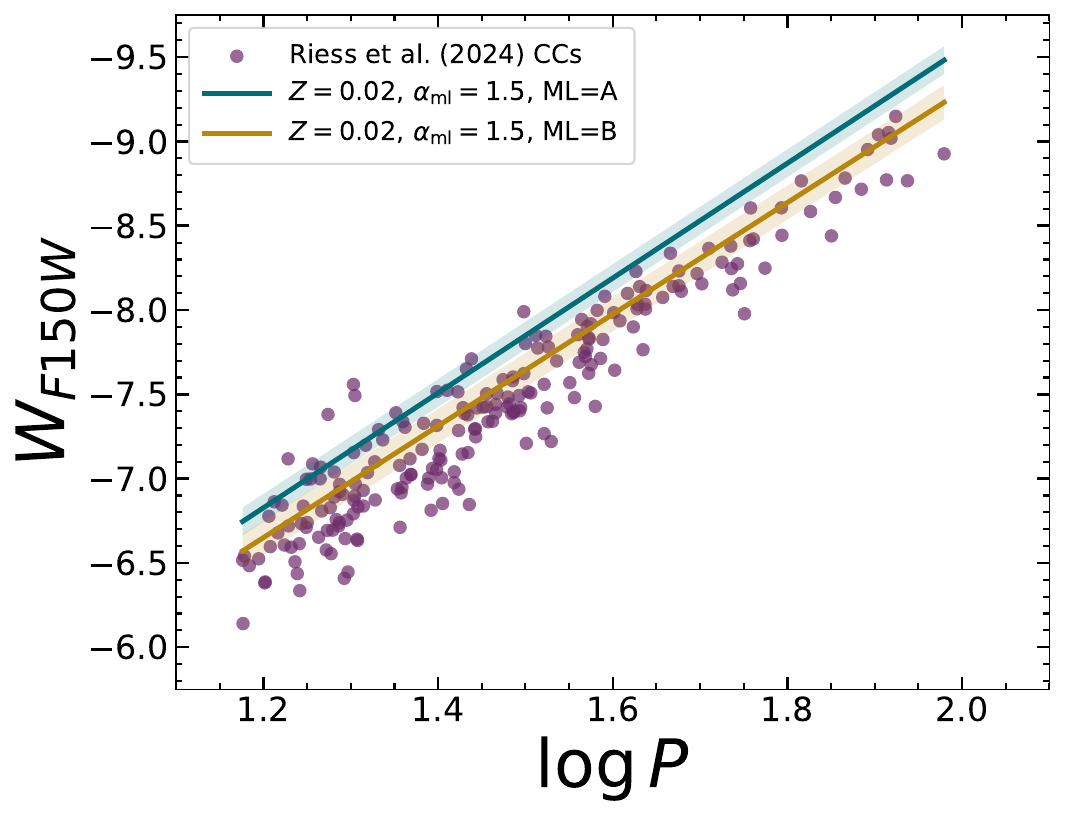}
\vspace{0.03cm}
\includegraphics[width=0.8\columnwidth]{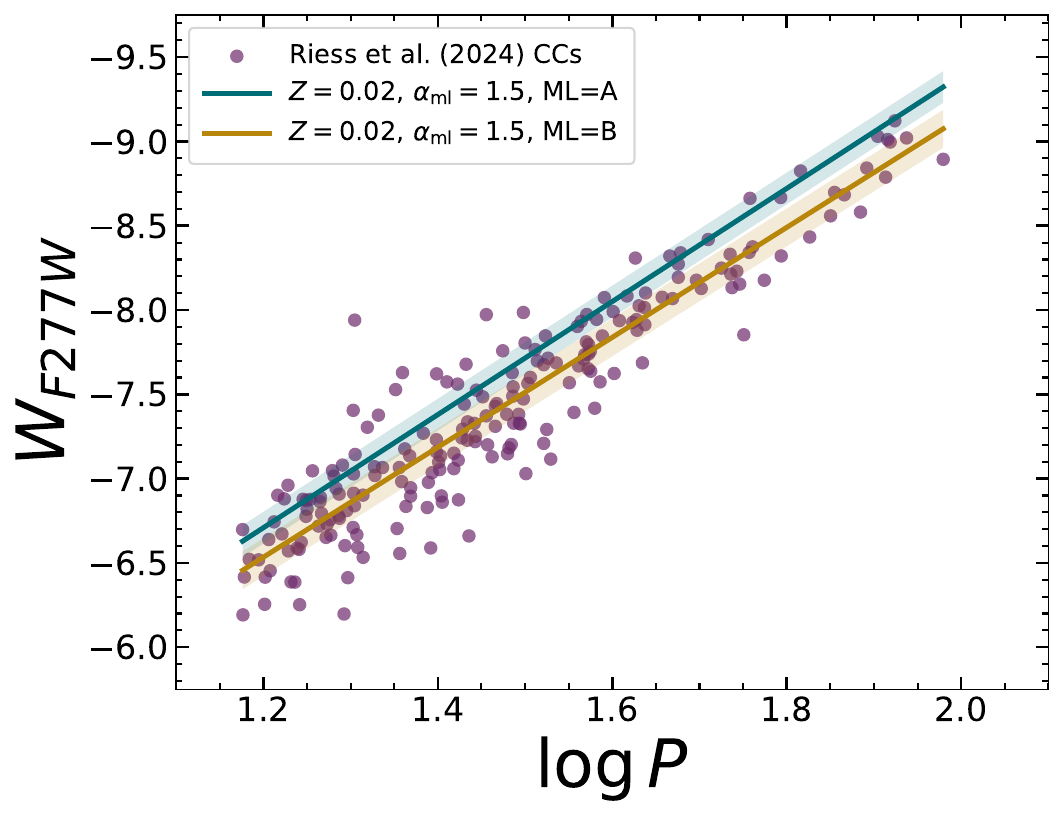}
\caption{Same as Figure~\ref{fig:ngc4258_pw}, but for NGC~5584.
}
\label{fig:ngc5584_pw}
\end{figure}

\begin{figure}[t]
\centering
\includegraphics[width=0.8\columnwidth]{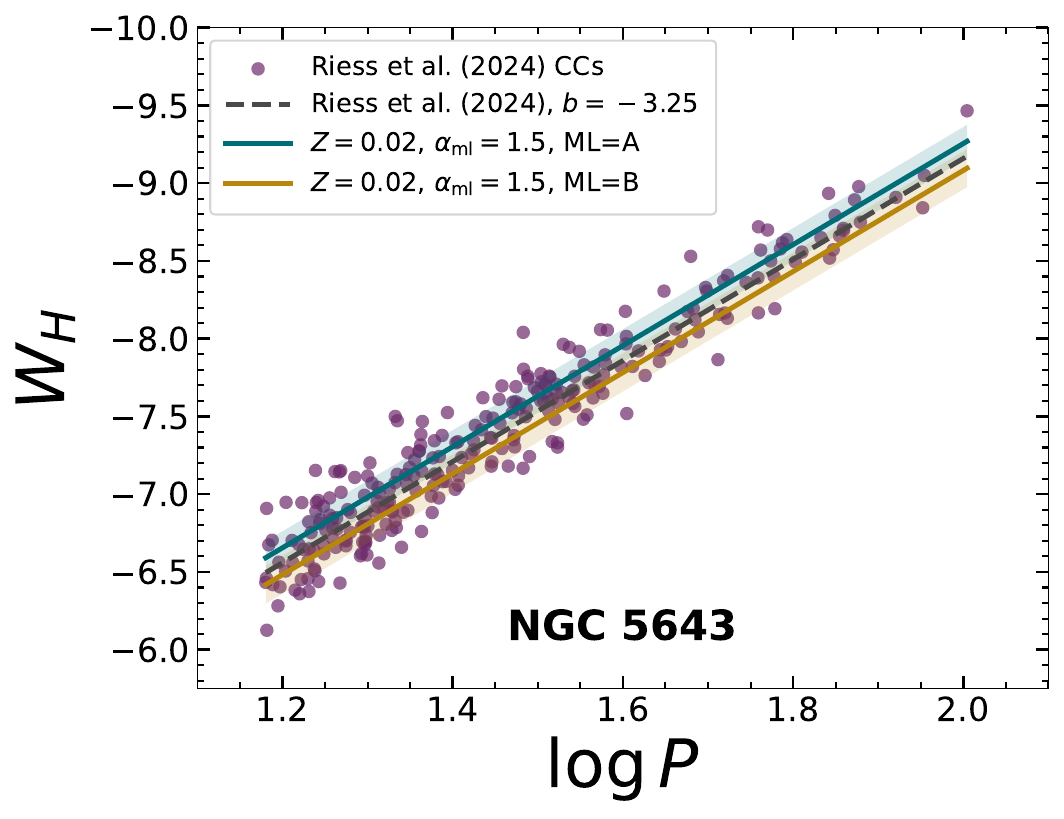}
\vspace{0.03cm}
\includegraphics[width=0.8\columnwidth]{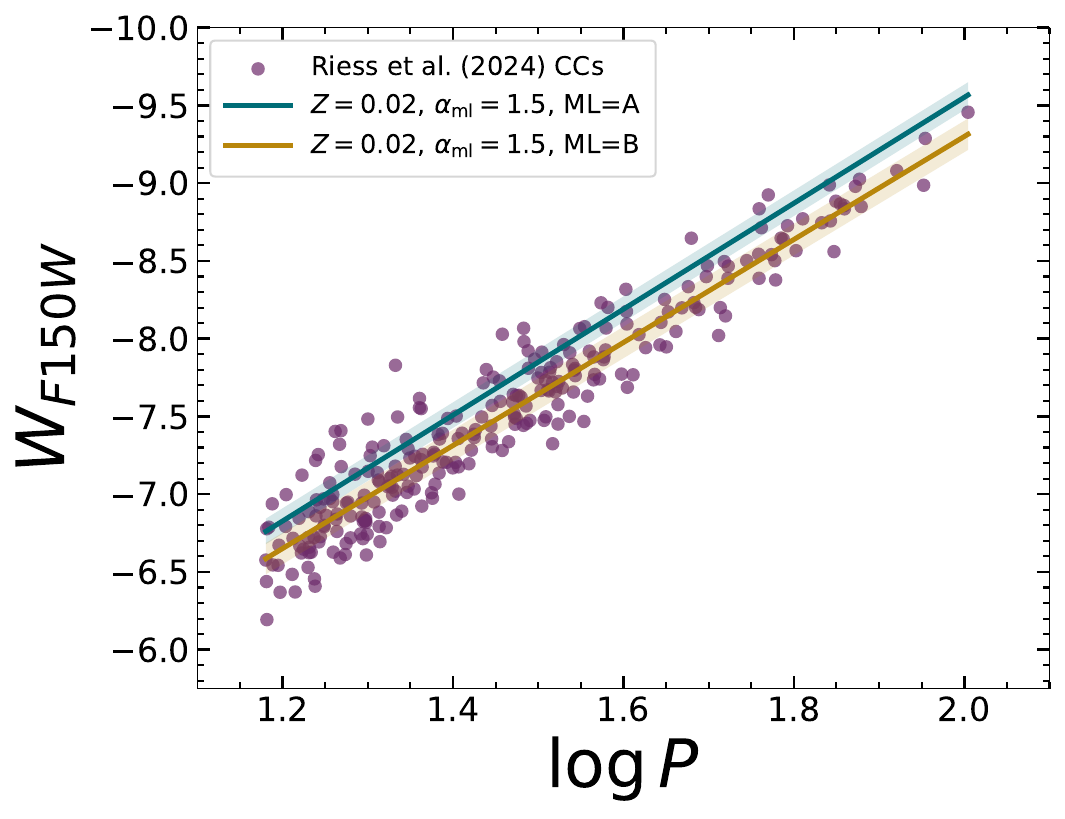}
\vspace{0.03cm}
\includegraphics[width=0.8\columnwidth]{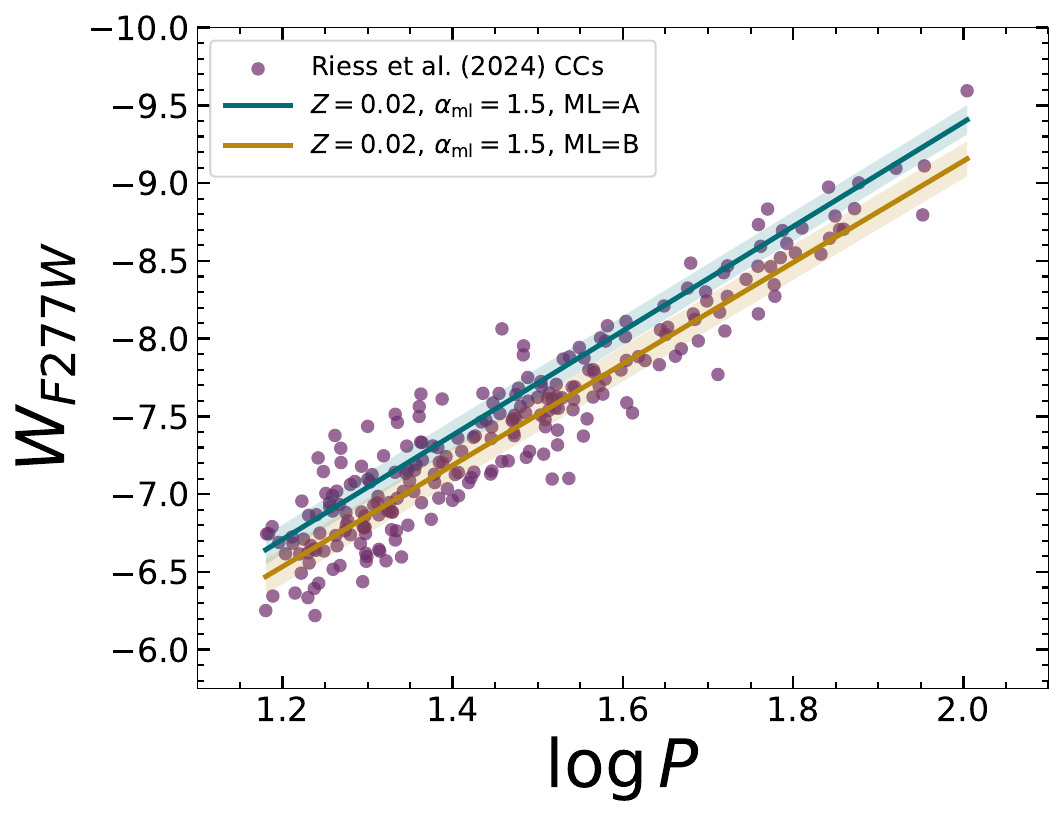}
\caption{Same as Figure~\ref{fig:ngc4258_pw}, but for NGC~5643.
}
\label{fig:ngc5643_pw}
\end{figure}

\begin{figure}[t]
\centering
\includegraphics[width=0.8\columnwidth]{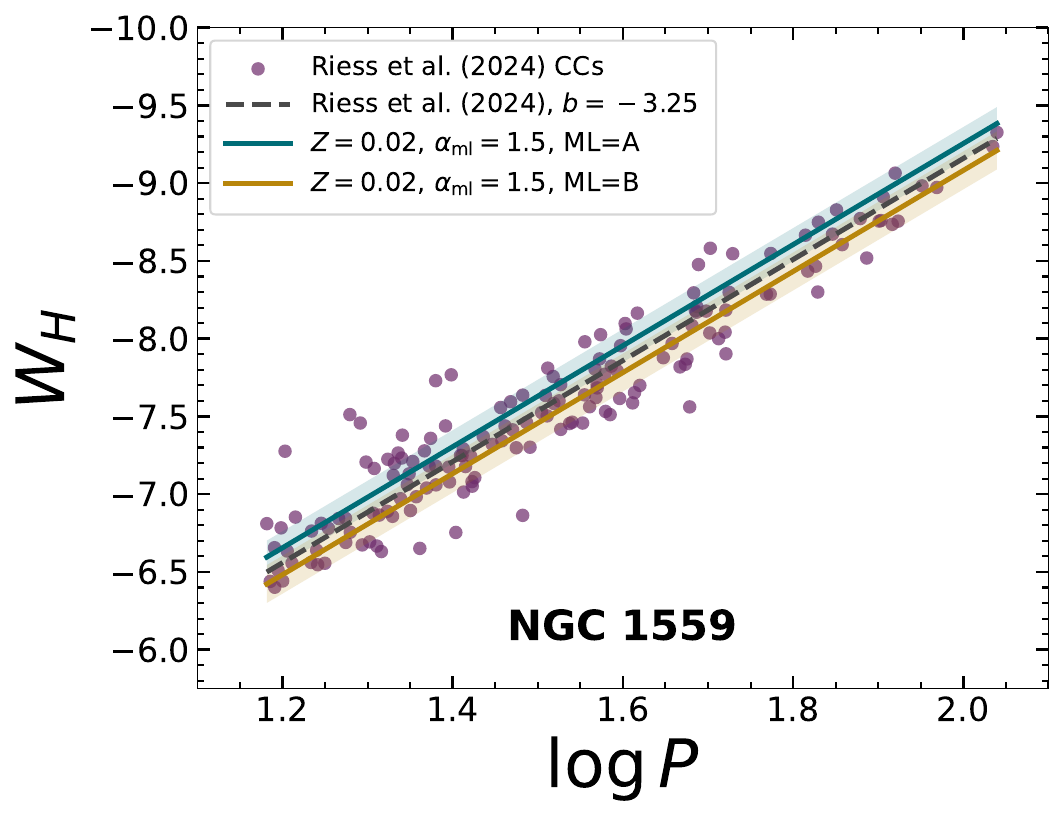}
\vspace{0.03cm}
\includegraphics[width=0.8\columnwidth]{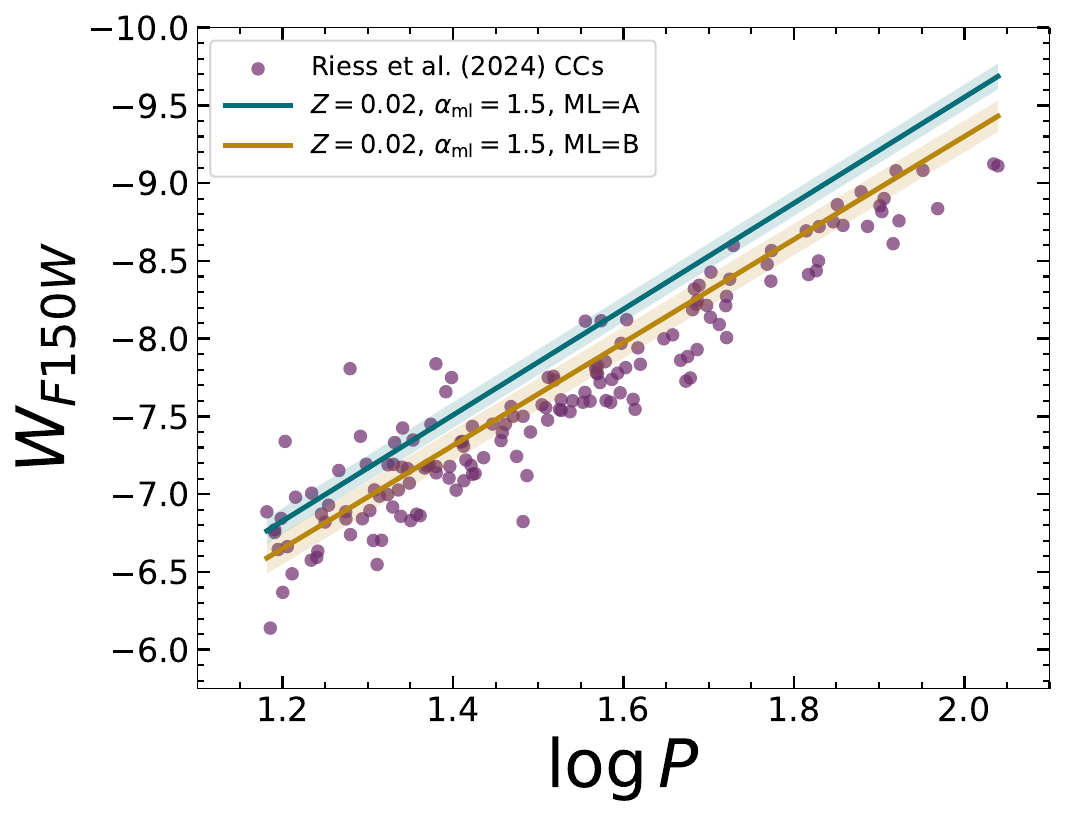}
\vspace{0.03cm}
\includegraphics[width=0.8\columnwidth]{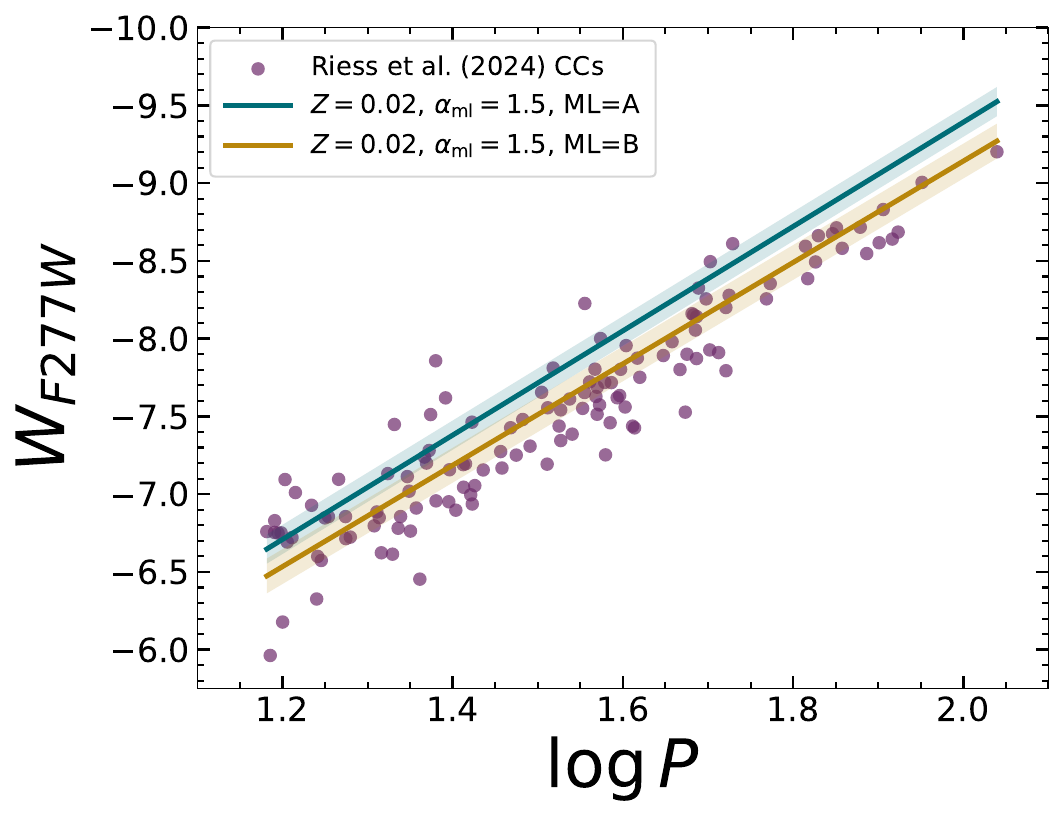}
\caption{Same as Figure~\ref{fig:ngc4258_pw}, but for NGC~1559.}
\label{fig:ngc1559_pw}
\end{figure}

\begin{figure}[t]
\centering
\includegraphics[width=0.8\columnwidth]{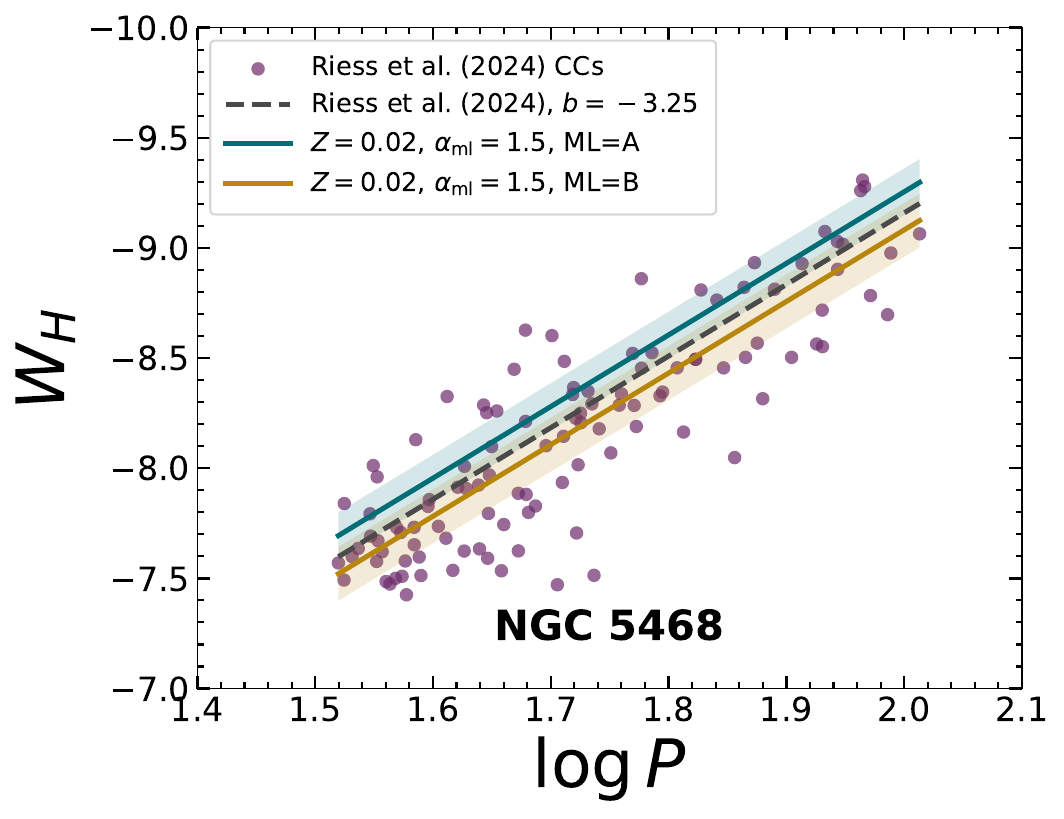}
\vspace{0.03cm}
\includegraphics[width=0.8\columnwidth]{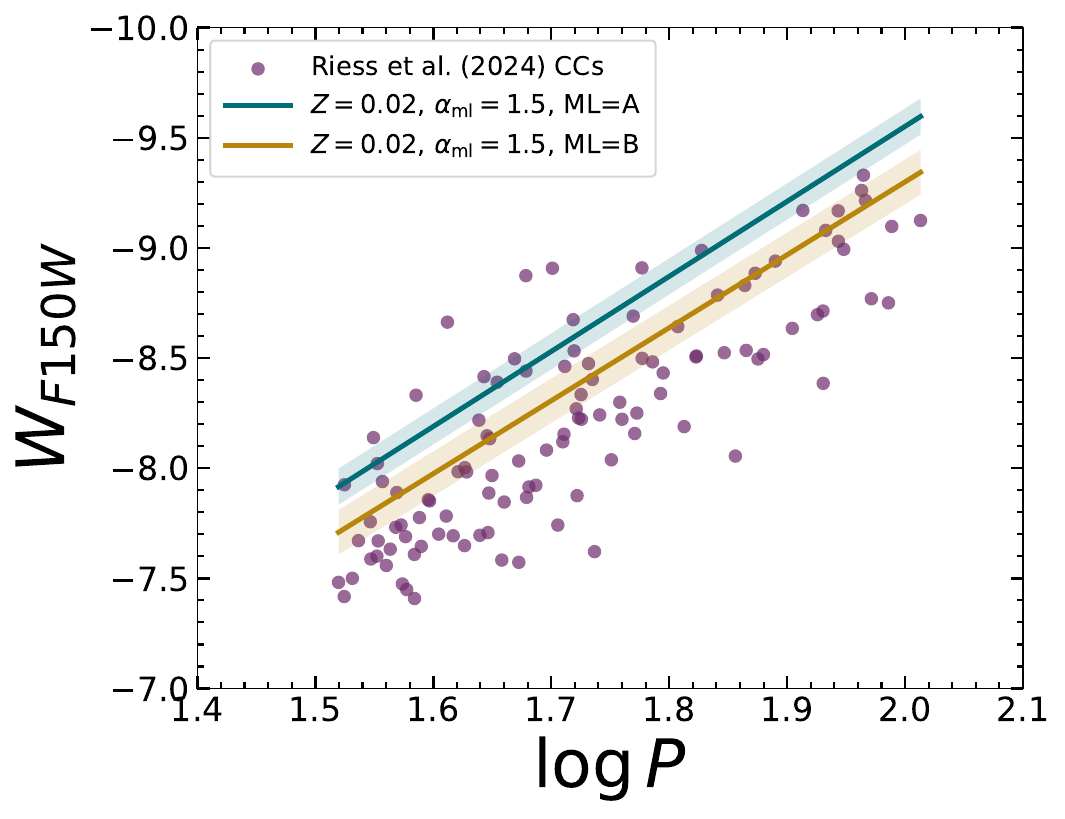}
\vspace{0.03cm}
\includegraphics[width=0.8\columnwidth]{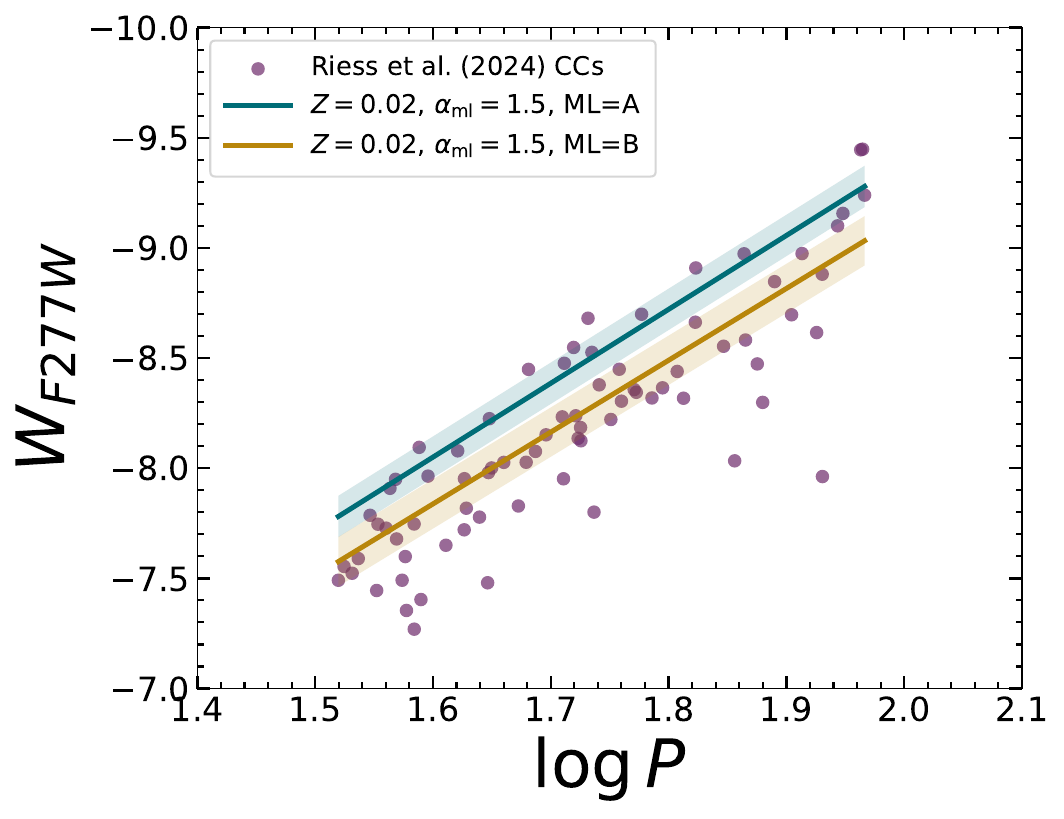}
\caption{Same as Figure~\ref{fig:ngc4258_pw}, but for NGC~5468.}
\label{fig:ngc5468_pw}
\end{figure}

\begin{figure}[t]
\centering
\includegraphics[width=0.8\columnwidth]{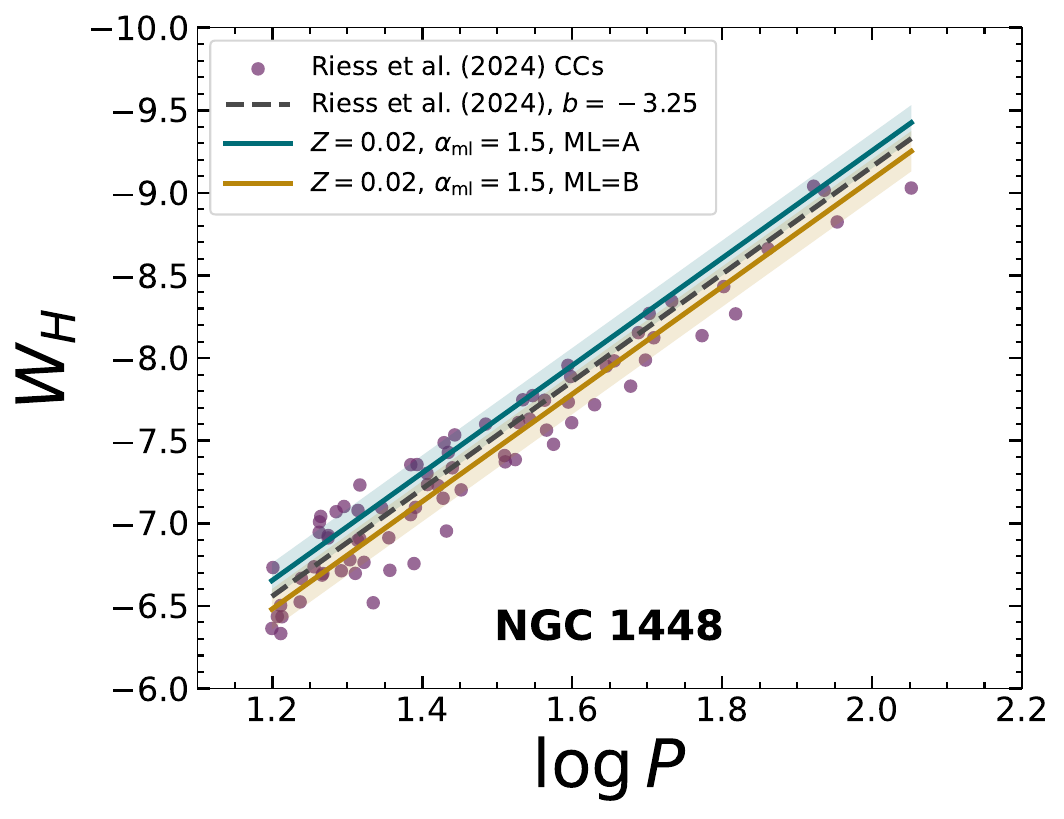}
\vspace{0.03cm}
\includegraphics[width=0.8\columnwidth]{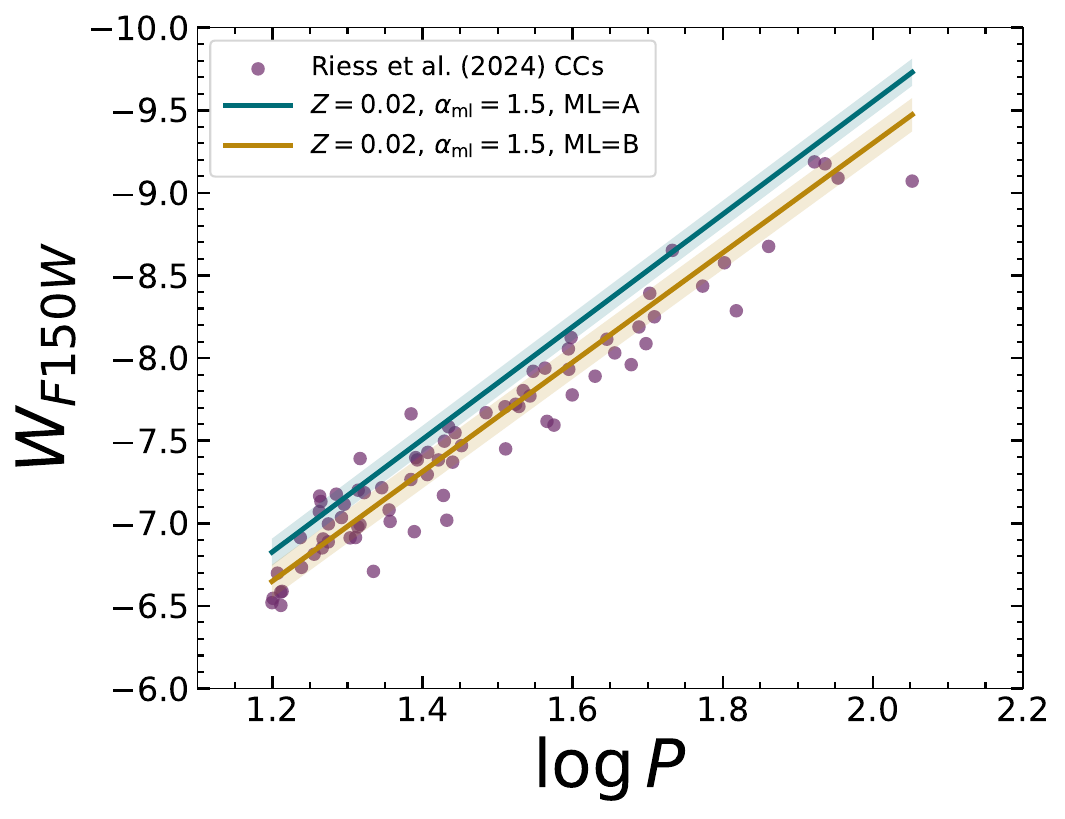}
\vspace{0.03cm}
\includegraphics[width=0.8\columnwidth]{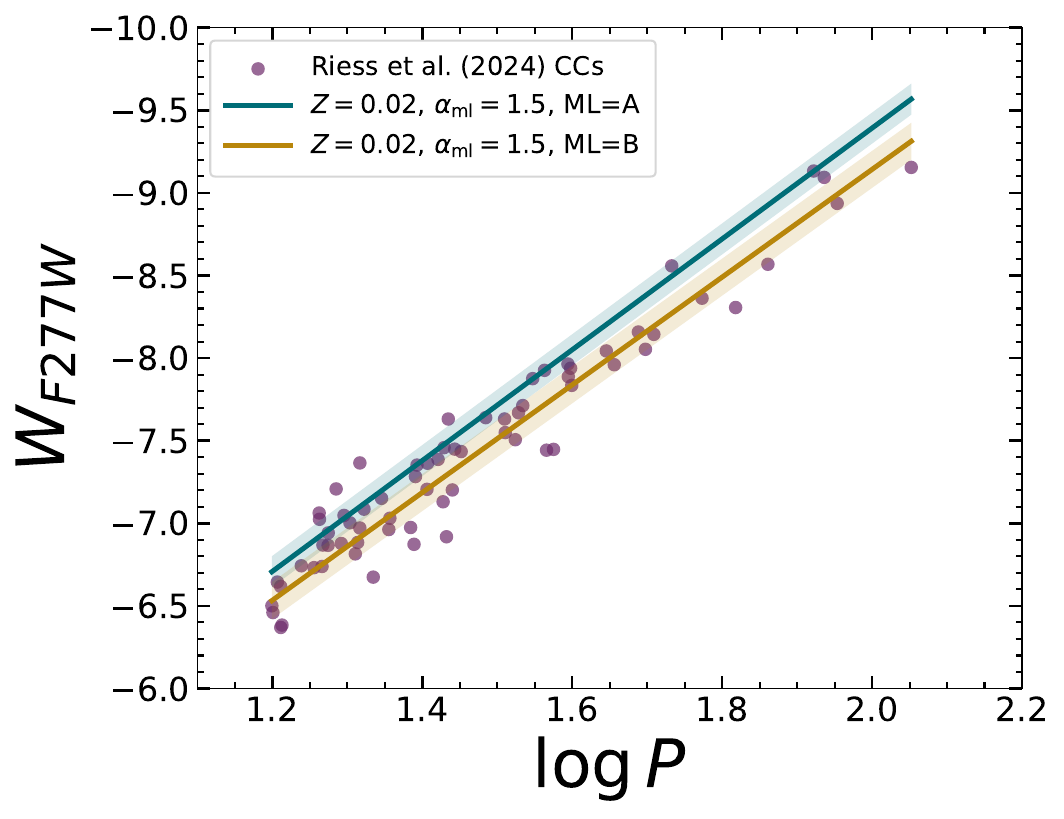}
\caption{Same as Figure~\ref{fig:ngc4258_pw}, but for NGC~1448.}
\label{fig:ngc1448_pw}
\end{figure}

\end{document}